\documentclass[aps,11pt,english,tightenlinesletterpaper,secnumarabic,nofootinbib,notitlepage]{revtex4-1}

\usepackage{graphicx} 	
\usepackage[justification=justified, singlelinecheck=off, compatibility=false]{caption}
\usepackage{subcaption}
\usepackage{amsmath}	
\usepackage{bm}		
\usepackage{amssymb}	
\usepackage{color} 		
\usepackage{times} 		
\usepackage{tensor} 	
\usepackage{comment}	
\usepackage{float}
\usepackage{hyperref} 
\hypersetup{
    colorlinks=true,       
    linkcolor=red,          
    citecolor=blue,        
    filecolor=magenta,      
    urlcolor=blue           
}


\usepackage{myterms}

\newcommand{\HP}{\text{\sc hp}}
\newcommand{\GKM}{\text{\sc gkm}}

\newcommand{\Higgs}{\Phi}
\newcommand{\Higgsp}{\Phi^{+}}

\newcommand{\Xa}{\hat{X}}
\newcommand{\Xb}{X}

\newcommand{\sw}{s_{w}}
\newcommand{\cw}{c_{w}}
\newcommand{\tw}{t_{w}}
\newcommand{\seps}{s_{\epsilon}}
\newcommand{\ceps}{c_{\epsilon}}
\newcommand{\teps}{t_{\epsilon}}
\newcommand{\sxi}{s_{\zeta}}
\newcommand{\cxi}{c_{\zeta}}
\newcommand{\txi}{t_{\zeta}}

\newcommand{\gbar}{\bar{g}}
\newcommand{\gp}{g^{\prime}}

\newcommand{\gX}{g_{\text{\sc x}}}
\newcommand{\gZ}{g_{\text{\sc z}}}

\newcommand{\qX}{q_{\text{\sc x}}}

\newcommand{\eps}{\epsilon}

\newcommand{\gPhipA}{g^{{\text{\tiny $\Higgs^{\text{\tiny +}}$}}}_{\text{\sc a}}}
\newcommand{\gPhipZ}{g^{{\text{\tiny $\Higgs^{\text{\tiny +}}$}}}_{\text{\sc z}}}
\newcommand{\gPhipX}{g^{{\text{\tiny $\Higgs^{\text{\tiny +}}$}}}_{\text{\sc x}}}

\newcommand{\gHA}{g^{{\text{\sc h}}}_{\text{\sc a}}}
\newcommand{\gHZ}{g^{{\text{\sc h}}}_{\text{\sc z}}}
\newcommand{\gHX}{g^{{\text{\sc h}}}_{\text{\sc x}}}

\newcommand{\gSA}{g^{{\text{\sc s}}}_{\text{\sc a}}}
\newcommand{\gSZ}{g^{{\text{\sc s}}}_{\text{\sc z}}}
\newcommand{\gSX}{g^{{\text{\sc s}}}_{\text{\sc x}}}

\newcommand{\yQ}{y_{\text{\sc q}}}
\newcommand{\yuR}{y_{u_R}}
\newcommand{\ydR}{y_{d_R}}
\newcommand{\yL}{y_{\text{\sc l}}}
\newcommand{\yeR}{y_{e_R}}
\newcommand{\yi}{y_{i}}

\newcommand{\giA}{g^{i}_{\text{\sc a}}}

\newcommand{\guLA}{g^{\tiny u_L}_{\text{\sc a}}}
\newcommand{\guLZ}{g^{\tiny u_L}_{\text{\sc z}}}
\newcommand{\guLX}{g^{\tiny u_L}_{\text{\sc x}}}
\newcommand{\guRA}{g^{\tiny u_R}_{\text{\sc a}}}
\newcommand{\guRZ}{g^{\tiny u_R}_{\text{\sc z}}}
\newcommand{\guRX}{g^{\tiny u_R}_{\text{\sc x}}}

\newcommand{\gdLA}{g^{\tiny d_L}_{\text{\sc a}}}
\newcommand{\gdLZ}{g^{\tiny d_L}_{\text{\sc z}}}
\newcommand{\gdLX}{g^{\tiny d_L}_{\text{\sc x}}}
\newcommand{\gdRA}{g^{\tiny d_R}_{\text{\sc a}}}
\newcommand{\gdRZ}{g^{\tiny d_R}_{\text{\sc z}}}
\newcommand{\gdRX}{g^{\tiny d_R}_{\text{\sc x}}}

\newcommand{\geLA}{g^{\tiny e_L}_{\text{\sc a}}}
\newcommand{\geLZ}{g^{\tiny e_L}_{\text{\sc z}}}
\newcommand{\geLX}{g^{\tiny e_L}_{\text{\sc x}}}
\newcommand{\geRA}{g^{\tiny e_R}_{\text{\sc a}}}
\newcommand{\geRZ}{g^{\tiny e_R}_{\text{\sc z}}}
\newcommand{\geRX}{g^{\tiny e_R}_{\text{\sc x}}}

\newcommand{\gnLA}{g^{\tiny \nu_L}_{\text{\sc a}}}
\newcommand{\gnLZ}{g^{\tiny \nu_L}_{\text{\sc z}}}
\newcommand{\gnLX}{g^{\tiny \nu_L}_{\text{\sc x}}}

\newcommand{\gdmA}{g^{\tiny \Psi}_{\text{\sc a}}}
\newcommand{\gdmZ}{g^{\tiny \Psi}_{\text{\sc z}}}
\newcommand{\gdmX}{g^{\tiny \Psi}_{\text{\sc x}}}

\newcommand{\thetauL}{\theta_{\tiny u_L}}
\newcommand{\thetauR}{\theta_{\tiny u_R}}

\newcommand{\thetanL}{\theta_{\tiny \nu_L}}

\newcommand{\gHstr}{g^{{\text{\sc h}}}_{\rm str}}
\newcommand{\gZstr}{g^{{\text{\sc z}}}_{\rm str}}

\newcommand{\Hpro}{{\rm h}}
\newcommand{\Spro}{{\rm s}}
\newcommand{\Zpro}{{\rm z}}
\newcommand{\Xpro}{{\rm x}}

\begin{document}

\title{Dark Strings and their Couplings to the Standard Model}

\date{\today}

\author{Jeffrey M. Hyde}
\email{jmhyde@asu.edu}
\author{Andrew J. Long}
\email{andrewjlong@asu.edu}
\author{Tanmay Vachaspati}
\email{tvachasp@asu.edu}
\affiliation{Physics Department, Arizona State University, Tempe, Arizona 85287, USA.}

\begin{abstract}
We consider the Standard Model extended by a hidden sector $\U{1}_X$ symmetry that is
spontaneously broken at the TeV scale by the vacuum expectation value of an additional
scalar field. We study ``dark string'' solutions in this model and their properties due
to the Higgs portal and gauge kinetic mixing operators. We find that dark strings effectively
interact with Higgs and $Z$ bosons by linear couplings, and with leptons and baryons via 
Aharonov-Bohm couplings, thus possibly leading to new cosmological constraints on dark 
matter models with spontaneously broken extra $\U{1}$ symmetry factors.

\vspace{1cm}

\end{abstract}

\maketitle

\setlength{\parindent}{20pt}

\setlength{\parskip}{-0.20cm}
\begingroup
\hypersetup{linkcolor=black}
\tableofcontents
\endgroup
\setlength{\parskip}{0.2cm}

\section{Introduction}\label{sec:Introduction}
\numberwithin{equation}{section}

Many compelling extensions of the Standard Model (SM) require additional gauged $\U{1}_X$
factors that are spontaneously broken giving rise to massive vector bosons.  
The high energy physics community has been studying the phenomenology of these models for years while collider experiments have been searching for the so-called $Z^{\prime}$ at energies up to $O({\rm TeV})$ (see, \eg, the reviews \cite{Leike:1998wr, Langacker:2008yv}).  
Similar models have recently attracted attention in the dark matter community as well.  
In this context it is commonly assumed that the fields that transform under the SM gauge 
group are singlets under the $\U{1}_X$ and vice versa.  
Such a theory decomposes into a visible sector (SM fields) and a hidden or dark sector (fields charged under $\U{1}_X$).  
The massive vector boson may either play the role of dark matter itself \cite{Kanemura:2010sh, Lebedev:2011iq, Farzan:2012hh, Baek:2012se}\footnote{If the gauge symmetry is non-Abelian the massive vector may still be the dark matter \cite{Hambye:2008bq}, but a topologically stable cosmic string solution is not guaranteed to exist.  }, or it may act as a mediator between the visible and the dark sectors  \cite{ArkaniHamed:2008qp, Cassel:2009pu, Chun:2010ve, Chu:2011be}\footnote{If the $\U{1}_X$ is unbroken, the massless force carrier is known as a dark photon \cite{Dobrescu:2004wz, Ackerman:2008gi}. In this case the model has no string solution. }.  
In these types of models, the breaking of the $\U{1}_X$ during a cosmological phase transition is invariably accompanied by the formation of a unique kind of cosmic string, known as a ``dark string'' \cite{Vachaspati:2009jx}.  

The presence of these dark strings in our universe has largely been overlooked.  
The tension, which is on the order of the symmetry breaking scale $\mu \sim \TeV^2$, is far too small for dark strings to have any detectable gravitational effect on the cosmic microwave background \cite{Ade:2013xla} or pulsar timing \cite{Blanco-Pillado:2013qja}, which typically provide the strongest constraints on GUT-scale strings \cite{Battye:2010xz} .  
However, as we will see below, the fields that compose the dark sting have very specific couplings to the SM fields, and therefore they are able to radiate and scatter on SM particles.  
The presence of these cosmic dark strings in our universe can, therefore, have observable consequences and yield constraints on model building that are as yet largely unexplored.

The structure of dark strings was first studied in Refs.~\cite{Vachaspati:2009jx}~and~\cite{Brihaye:2009fs}.  
Our analysis expands upon that work in a number of ways:  (i) we retain the complete electroweak gauge sector, specifically, we do not work in the semilocal limit $\sin^2 \theta_w = 1$, where
$\theta_w$ is the weak mixing angle as in \cite{Brihaye:2009fs}; 
(ii) we restrict the parameter space using the measured value of the Higgs boson mass $M_H \approx 125 \GeV$ \cite{ATLAS:2012gk, CMS:2012gu}, which had not been discovered at the time of the previous work; (iii) we include the interaction between the Higgs field and the scalar field responsible for the formation of the string [see \eref{eq:LHP}]; (iv) we do not necessarily assume that the gauge-kinetic mixing is small ($\sin \epsilon \ll 1$; see below); and (v) we calculate, for the first time, the effective couplings of the dark string to the SM fields.  
Understanding the structure of the dark string and its couplings to SM fields, in particular, are important in evaluating the cosmological signatures of dark strings.  

Our analysis will focus on the smallest extension of the SM that contains a spontaneously broken, gauged Abelian symmetry.  
Specifically, we introduce a complex scalar field $S$ charged under the Abelian symmetry group $\U{1}_X$, which has $\hat{X}^{\mu}$ as its vector potential; collectively, these fields will be referred to as the dark sector.  
After $S$ acquires a vacuum expectation value, the mass for $\hat{X}^{\mu}$ is generated.  
This model is particularly interesting because the symmetries forbid all but two renormalizable, tree-level interactions between the SM and hidden sector fields.  
These are the Higgs portal (HP) operator \cite{Patt:2006fw}
\begin{align}\label{eq:LHP}
	\mathcal{L}_{\HP} = - \alpha \Higgs^{\dagger} \Higgs S^{\ast} S
\end{align}
where $\Phi$ is the SM Higgs doublet, and the gauge kinetic mixing (GKM) 
operator \cite{Holdom:1985ag, Foot:1991kb, Babu:1997st}
\begin{align}\label{eq:LGKM}
	\mathcal{L}_{\GKM} = - \frac{\sin \epsilon}{2} \hat{X}_{\mu \nu} Y^{\mu \nu} 
\end{align}
where $\hat{X}_{\mu \nu}$ and $Y_{\mu \nu}$ are the field strength tensors for the $\U{1}_X$ and $\U{1}_Y$ hypercharge.  
Vacuum stability considerations bound $\abs{\alpha}$ from above [see \sref{sub:ScalarSector}], and the avoidance of ghosts requires $\abs{\sin \epsilon} < 1$.  
For the sake of generality, we will study this model with $\alpha, \epsilon \neq 0$.  
However, note that in this case neither the $S$ nor the $\hat{X}^{\mu}$ field is stable.  
The model must be extended if it is to include a stable dark matter candidate\footnote{After 
electroweak symmetry breaking, $S$ mixes with the Higgs and thereby acquires 
all of its interactions with the SM fields, which opens new decay channels.  Similarly, $\hat{X}^{\mu}$ 
mixes with the Z-boson.  If the mass scale in the hidden sector is very low or the couplings very 
small, then the dark matter can be metastable.  Such models will also contain dark strings.  In this paper, however, we will focus on strings with energy scales higher than the electroweak scale and arbitrary couplings.  
}.
 Alternatively, imposing a discrete (reflectional) symmetry on $\hat{X}^{\mu}$ enforces 
 $\sin \epsilon = 0$ \cite{Kanemura:2010sh, Lebedev:2011iq, Farzan:2012hh, Baek:2012se}.  

The interaction in $\Lcal_{\HP}$ gives rise to a mixing between the Higgs and the singlet scalar, and therefore it is constrained in light of the Higgs discovery \cite{Carmi:2012yp}.  
The interaction in $\Lcal_{\GKM}$ is tightly constrained by a number of observables at low energies giving roughly (see \cite{Hook:2010tw} for a review)
\begin{align}\label{eq:epsbound}
	\abs{\sin \epsilon} < O(10^{-3})
	\qquad {\rm for} \qquad 
	M_X \lesssim \TeV \per
\end{align}
However, it is important to recognize that the model is yet unconstrained if the masses are large, $M_S \sim M_X > O({\rm TeV})$, where laboratory tests have not yet explored.  
For the sake of generality, we will not make any {\it a priori} assumptions about the scale of symmetry breaking in our analytic analysis, and in our numerical analysis we will focus on $M_S \sim M_X > M_H$  allowing $\sin \epsilon = O(1)$.  
The small $\epsilon$ expansion of various parameters may be found in \aref{app:Smallepsilon}.

After setting up the model in \sref{sec:Model}, we diagonalize the gauge sector and derive the equations of motion relevant for a string. 
In \sref{sec:Properties} we find the dark string solution and calculate the effective couplings of the string to the SM fields in terms of the Higgs portal and gauge kinetic mixing parameters.  
The SM Higgs interacts with the string and thus we also take into account the possibility that 
it winds around the string. However the lightest string is obtained when only the dark scalar field winds and so we focus on more detailed properties of these strings, especially their three
types of  interactions with SM particles.
Fermions of the SM can have Aharonov-Bohm couplings to the dark string if there is 
gauge-kinetic mixing between the hypercharge and dark $\U{1}$'s.
The SM Higgs can have a non-trivial interaction in the presence of a ``Higgs portal'' coupling -- 
a quartic interaction between the Higgs and the dark scalar field.  
The $Z$ gauge field also has a non-trivial profile on the string because of the gauge-kinetic 
mixing. Each of these interactions is potentially relevant to the cosmological evolution
of the dark string network. 
We summarize our findings in \sref{sec:Conclusion}.  
\aref{app:Smallepsilon} contains a list of variables, defined in the main body of the text, which 
have been expanded in the limit that the GKM coupling is small, \ie, $\sin \epsilon \ll 1$.

\section{The Dark String Model}\label{sec:Model}
\numberwithin{equation}{subsection}

In this section we introduce the model.  
We focus on the gauge sector first and the role of the GKM operator, and then we turn to the scalar sector and the HP operator.  
In the third subsection we derive the string equations and discuss the boundary conditions.  

\subsection{Gauge Sector}\label{sub:GaugeSector}

We consider an extension of the SM electroweak sector that adds a complex scalar field $S(x)$ charged under a new gauge group, $\U{1}_X$, that has ${\Xa}^{\mu}(x)$ as its vector potential.  
In general, one can only write two renormalizable interactions between the SM and the dark sector:  the Higgs portal operator, 
$\Higgs^{\dagger} \Higgs S^{\ast} S$, and the gauge kinetic mixing operator, ${\Xa}_{\mu \nu} Y^{\mu \nu}$.  
The Lagrangian that defines this model is 
\begin{align}\label{eq:L1}
	\mathcal{L} 
	= \abs{ D_{\mu} \Higgs }^2
	+ \abs{ D_{\mu} S }^2
	- U(\Higgs,S) 
	- \frac{1}{4} \sum_{a=1,2,3} \bigl( W_{\mu \nu}^{a} \bigr)^2
	- \frac{1}{4} \bigl( Y_{\mu \nu} \bigr)^2
	- \frac{1}{4} \bigl( {\Xa}_{\mu \nu} \bigr)^2
	- \frac{\seps}{2} {\Xa}_{\mu \nu} Y^{\mu \nu} 
\end{align}
plus the remaining terms in the SM Lagrangian, which are unmodified and not written explicitly here.  
The parameter $\seps \equiv \sin \eps$ with $-\pi/2 \leq \eps \leq \pi/2$ controls the strength of the gauge kinetic mixing.  
The covariant derivatives are given by 
\begin{align}\label{eq:covar_deriv1}
\begin{array}{l}
	D_{\mu} \Higgs = \bigl( \partial_{\mu} - i \frac{g}{2} \, \sigma^{a} W_{\mu}^a - i \frac{g^{\prime}}{2} Y_{\mu} \bigr) \Higgs \\
	D_{\mu} S = \bigl( \partial_{\mu} - i \frac{\gX}{2} {\Xa}_{\mu} \bigr) S
\end{array} 
\end{align}
where $\Higgs = (\Higgsp \, , \, H)^T$ is the Higgs doublet.  
The scalar potential is
\begin{align}\label{eq:U1}
	U(\Higgs , S) = \lambda \bigl( \Higgs^{\dagger} \Higgs - \eta^2 \bigr)^2 + \kappa \bigl( S^{\ast} S - \sigma^2 \bigr)^2 + \alpha \bigl( \Higgs^{\dagger} \Higgs - \eta^2 \bigr) \bigl( S^{\ast} S - \sigma^2 \bigr) \com
\end{align}
and the parameter $\alpha$ is called the ``Higgs portal coupling'' as it is the gateway
for interactions between the SM and dark sectors.  
This potential induces the vacuum expectation values
\begin{align}\label{eq:VEVs}
	\langle \Higgs \rangle = ( 0 \, , \, \eta )^T
	\qquad \text{and} \qquad 
	\langle S \rangle = \sigma
\end{align}
with $\eta = v / \sqrt{2} \approx 174 \GeV$ (see \sref{sub:ScalarSector} for an extended discussion of the vacuum structure).  
The parameter $\lambda$ can be exchanged for the Higgs boson mass, and we are left with five free parameters: $\alpha$, $\kappa$, $\sigma$, $\gX$, and $\seps$.  

The Lagrangian \eref{eq:L1} gives rise to the following field equations:
\begin{align}\label{eq:fieldeqns1}  
\begin{array}{l}
	\left( D_{\nu} W^{\nu \mu} \right)^a = \frac{1}{2} g J_{\Higgs}^{a \, \mu} \\
	\partial_{\nu} Y^{\nu \mu} - \seps \, \partial_{\nu} {\Xa}^{\nu \mu} = \frac{1}{2} g^{\prime} J_{\Higgs}^{\mu} \\
	\partial_{\nu} {\Xa}^{\nu \mu} - \seps \, \partial_{\nu} Y^{\nu \mu} = \frac{1}{2} \gX J_{S}^{\mu} \\
	D_{\mu} D^{\mu} \Higgs = - 2 \lambda \left( \Higgs^{\dagger} \Higgs - \eta^2 \right) \Higgs - \alpha \left( S^{\ast} S - \sigma^2 \right) \Higgs \\
	D_{\mu} D^{\mu} S = - 2 \kappa \left( S^{\ast} S - \sigma^2 \right) S - \alpha \left( \Higgs^{\dagger} \Higgs - \eta^2 \right) S
\end{array}
\end{align}
where the currents are defined as 
\begin{align}\label{eq:Jdef}
\begin{array}{l}
	J_{\Higgs}^{a \, \mu} \equiv i \left( \left( D^{\mu} \Higgs \right)^{\dagger} \sigma^{a} \Higgs - \Higgs^{\dagger} \sigma^{a} D^{\mu} \Higgs \right) \\
	J_{\Higgs}^{\mu} \equiv i \left( \left( D^{\mu} \Higgs \right)^{\dagger} \Higgs - \Higgs^{\dagger} D^{\mu} \Higgs \right) \\
	J_{S}^{\mu} \equiv i \left( S D^{\mu} S^{\ast} - S^{\ast} D^{\mu} S \right)
\end{array}
\end{align}
and $\left( D_{\nu} W^{\mu \nu} \right)^a \equiv \partial_{\nu} W^{a \, \mu \nu} + g \epsilon^{abc} W_{\nu}^b W^{c \, \mu \nu}$.  
The presence of the $O(\seps)$ terms in \eref{eq:fieldeqns1} implies that both gauge fields $Y^{\mu}$ and ${\Xa}^{\mu}$ are sourced when either $J_{\Higgs}^{\mu}$ or $J_{S}^{\mu}$ is nonzero.  

It will be convenient to move to a basis in which the GKM term is absent from the Lagrangian.  
This could be accomplished by merely rotating between the $\U{1}$ gauge fields, $Y^{\mu}$ and ${\Xa}^{\mu}$, as was done in previous studies of the dark string \cite{Brihaye:2009fs, Vachaspati:2009jx}.  
However, in order to connect with the low energy observables, we would like to choose the basis that coincides with the mass eigenstates after electroweak symmetry breaking.  
In order to identify the appropriate basis, we insert the vacuum expectation values \eref{eq:VEVs} into the Lagrangian \eref{eq:L1} to obtain
\begin{align}\label{eq:L2}
	\mathcal{L} \bigr|_{\rm vevs} = & \ 
	m_W^2 \left| \frac{W^{1}_{\mu} - i W^{2}_{\mu}}{\sqrt{2}} \right|^2
	+ \frac{1}{2} m_Z^2 \bigl( \cw W_{\mu}^{3} - \sw Y_{\mu} \bigr)^2 
	+ \frac{1}{2} m_{X}^2 \bigl( {\Xa}_{\mu} \bigr)^2 \nn 
	&- \frac{1}{4} \sum_{a=1,2,3} W_{\mu \nu}^{a} W^{a \, \mu \nu} 
	- \frac{1}{4} Y_{\mu \nu} Y^{\mu \nu} 
	- \frac{1}{4} {\Xa}_{\mu \nu} {\Xa}^{\mu \nu} 
	- \frac{\seps}{2} \, {\Xa}_{\mu \nu} Y^{\mu \nu} 
\end{align}
where 
\begin{align}\label{eq:m_defs}
	m_W \equiv \frac{g \eta}{\sqrt{2}}
	\qquad , \qquad
	m_Z \equiv \frac{\gbar \eta}{\sqrt{2}}
	\qquad , \qquad 
	m_X \equiv \frac{\gX \sigma}{\sqrt{2}} \com
\end{align}
and the weak mixing angle is defined as usual: $\sw \equiv \sin \theta_w = g^{\prime} / \gbar$ and $\cw \equiv \cos \theta_w = g / \gbar$ with $\gbar \equiv \sqrt{g^2 + g^{\prime \, 2}}$.  
Both the kinetic and the mass terms of the Lagrangian, \eref{eq:L2}, can be diagonalized by the transformation 
\begin{align}\label{eq:field_trans}
	\begin{pmatrix} W^{1}_{\mu} \\ W^{2}_{\mu} \end{pmatrix} = \begin{pmatrix} \frac{1}{\sqrt{2}} & \frac{1}{\sqrt{2}} \\ \frac{i}{\sqrt{2}} & -\frac{i}{\sqrt{2}} \end{pmatrix} \begin{pmatrix} W^{+}_{\mu} \\ W^{-}_{\mu} \end{pmatrix}
	\qquad \text{and} \qquad 
	\begin{pmatrix} Y_{\mu} \\ W^3_{\mu} \\ {\Xa}_{\mu} \end{pmatrix} = {\bf M} \begin{pmatrix} A_{\mu} \\ Z_{\mu} \\ {\Xb}_{\mu} \end{pmatrix}
\end{align}
where
\begin{align}\label{eq:Mdef}
{\bf M} = \begin{pmatrix} 	\cw & - \sw \cxi - \teps \sxi  & \sw \sxi - \teps \cxi \\
	\sw & \cw \cxi & - \cw \sxi \\
	0 & \sxi / c_{\epsilon} & \cxi / c_{\epsilon}
\end{pmatrix} \per
\end{align}
We continue to use the shorthand $s_{\theta} = \sin \theta$, $c_{\theta} = \cos \theta$, and $t_{\theta} = \tan \theta$ for $\theta = \epsilon, \zeta$.  
The angle $\zeta$ falls in the range $-\pi/4 < \zeta < \pi/4$, and its value is given by 
\begin{align}\label{eq:def_tan2xi}
	\tan 2 \zeta &= \frac{-2 \sw \seps c_{\epsilon} }{(R^2 -1) + \seps^2 (1 + \sw^2)} \per
\end{align}
Here we have defined $R \equiv m_X / m_Z$, and we will assume $R > 1$.  
Note that ${\bf M}$ consists of a rotation and a rescaling, otherwise known as a principal axis transformation.  

After performing the transformation in \eref{eq:field_trans}, the full Lagrangian becomes
\begin{align}\label{eq:L3}
	\mathcal{L} 
	=& \abs{ D_{\mu} \Higgs }^2
	+ \abs{ D_{\mu} S }^2
	- U(\Higgs,S) 
	- \frac{1}{2} W_{\mu \nu}^{-} W^{+ \, \mu \nu}  \nn
	&- \frac{1}{4} A_{\mu \nu} A^{\mu \nu} 
	- \frac{1}{4} Z_{\mu \nu} Z^{\mu \nu} 
	- \frac{1}{4} {\Xb}_{\mu \nu} {\Xb}^{\mu \nu} 
	+ \mathcal{L}_{\rm int} 
\end{align}
where we have written each of the field strength tensors in the form $K_{\mu \nu} = \partial_{\mu} K_{\nu} - \partial_{\nu} K_{\mu}$ for $K = W^-, W^+, A, Z,$ and ${\Xb}$.  
The term $\mathcal{L}_{\rm int}$ corresponds to interactions among the gauge fields, which are at least second order in $W^{\pm}$.  
As we discuss below, we can consistently set $W^{\pm} = 0$ for our dark string analysis and neglect these terms.  
The scalar field covariant derivatives now become
\begin{align}\label{eq:covar_deriv2}
	D_{\mu} \Higgs 
	& = \begin{pmatrix} 
	D_{\mu} \Higgsp - i \frac{g}{\sqrt{2}} W^{+}_{\mu} H \\
	D_{\mu} H - i \frac{g}{\sqrt{2}} W^{-}_{\mu} \Higgsp 
	\end{pmatrix} \nn
	D_{\mu} S 
	&= \Bigl( \partial_{\mu} - i ( \gSA A_{\mu}  + \gSZ Z_{\mu}  + \gSX {\Xb}_{\mu} ) \Bigr) S
\end{align}
where we have defined 
\begin{align}\label{eq:covar_deriv3}
\begin{array}{l}
	D_{\mu} \Higgsp \equiv \left( \partial_{\mu} - i ( \gPhipA A_{\mu} + \gPhipZ Z_{\mu} + \gPhipX {\Xb}_{\mu} ) \right) \Higgsp \\
	D_{\mu} H \equiv \left( \partial_{\mu} - i (\gHA A_{\mu} + \gHZ Z_{\mu} + \gHX {\Xb}_{\mu} ) \right) H
\end{array} \per
\end{align}
The couplings are found to be
\begin{align}\label{eq:scalar_gauge_couplings}
\begin{array}{lclcl}
	\gPhipA = e 
	& \qquad & 
	\gPhipZ = \cxi \frac{e}{2} \bigl( \frac{1}{\tw} - \tw \bigr) - \sxi \frac{e}{2} \frac{\teps}{\cw} 
	& \qquad & 
	\gPhipX = - \cxi \frac{e}{2} \frac{\teps}{\cw} - \sxi \frac{e}{2} \bigl( \frac{1}{\tw} - \tw \bigr) \\
	\gHA = 0
	& \qquad & 
	\gHZ = -\cxi \frac{e}{2} \frac{1}{\sw \cw} - \sxi \frac{e}{2} \frac{\teps}{\cw}
	& \qquad & 
	\gHX = - \cxi \frac{e}{2} \frac{\teps}{\cw} +  \sxi \frac{e}{2}\frac{1}{\sw \cw}  \\
	\gSA = 0 
	& \qquad &
	\gSZ = \sxi \frac{\gX}{2} \frac{1}{c_{\epsilon}} 
	& \qquad &
	\gSX = \cxi \frac{\gX}{2} \frac{1}{c_{\epsilon}}
\end{array}
\end{align}
where $e = g \, \sw = \gp \cw = \gbar \sw \cw$ is the electromagnetic coupling constant.  

Now one can see the consequences of the GKM operator.  
As reflected in the nonzero couplings $\gPhipX$, $\gHX$, and $\gSZ$, the Higgs acquires an interaction with the mass eigenstate ${\Xb}$ boson, and similarly the $S$ interacts with the $Z$ boson.  
However, the vanishing of $\gHA$ and $\gSA$ implies that the GKM does not induce a coupling between the photon and the electromagnetically neutral scalars; this is a consequence of the residual electromagnetic gauge invariance.  

\begin{figure}[t]
\vspace{0.2in}
	\includegraphics[width=0.48\textwidth]{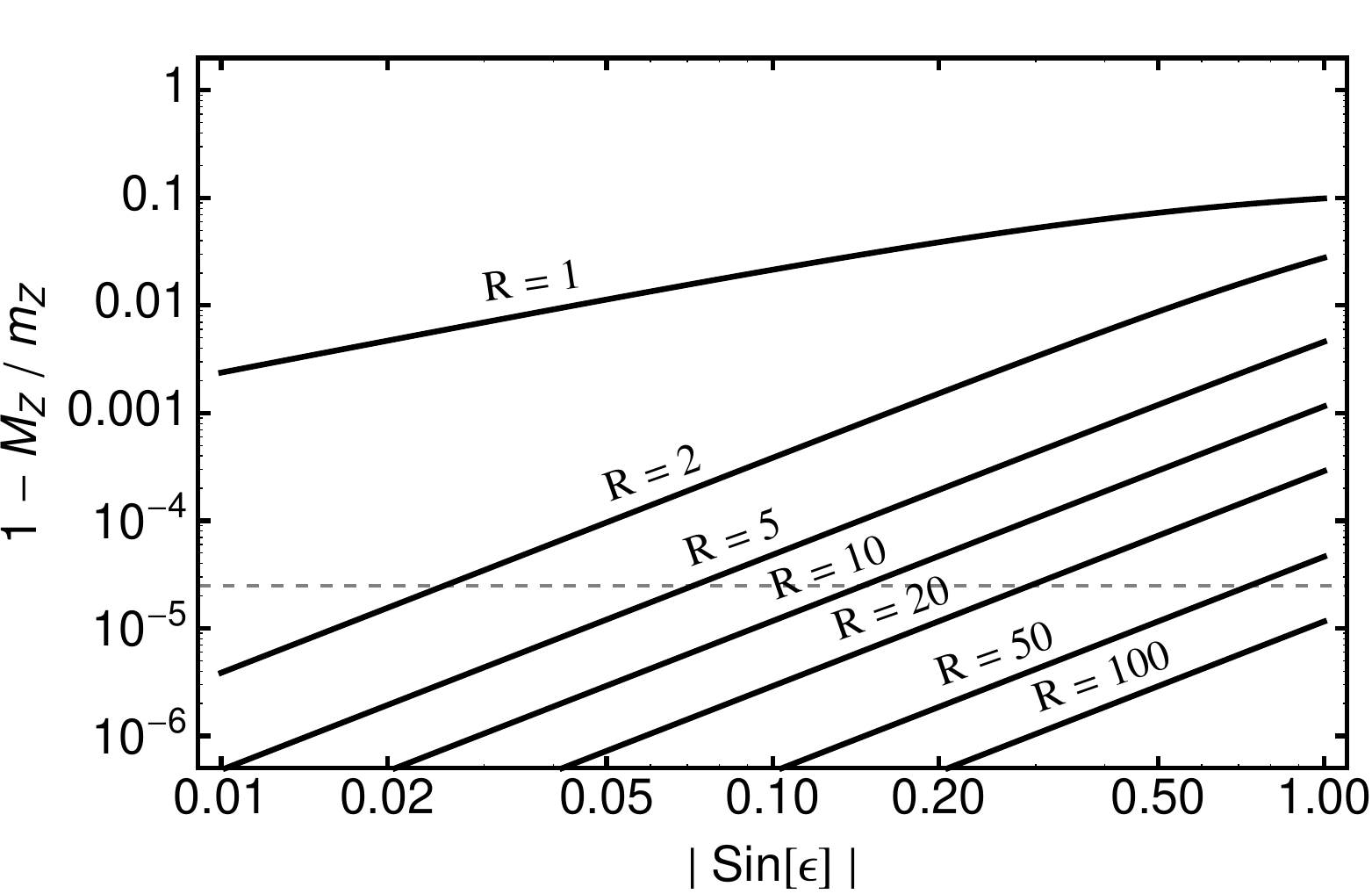} \hfill
	\includegraphics[width=0.48\textwidth]{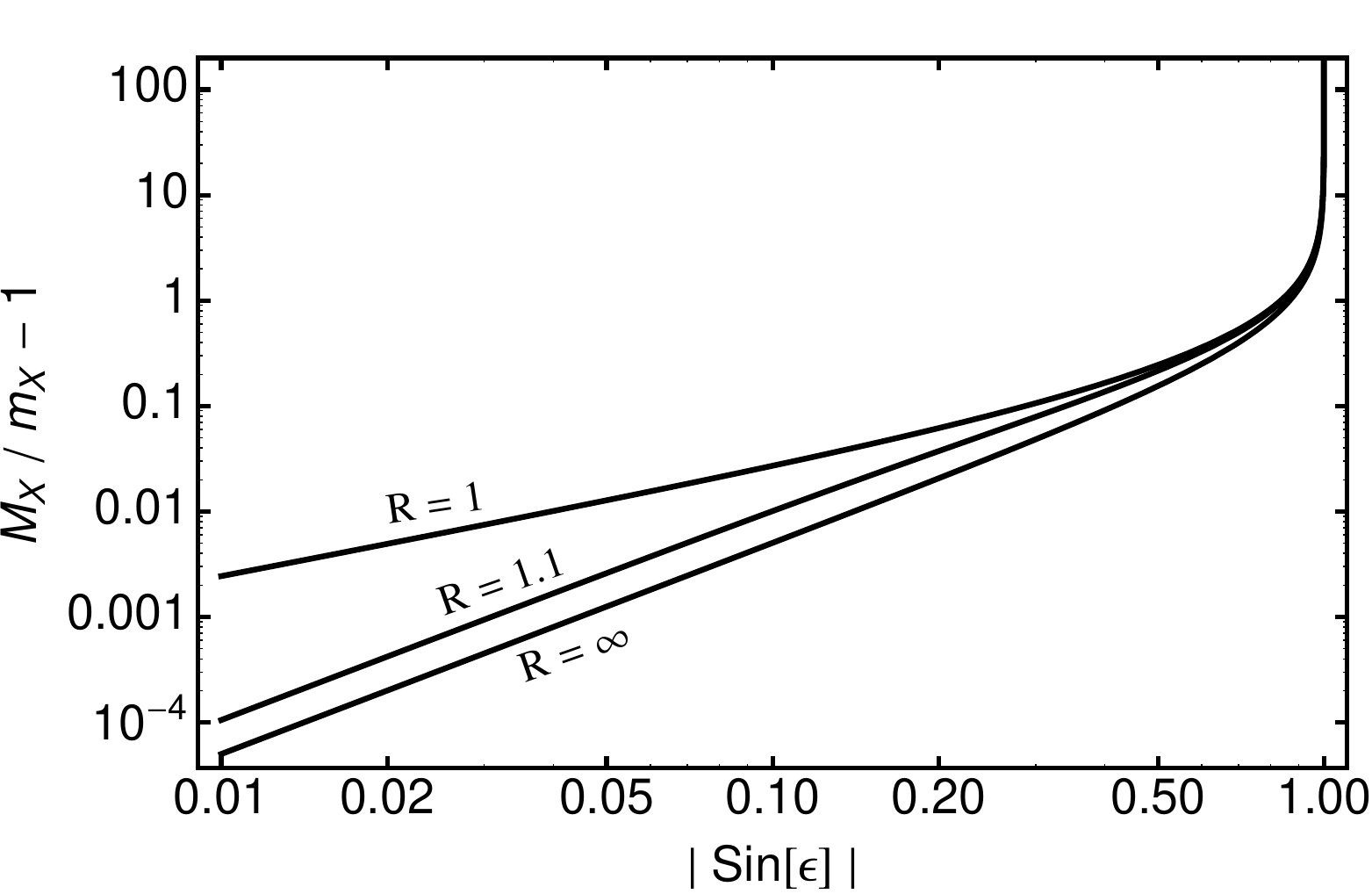} 
	\caption{\label{fig:gaugespec}
	The spectrum of gauge bosons, given by \eref{eq:gauge_evals}, for various values of $\sin \epsilon$ and $R = m_X / m_Z$.  We have fixed $g = 0.654$, $g^{\prime} = 0.359$, and $\eta = 174 \GeV$.  
	}
\end{figure}

After electroweak symmetry breaking, see \eref{eq:VEVs}, the gauge fields acquire masses
\begin{align}\label{eq:L4}
	\mathcal{L} \bigr|_{\rm vevs} \ni &\
	M_W^2 W_{\mu}^{+} W^{- \, \mu}
	+ \frac{1}{2} M_{A}^2 (A_{\mu})^2 
	+ \frac{1}{2} M_Z^2 (Z_{\mu})^2 
	+ \frac{1}{2} M_{X}^2 ({\Xb}_{\mu})^2
\end{align}
with the spectrum
\begin{align}\label{eq:gauge_evals}
	M_{W}^2 &= m_W^2 \nn
	M_{A}^2 &= 2 (\gHA)^2 \eta^2 + 2 (\gSA)^2 \sigma^2 = 0  \nn
	M_{Z}^2 &= 2 (\gHZ)^2 \eta^2 + 2 (\gSZ)^2 \sigma^2 = m_{Z}^2 \left( 1 + \sw \txi \teps \right) \nn
	M_{X}^2 &= 2 (\gHX)^2 \eta^2 + 2 (\gSX)^2 \sigma^2 = \frac{m_{X}^2}{c_{\epsilon}^2(1+\sw \txi \teps )}
\end{align}
Once again, the massless photon is a sign of the residual gauge invariance.  
As can be seen in \eref{eq:def_tan2xi}, the angles $\zeta$ and $\epsilon$ always have opposite signs, and therefore one has in general $M_Z < m_Z$ and $M_{X} > m_X$.  
The $Z$ and $X$ boson masses are plotted in \fref{fig:gaugespec}.  
Over most of the parameter range, these masses are well approximated as $M_Z \approx m_Z$ and $M_X \approx m_X \approx R M_Z$.  
To provide a reference point, we also show (on the left panel) the relative error bar on the measured $Z$ boson mass, $\delta M_Z / M_Z \simeq 2.3 \times 10^{-5}$ \cite{Beringer:2012zz}, as a dashed line. 
Roughly speaking, the parameter range above the dashed line is excluded, or conversely, $\seps$ becomes unconstrained in the decoupling limit $R \gg 1$.  
However, to rigorously ascertain if a model is excluded, all available observables should be folded in together (see, \eg, \cite{Hook:2010tw}).  
Since it is not the goal of this paper to impose phenomenological constraints, we will reserve that discussion for a future work.  

We can now calculate the Euler-Lagrange equations for the diagonalized Lagrangian, \eref{eq:L3}.  
Since we are only interested in string solutions, it is prudent at this point to recognize that because $\Higgsp$ does not acquire a vev, we can consistently set $\Higgsp = W^{\pm}_{\mu} = A_{\mu} = 0$.  
That is, these fields are not sourced by the nontrivial profiles of the remaining scalar and gauge fields.  
Then the remaining field equations become
\begin{align}\label{eq:fieldeqns2}
\begin{array}{l}
	\partial_{\nu} Z^{\nu \mu} 
	= \gHZ J_{H}^{\mu} + \gSZ J_{S}^{\mu} \\
	\partial_{\nu} {\Xb}^{\nu \mu} 
	= \gHX J_{H}^{\mu} + \gSX J_{S}^{\mu} \\
	D_{\mu} D^{\mu} H = - 2 \lambda \left( H^{\ast} H - \eta^2 \right) H - \alpha \left( S^{\ast} S - \sigma^2 \right) H \\
	D_{\mu} D^{\mu} S = - 2 \kappa \left( S^{\ast} S - \sigma^2 \right) S - \alpha \left( H^{\ast} H - \eta^2 \right) S
\end{array}
\end{align}
where the currents are given by 
\begin{align}\label{eq:J2}
\begin{array}{l}
	J_{H}^{\mu} \equiv i \bigl( H (D^{\mu} H)^{\ast} - H^{\ast} D^{\mu} H \bigr) \\
	J_{S}^{\mu} \equiv i \bigl( S (D^{\mu} S)^{\ast} - S^{\ast} D^{\mu} S \bigr)
\end{array} \com
\end{align} 
and the covariant derivatives are given by \eref{eq:covar_deriv2}.  
These field equations will be used in \sref{sub:DarkString} to obtain the string equations.  

\subsection{Scalar Sector}\label{sub:ScalarSector}

After symmetry breaking, both the fields $H$ and $S$ acquire vevs.  
The three SM would-be Goldstone bosons, $\Phi^{+}$ and $a_H = {\rm Arg}[H]$, and the fourth would-be Goldstone boson, $a_S = {\rm Arg}[S]$, are eaten leaving only two massive scalars, $\bar{h} = \sqrt{2} (\abs{H} - \eta)$ and $\bar{s} = \sqrt{2} ( \abs{S} - \sigma )$.  
The Higgs portal operator allows these scalars to mix.  

The scalar fields can be parametrized as $H = (\eta + \bar{h} / \sqrt{2}) \exp{i a_H}$ and $S = (\sigma + \bar{s} / \sqrt{2} )  \exp{i a_S}$.  
After defining
\begin{align}
	m_H \equiv \sqrt{4 \lambda \eta^2}
	\qquad {\rm and} \qquad 
	m_S \equiv \sqrt{4 \kappa \sigma^2} \com
\end{align}
the scalar potential becomes 
\begin{align}\label{eq:U_bilinear}
	U \ni \frac{1}{2} \begin{pmatrix} \bar{h} & \bar{s} \end{pmatrix} \begin{pmatrix} 
	m_H^2 & 2 \alpha \eta \sigma \\ 2 \alpha \eta \sigma & m_S^2
	\end{pmatrix} 
	\begin{pmatrix} \bar{h} \\ \bar{s} \end{pmatrix} 
\end{align}
plus higher order interactions.  
This mass matrix is diagonalized by 
\begin{align}
	\begin{pmatrix} \bar{h} \\ \bar{s} \end{pmatrix} = \begin{pmatrix} \cos \theta & \sin \theta \\ - \sin \theta & \cos \theta \end{pmatrix} \begin{pmatrix} \phi_{H} \\ \phi_{S} \end{pmatrix}
\end{align}
where the mixing angle, $- \pi/4 < \theta < \pi/4$, is given by 
\begin{align}\label{eq:mixing}
	\tan 2 \theta = \frac{4\alpha \eta \sigma}{m_S^2 - m_H^2} \com
\end{align}
and the eigenstates $\phi_H$ and $\phi_S$ have masses
\begin{align}\label{eq:eigenvalues}
	M_H^2 & = m_H^2 - \left( m_S^2 - m_H^2 \right) \frac{\sin^2 \theta}{\cos 2 \theta} \\
	M_S^2 & = m_S^2 + \left( m_S^2 - m_H^2 \right) \frac{\sin^2 \theta}{\cos 2 \theta} \com
\end{align}
respectively.  
We will assume that $M_S > M_H$ (equivalently, $m_S > m_H$) and that $M_H \approx 125 \GeV$ is the mass of the Higgs boson measured by the LHC.  

The mixing angle can also be written as 
\begin{align}
	\tan 2 \theta = \frac{\alpha}{\alpha_0} \frac{8 m_H m_S}{m_S^2 - m_H^2} 
\end{align}
where $\alpha_0 \equiv \sqrt{4 \lambda \kappa}$.  
To ensure that the determinant of the mass matrix in \eref{eq:U_bilinear} is positive, we must have $\alpha < \alpha_0$.  
In the decoupling limit, $m_S \gg m_H$, the mixing angle becomes $\abs{\theta} \approx ( \abs{\alpha} / \alpha_{0} ) (4M_H / M_S) \ll 1$, and the eigenvalues become $M_H \approx m_H$ and $M_S \approx m_S$.  
In this limit, the heavy scalar $\phi_S \approx \bar{s}$ is decoupled from the SM Higgs $\phi_H \approx \bar{h}$.  
As we reduce the hierarchical ratio, $M_S / M_H$, the amount of mixing grows larger until it becomes maximal ($\theta = 45^{\circ}$) and $M_S / M_H = 1$.  
Observations of the Higgs at the LHC constrain the mixing with a hidden sector scalar to be $\theta \lesssim 40^{\circ}$ \cite{Carmi:2012yp}.  
Since, for the present study, we are not interested in rigorously applying observational constraints, we will simply take $M_H = 125 \GeV$ and require $M_S > M_H$.  
The scalar self-couplings are then determined by 
\begin{align}\label{eq:lamkap}
	\lambda & = \frac{M_H^2}{4\eta^2} + \frac{M_S^2 - M_H^2}{8 \eta^2} \left( 1 - \sqrt{1 - \left( \frac{4\alpha \eta \sigma}{M_S^2 - M_H^2} \right)^2 } \right) \nn
	\kappa & = \frac{M_S^2}{4\sigma^2} - \frac{M_S^2 - M_H^2}{8 \sigma^2} \left( 1 - \sqrt{1 - \left( \frac{4\alpha \eta \sigma}{M_S^2 - M_H^2} \right)^2 } \right) 
\end{align}
provided that
\begin{align}\label{eq:alpha_max}
	\abs{\alpha} < \alpha_{\rm max} \equiv \frac{M_S^2 - M_H^2}{4 \eta \sigma} \per
\end{align}
Note that \eref{eq:alpha_max} subsumes the previous bound, 
$\alpha < \alpha_0 = \sqrt{4\lambda\kappa}$~, 
because \eref{eq:lamkap} gives 
$\alpha_0 = \abs{\alpha} \sqrt{1 + (M_S M_H/ 4 \eta \sigma)^2} > \abs{\alpha}$.  

In order to discuss the string solutions below, it will be useful here to identify the extrema of the scalar potential \eref{eq:U1}.  
We set $\Higgs^+ = 0$ and solve the two equations $\partial U / \partial H = \partial U / \partial S = 0$.  
There are four solutions with both $H$ and $S$ nonnegative:  
\begin{align}\label{eq:Uextrema}
\begin{array}{lclcl}
	H = \eta \ &{\rm ,}&\ S = \sigma \quad&\Rightarrow&\quad {\rm minimum} \\
	H = 0 \ &{\rm ,}&\ S = 0 \quad&\Rightarrow&\quad {\rm maximum} \\
	H = H_0 \ &{\rm ,}&\ S = 0 \quad&\Rightarrow&\quad {\rm saddle \ \ point} \\
	H = 0 \ &{\rm ,}&\ S = S_0 \quad&\Rightarrow&\quad {\rm saddle \ \ point} 
\end{array}
\end{align}
where
\begin{align}\label{eq:H0S0_def}
\begin{array}{l}
	H_0 \equiv \eta \sqrt{1 + \frac{\alpha \sigma^2}{2 \lambda \eta^2}} \\
	S_0 \equiv \sigma \sqrt{1 + \frac{\alpha \eta^2}{2 \kappa \sigma^2}} 
\end{array} \per
\end{align}
For the case $\alpha < 0$, the saddle point solutions do not exist if $\abs{\alpha} > 2 \kappa \sigma^2 / \eta^2$.

\subsection{Dark String Ansatz}\label{sub:DarkString}

Let us now derive the equations for the dark string.  
We will work in cylindrical coordinates, $\rho = \sqrt{x^2 + y^2}$ and $\varphi = \arctan(y / x)$, and we will use the dimensionless radial coordinate $\xi = \rho / \rho_0$ where $\rho_0 = 1/\sigma$.  
Seeking the straight, static dark string solution, we take the ansatz\footnote{This corresponds to $Z_{\mu} dx^{\mu} = \Zpro \, d \varphi$ and $X_{\mu} d x^{\mu} = \Xpro \, d \varphi$.  }
\begin{align}\label{eq:Ansatz}
	&\Higgsp(x) = 0 \com
	& 
	&H(x) = \eta \, \Hpro(\xi) e^{i n \varphi} \com
	&
	&Z_{\mu}(x) = \frac{1}{\rho_0} \frac{\Zpro(\xi)}{\xi} V_{\mu}(\varphi) \com \nn
	&W^{\pm}_{\mu} = A_{\mu} = 0 \com
	&
	&S(x) = \sigma \, \Spro(\xi) e^{i m \varphi} \com
	&
	&{\Xb}_{\mu}(x) = \frac{1}{\rho_0} \frac{\Xpro(\xi)}{\xi} V_{\mu}(\varphi) \com
\end{align}
where $n, m \in \mathbb{Z}$ and $\Hpro, \Spro, \Zpro, \Xpro \in \mathbb{R}$ 
and  $V_{\mu} \equiv \rho \partial_{\mu} \varphi = \bigl\{ 0 \, , \, - \sin \varphi \, , \, \cos \varphi \, , \, 0 \bigr\}$.  
With this ansatz, 
the currents in \eref{eq:J2} become
\begin{align}\label{eq:currents_Ansatz}
	J_{H}^{\mu} = \frac{2 \eta^{2}}{\rho_0} \frac{\Hpro^2 C_{H}}{\xi} V^{\mu} 
	\qquad {\rm and} \qquad 
	J_{S}^{\mu} = \frac{2 \sigma^2}{\rho_0} \frac{\Spro^2 C_S}{\xi} V^{\mu}
\end{align}
where
\begin{align}\label{eq:C_def}
\begin{array}{l}
	C_{H}(\xi) \equiv n - \gHZ \Zpro(\xi) - \gHX \Xpro(\xi) \\
	C_{S}(\xi) \equiv m - \gSZ \Zpro(\xi) - \gSX \Xpro(\xi) 
\end{array} \per
\end{align}  
The field equations in \eref{eq:fieldeqns2} become
\begin{subequations}\label{eq:darkstring_eqns}
\begin{align}
\label{eq:eom_1}
	\left( \frac{\Zpro^{\prime}}{\xi} \right)^{\prime} =
	& - 2 \gHZ (\rho_0 \eta)^2 \frac{\Hpro^2 C_H}{\xi} -2 \gSZ (\rho_0 \sigma)^2 \frac{\Spro^2 C_S}{\xi} \\
\label{eq:eom_2}
	\left( \frac{\Xpro^{\prime}}{\xi} \right)^{\prime} =
	& - 2 \gHX (\rho_0 \eta)^2 \frac{\Hpro^2 C_H}{\xi} -2 \gSX (\rho_0 \sigma)^2 \frac{\Spro^2 C_S}{\xi} \\
\label{eq:eom_3}
	(\xi \Hpro^{\prime})^{\prime} =
	&  C_{H}^2 \frac{\Hpro}{\xi} 
	- 2 \lambda (\rho_0 \eta)^2 \left(1 \! - \! \Hpro^2  \right) \xi \Hpro - \alpha (\rho_0 \sigma)^2 (1 \! - \! \Spro^2 ) \xi \Hpro \\
\label{eq:eom_4}
	(\xi \Spro^{\prime})^{\prime} =
	& C_S^2 \frac{\Spro}{\xi}  
	- 2 \kappa (\rho_0 \sigma)^2 (1 \! - \! \Spro^2)  \xi \Spro - \alpha (\rho_0 \eta)^2 (1 \! - \! \Hpro^2 ) \xi \Spro \per
\end{align}
\end{subequations}
Although we take $\rho_0 = 1/\sigma$, we have retained $\rho_0$ in these expressions so as to avoid confusion as to where the $\sigma$ enters explicitly as the VEV of $S$ and where it enters as our choice of the radial length scale.  
If we were to turn off both the GKM and HP operators by taking $\epsilon = \alpha = 0$, then we would regain the string equations for two, uncoupled Nielsen-Olesen strings of winding $n$ and $m$.  

The scalar field boundary conditions can be divided into three cases depending on which of the two winding parameters, $n$ and $m$, are nonzero.  
In each case, we must require $\Hpro(\infty) = \Spro(\infty) = 1$ at spatial infinity and that $H(x)$ and $S(x)$ are regular at the origin.  
The cases are: 
\begin{align}\label{eq:BC1b}
\left|
	\begin{array}{c}
	{\bf Case \, 1:}  \\
	\begin{array}{c}
	n \neq 0 \\
	m \neq 0 
	\end{array}
	\hspace{0.25cm}
	\Rightarrow
	\hspace{0.25cm}
	\begin{array}{l}
	\Hpro(0) = 0 \\
	\Hpro(\infty) = 1 \\
	\Spro(0) = 0 \\
	\Spro(\infty) = 1
	\end{array}
	\end{array}
\right|
\hspace{0.25cm}
\left|
	\begin{array}{c}
	{\bf Case \, 2:}  \\
	\begin{array}{c}
	n = 0 \\
	m \neq 0 
	\end{array}
	\hspace{0.25cm}
	\Rightarrow
	\hspace{0.25cm}
	\begin{array}{l}
	\Hpro'(0) = 0 \\
	\Hpro(\infty) = 1 \\
	\Spro(0) = 0 \\
	\Spro(\infty) = 1 
	\end{array}
	\end{array}
\right|
\hspace{0.25cm}
\left|
	\begin{array}{c}
	{\bf Case \, 3:}  \\
	\begin{array}{c}
	n \neq 0 \\
	m = 0 
	\end{array}
	\hspace{0.25cm}
	\Rightarrow
	\hspace{0.25cm}
	\begin{array}{l}
	\Hpro(0) = 0 \\
	\Hpro(\infty) = 1 \\
	\Spro'(0) = 0 \\
	\Spro(\infty) = 1 
	\end{array}
	\end{array}
\right|  \per
\end{align}
Case 3 resembles the SM semilocal and electroweak strings \cite{Achucarro:1999it}, which are not topological and therefore not stable.  
For this reason, we will focus on Cases 1 and 2.  
In Case 2 we have mixed Neumann and Dirichlet boundary conditions, and we do not expect $\Hpro(0) = 1$ in general.  
By considering the energetics, it is clear that $\Hpro(0) = 1$ will minimize the gradient contribution to the energy of the string.  
However, in terms of the potential energy, we expect that the value of the Higgs condensate at the core of the string will relax toward the saddle point at $H = H_0$ and $S =0$ [see \eref{eq:Uextrema}].  
In general we expect
\begin{align}\label{eq:scalarBC}
	{\bf Case \, 2:} \qquad 
	\begin{cases}
	h_0 < \Hpro(0) < 1 & \alpha < 0 \\
	1 < \Hpro(0) < h_0 & \alpha > 0 
	\end{cases}
\end{align}
where $h_0 \equiv H_0 / \eta = \sqrt{1 + \alpha \sigma^2 / (2 \lambda \eta^2)}$ and $H_0$ is given by \eref{eq:H0S0_def}.  

The gauge field boundary conditions are 
\begin{align}\label{eq:gaugeBC}
	\Zpro(0) = \Xpro(0) = 0 
	\qquad , \qquad
	\Zpro(\infty) = \frac{\gSX n - \gHX m}{\gSX \gHZ - \gHX \gSZ} 
	\qquad , \qquad
	\Xpro(\infty) = \frac{\gHZ m - \gSZ n}{\gSX \gHZ - \gHX \gSZ}  \per
\end{align}
These ensure that $Z_{\mu}(x)$ and $X_{\mu}(x)$ are regular at the origin and that at spatial infinity 
\begin{align}
	C_{H}(\infty) = C_{S}(\infty) = 0 \com
\end{align}
and the action is finite.  
An interesting consequence of the GKM is that both gauge fields have nontrivial profiles if either scalar field has a winding (either $n$ or $m$ is nonzero).  
This is evident in the limit $\seps \ll1$ where
\begin{align}\label{eq:gaugeBC_expand}
	\Zpro(\infty) &\approx - \frac{\sqrt{2} \eta}{m_Z} n - \frac{\sqrt{2} \sw R^2 \sigma}{m_X(R^2-1)} m \, \seps + O(\seps^2) \nn
	\Xpro(\infty) &\approx \frac{\sqrt{2} \sigma}{m_X} m - \frac{\sqrt{2} \sw \eta}{m_Z(R^2-1)} n \, \seps + O(\seps^2) \per
\end{align}
For example, taking $n=0$ and $m=1$ induces an $O(\seps)$ expectation value for the $Z$ field.

\section{Properties of the Dark String}\label{sec:Properties}
\numberwithin{equation}{subsection}

The dark string is the solution of the system of equations given by \eref{eq:darkstring_eqns} along with the boundary conditions in Eqs.~(\ref{eq:BC1b}) and (\ref{eq:gaugeBC}).  
We solve these equations numerically as described in \aref{app:numerical_method}.  

We calculate the dark string solution for various values of the model parameters:  $(n,m)$, $\alpha$, $\seps$, $\gX$, $\sigma$, and $M_S$ while fixing
$\eta = 174 \GeV$, $M_H = 125 \GeV$, $g = 0.654$, and $g^{\prime} = 0.359$ and using \eref{eq:lamkap} to determine $\lambda$ and $\kappa$.  
With this choice of parameters, the masses $M_Z$ and $M_X$ are given by \eref{eq:gauge_evals}.  
Although these masses depend upon $\seps$, it is typically the case that $M_Z \approx 91.2 \GeV$ and $M_X \approx m_X = \gX \sigma / \sqrt{2}$.  
Having obtained the dark string solution, we study its properties and couplings, which are discussed in the remainder of this section.  

\subsection{String Solution}\label{sub:StringSoln}

Generally, the strings with higher order windings, $(n,m)$ with $n,m > 1$, are unstable, and they will decay on a microscopic time scale into the lightest strings.  
The winding $m$ of the singlet scalar $S$ is topological by virtue of the $\U{1}_X$ symmetry, however the winding $n$ of the Higgs field is not topological -- just as in the case of the electroweak strings in the SM \cite{Achucarro:1999it}.    
This means that any $(n,m)$ string with $n \geq 1$ will fragment and decay into the $(0,1)$ string, which generally has a lower tension than the $(1,1)$ string.
We will focus on the properties of the $(0,1)$ string, but we will also compare against the $(1,1)$ string.

\begin{figure}[t!]
	\begin{subfigure}[h]{0.5\textwidth}
		\includegraphics[width=\textwidth]{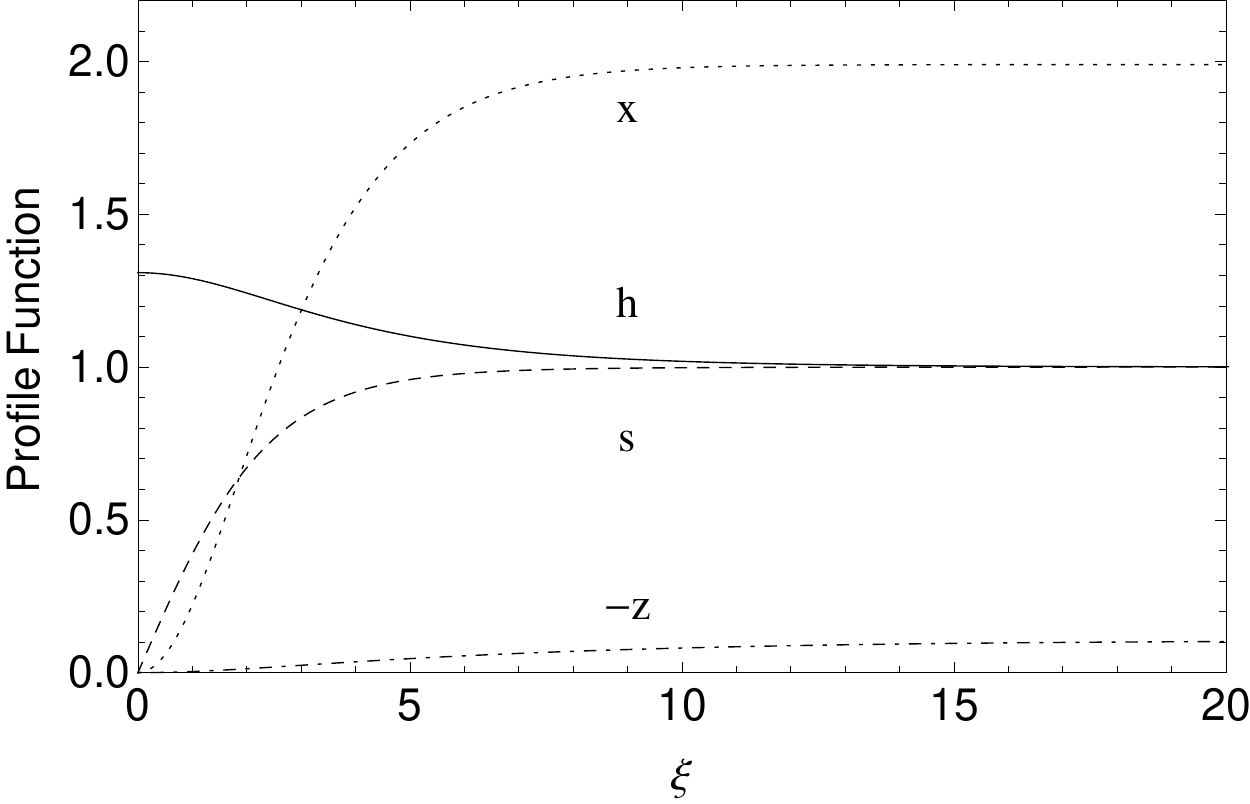}
		\caption{\label{fig:pr01}Profile functions for (0,1) string.}
	\end{subfigure}%
	\begin{subfigure}[h]{0.5\textwidth}
		\includegraphics[width=\textwidth]{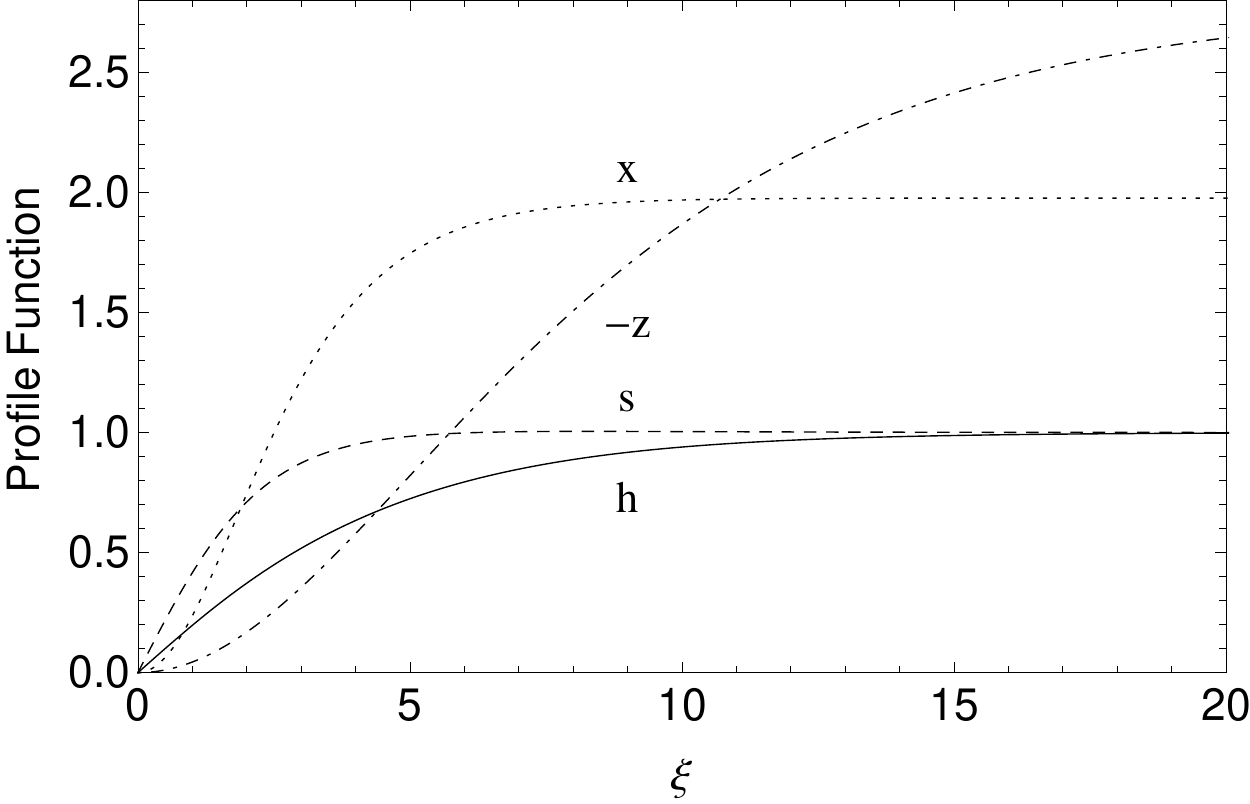}
		\caption{\label{fig:pr11}Profile functions for (1,1) string.}
	\end{subfigure}
	\begin{subfigure}[h]{0.5\textwidth}
		\includegraphics[width=\textwidth]{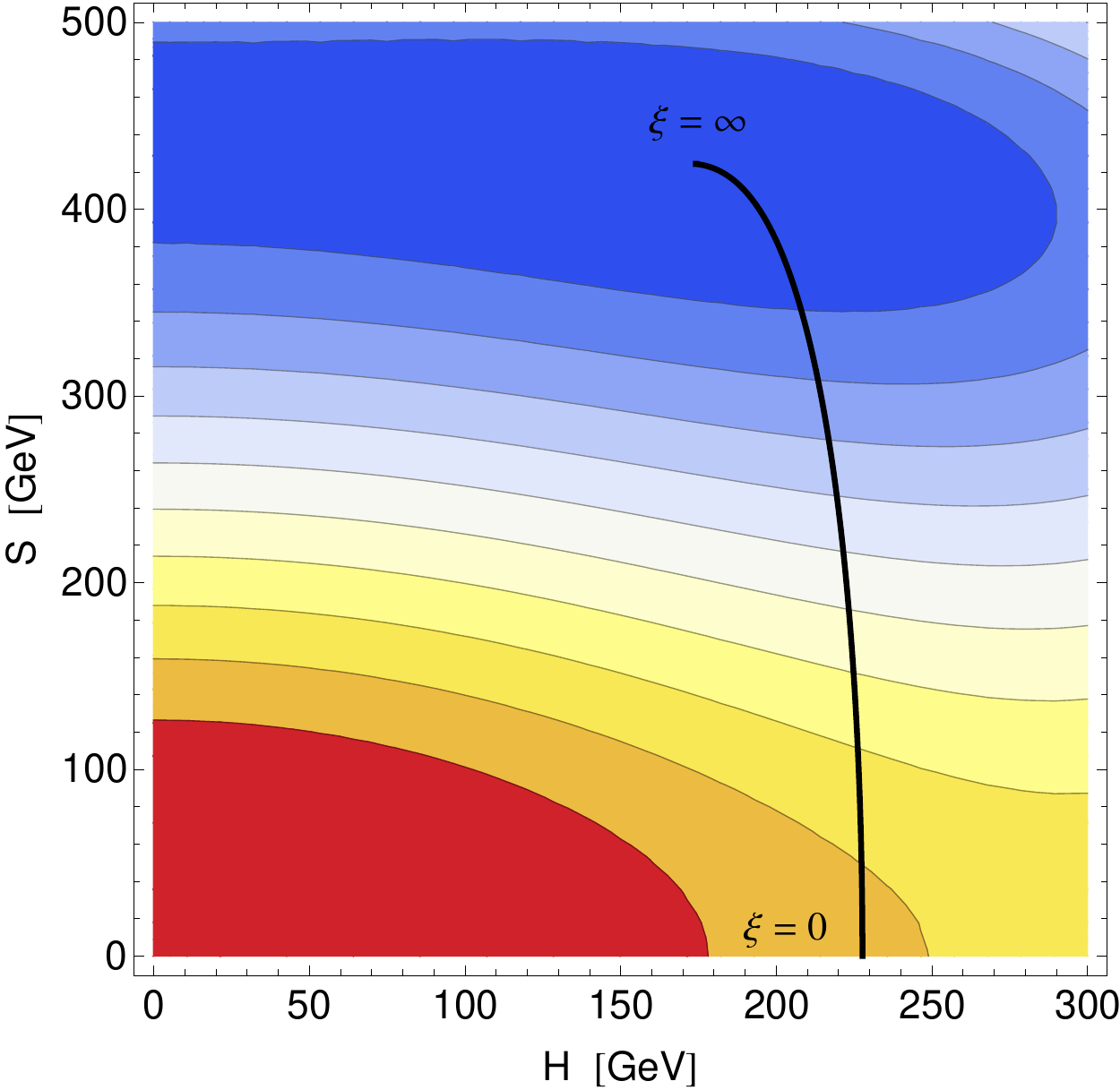}
		\caption{\label{fig:eg01}(0,1) string solution in (H,S) plane.}
	\end{subfigure}%
	\begin{subfigure}[h]{0.5\textwidth}
		\includegraphics[width=\textwidth]{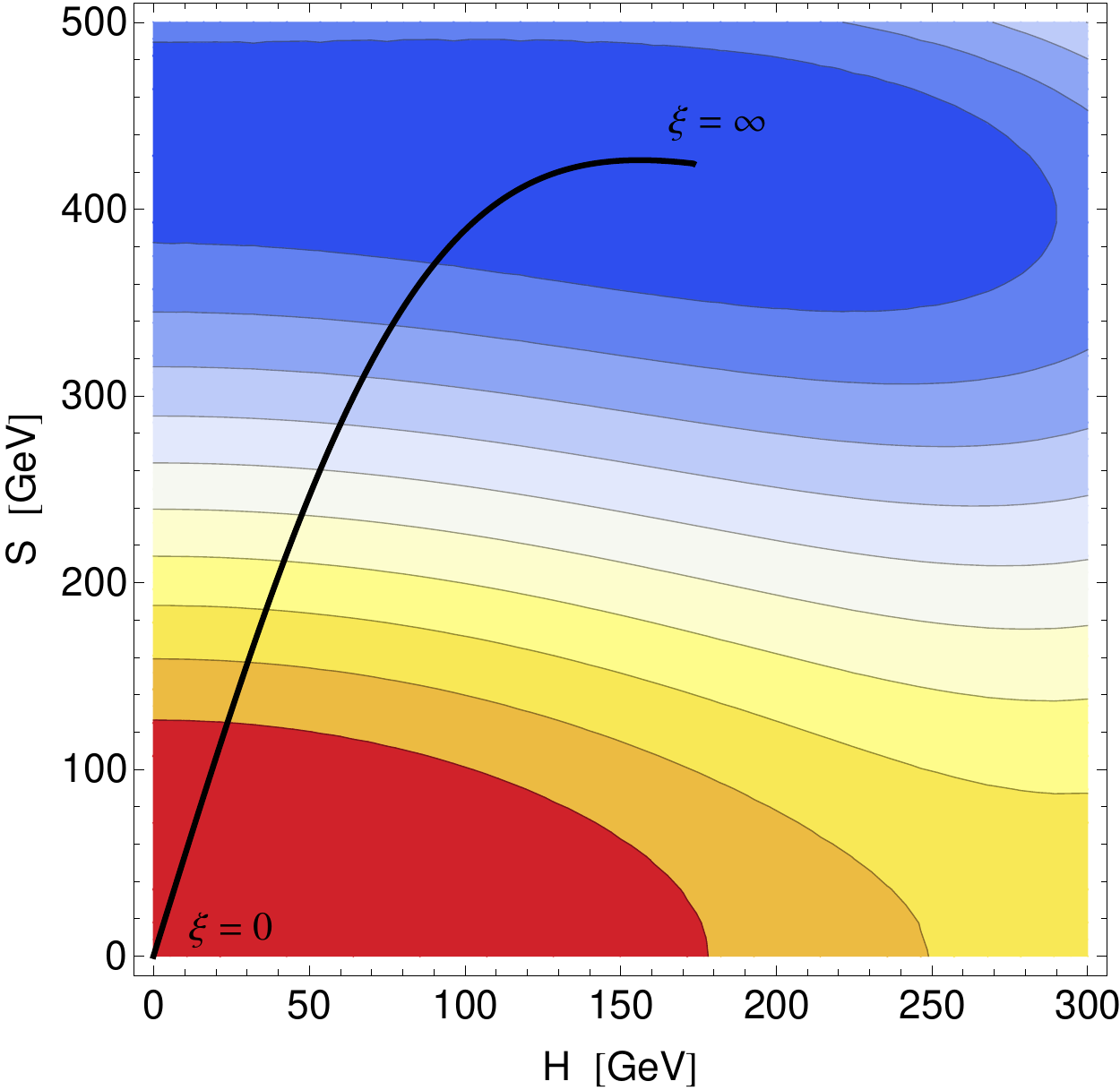}
		\caption{\label{fig:eg11}(1,1) string solution in (H,S) plane.}
	\end{subfigure}
	\caption{\label{fig:profiles}
	String solutions for $m_X =  M_S = \sigma / \sqrt{2} = 200 \GeV$, $\alpha = 0.1$, $\seps = 0.1$, and $\gX = 1$.  The bottom panels show the scalar potential, \eref{eq:U1}, where the blue (red) contours are lower (higher).  
	}
\end{figure}

\begin{figure}[t!]
	\begin{subfigure}[b]{0.5\textwidth}
		\includegraphics[width=\textwidth]{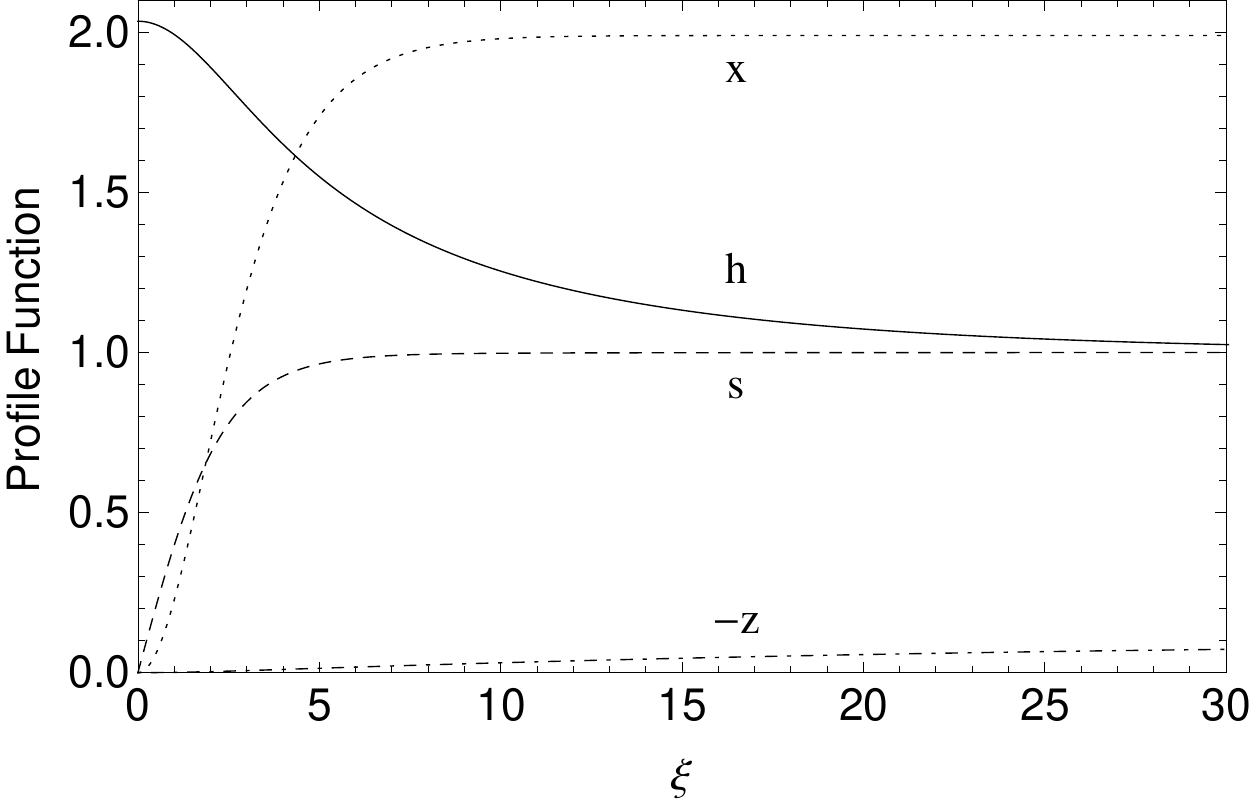}
		\caption{\label{fig:pr01_1tev}Profile functions for (0,1) string.}
	\end{subfigure}%
	\begin{subfigure}[b]{0.5\textwidth}
		\includegraphics[width=\textwidth]{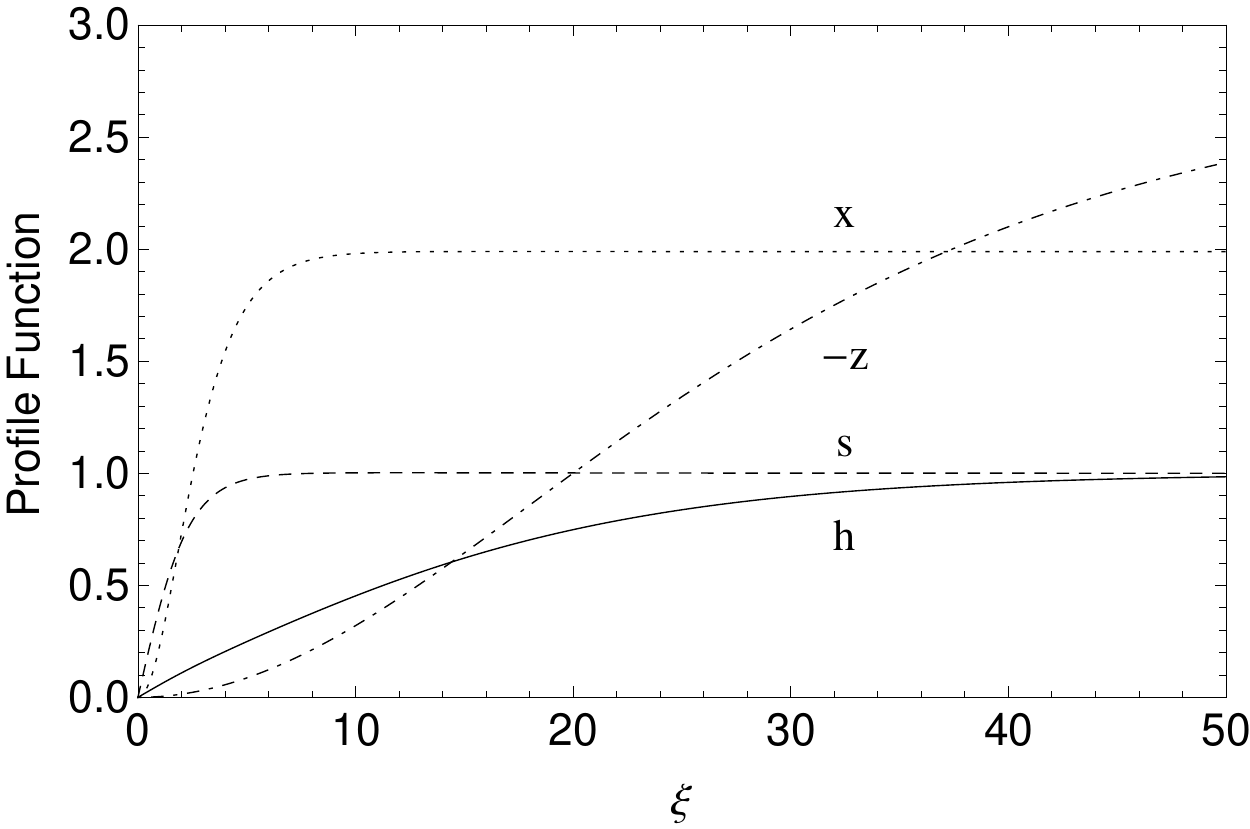}
		\caption{\label{fig:pr11_1tev}Profile functions for (1,1) string.}
	\end{subfigure}
	\begin{subfigure}[b]{0.5\textwidth}
		\includegraphics[width=\textwidth]{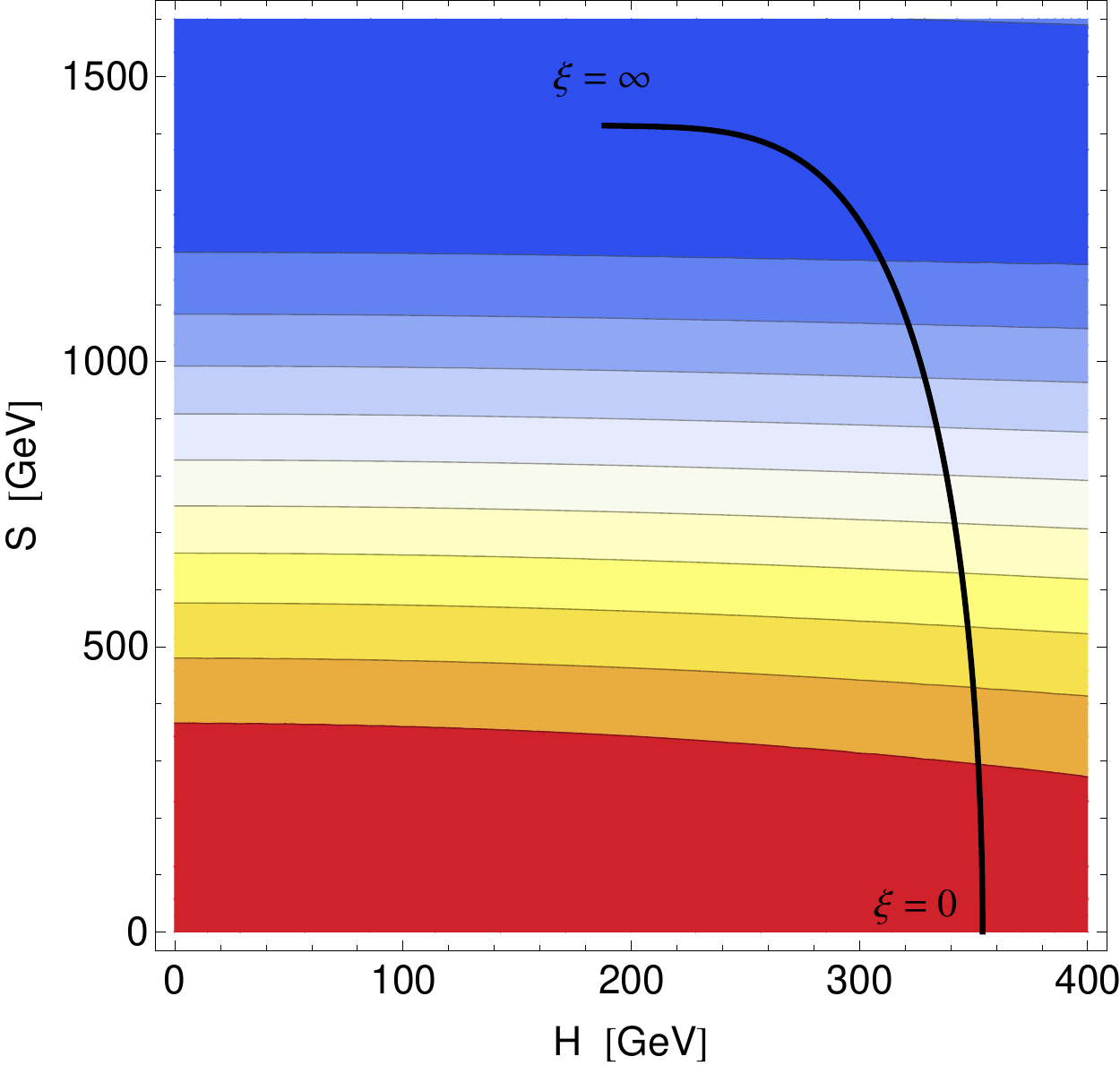}
		\caption{\label{fig:eg01_1tev} (0,1) string solution in (H,S) plane.}
	\end{subfigure}%
	\begin{subfigure}[b]{0.5\textwidth}
		\includegraphics[width=\textwidth]{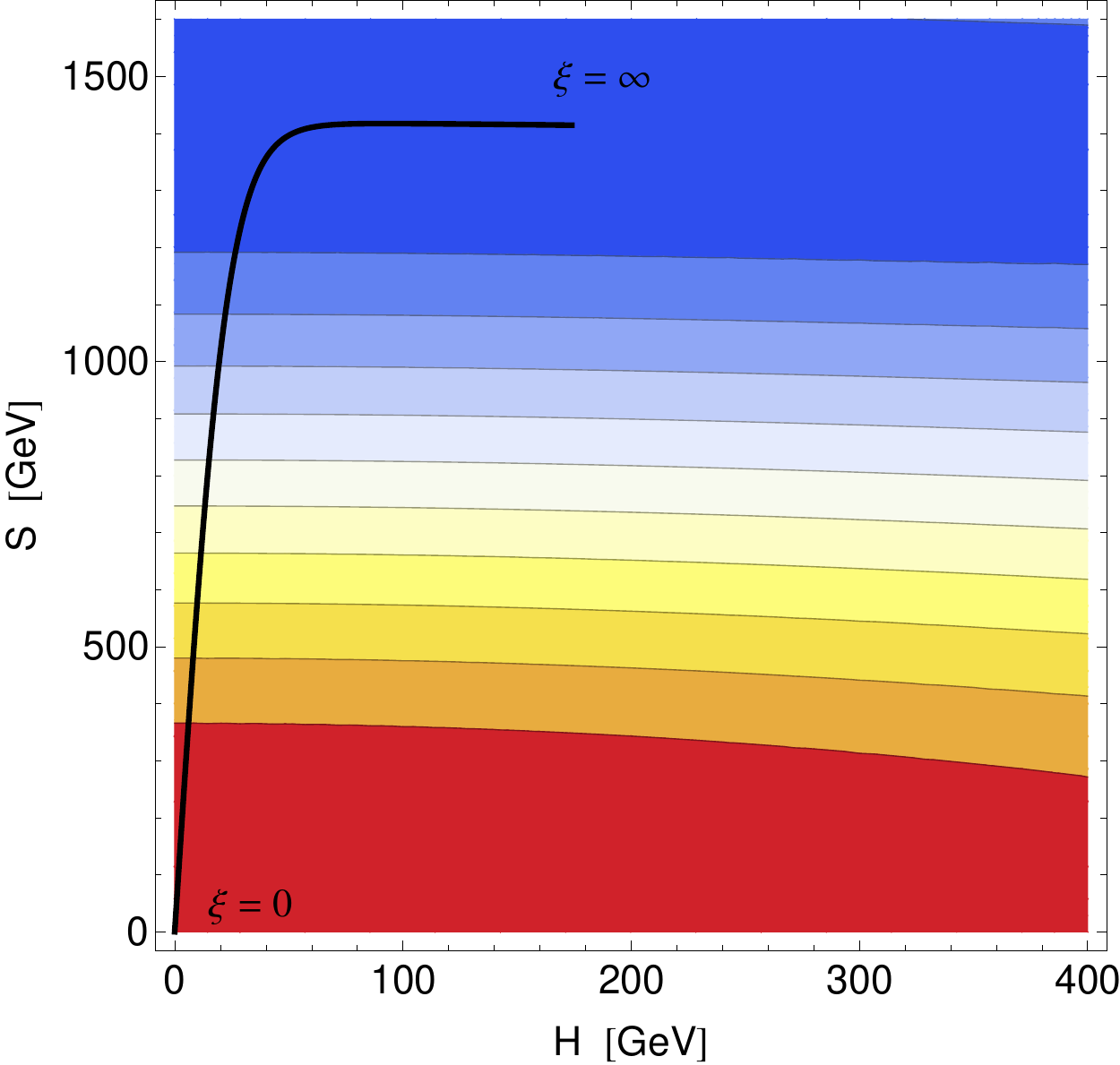}
		\caption{\label{fig:eg11_1tev} (1,1) string solution in (H,S) plane.}
	\end{subfigure}
	\caption{\label{fig:profiles_1tev}
	Same as \fref{fig:profiles} but for $m_X =  M_S = \sigma / \sqrt{2} = 1 \TeV$.  
	}
\end{figure}

\begin{figure}[t!]
	\begin{subfigure}[b]{0.5\textwidth}
		\includegraphics[width=\textwidth]{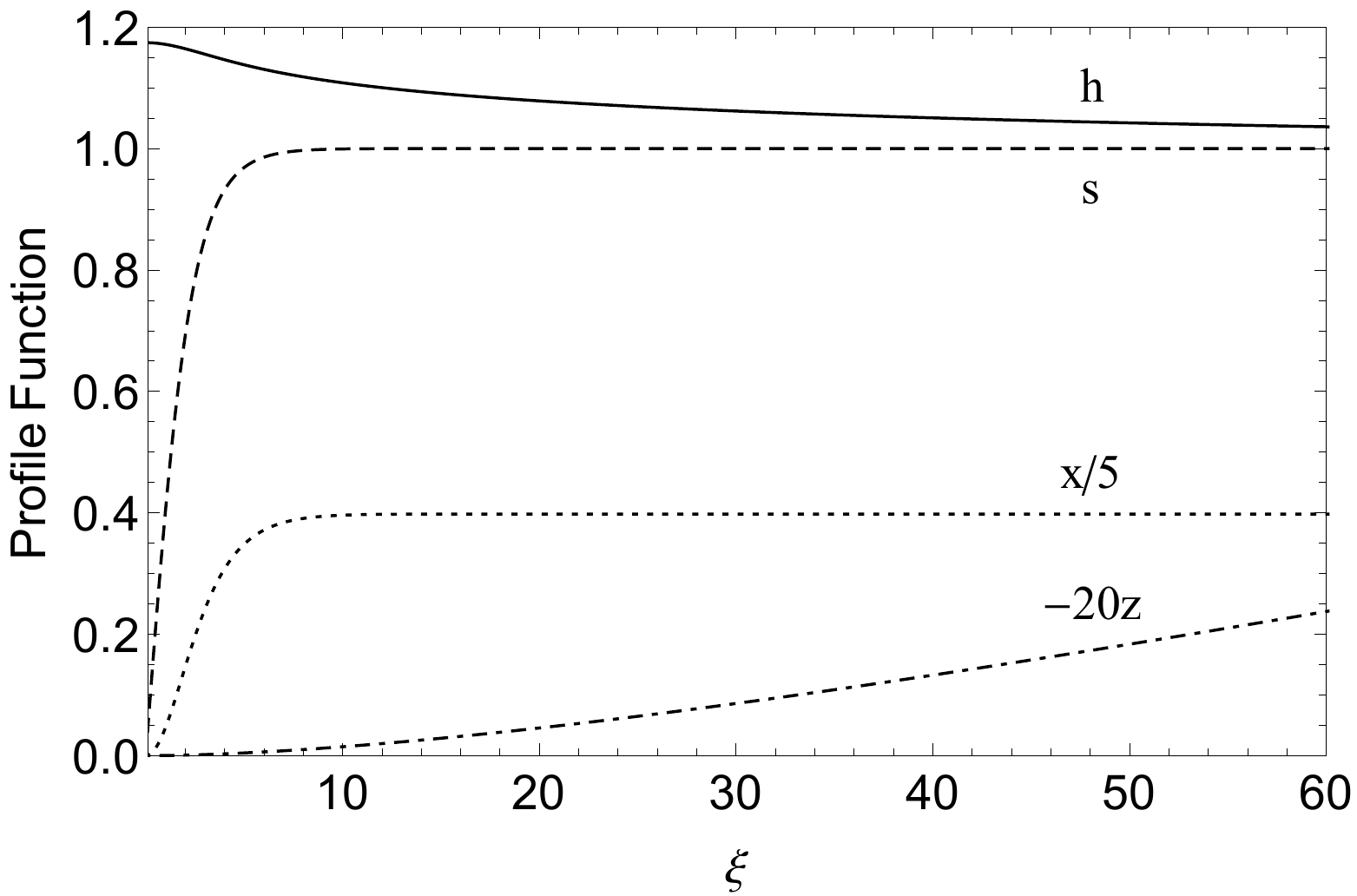}
		\caption{\label{fig:pr01_10tev}Profile functions for (0,1) string.}
	\end{subfigure}%
	\begin{subfigure}[b]{0.5\textwidth}
		\includegraphics[width=\textwidth]{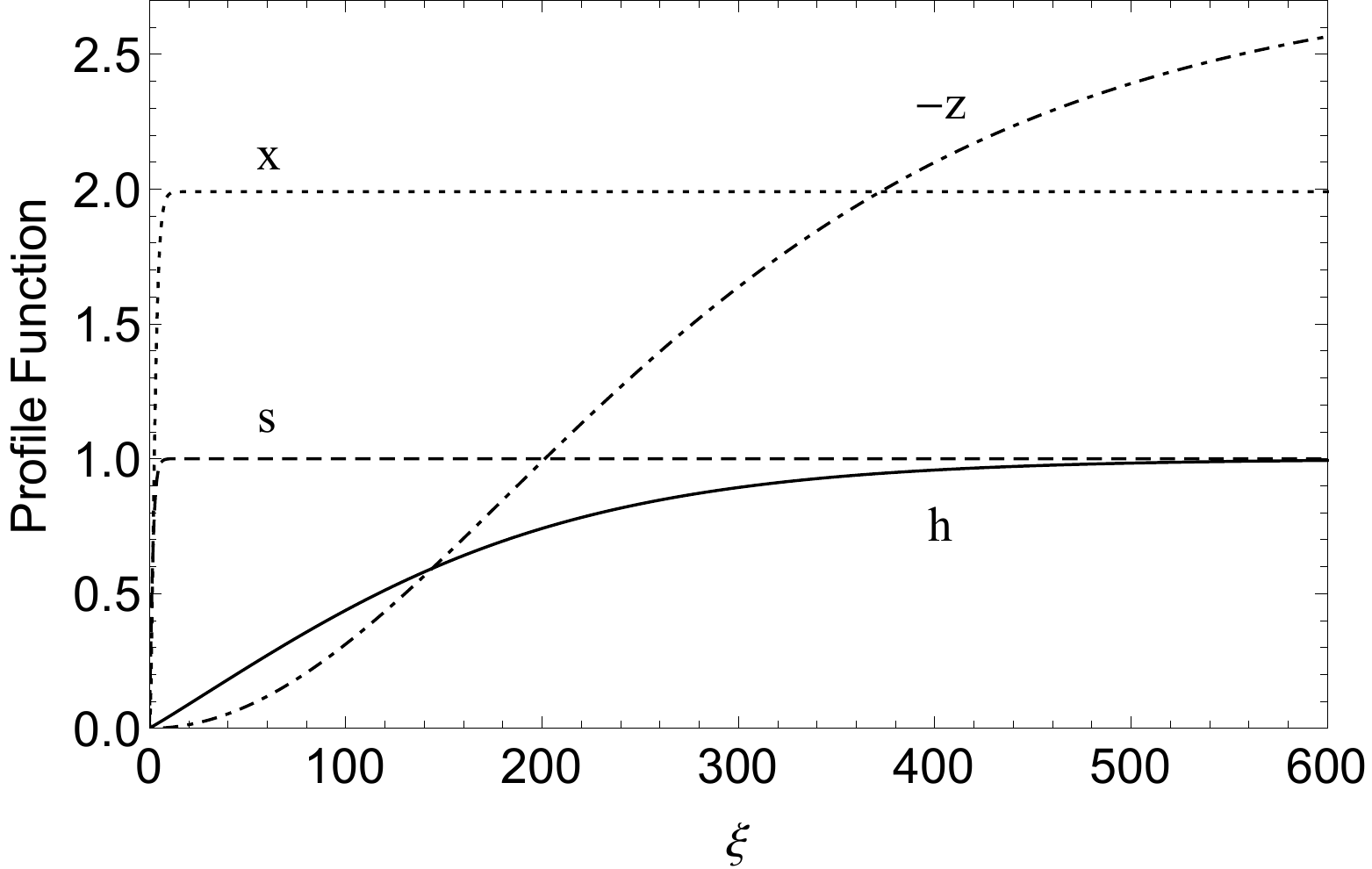}
		\caption{\label{fig:pr11_10tev}Profile functions for (1,1) string.}
	\end{subfigure}
	\begin{subfigure}[b]{0.5\textwidth}
		\includegraphics[width=\textwidth]{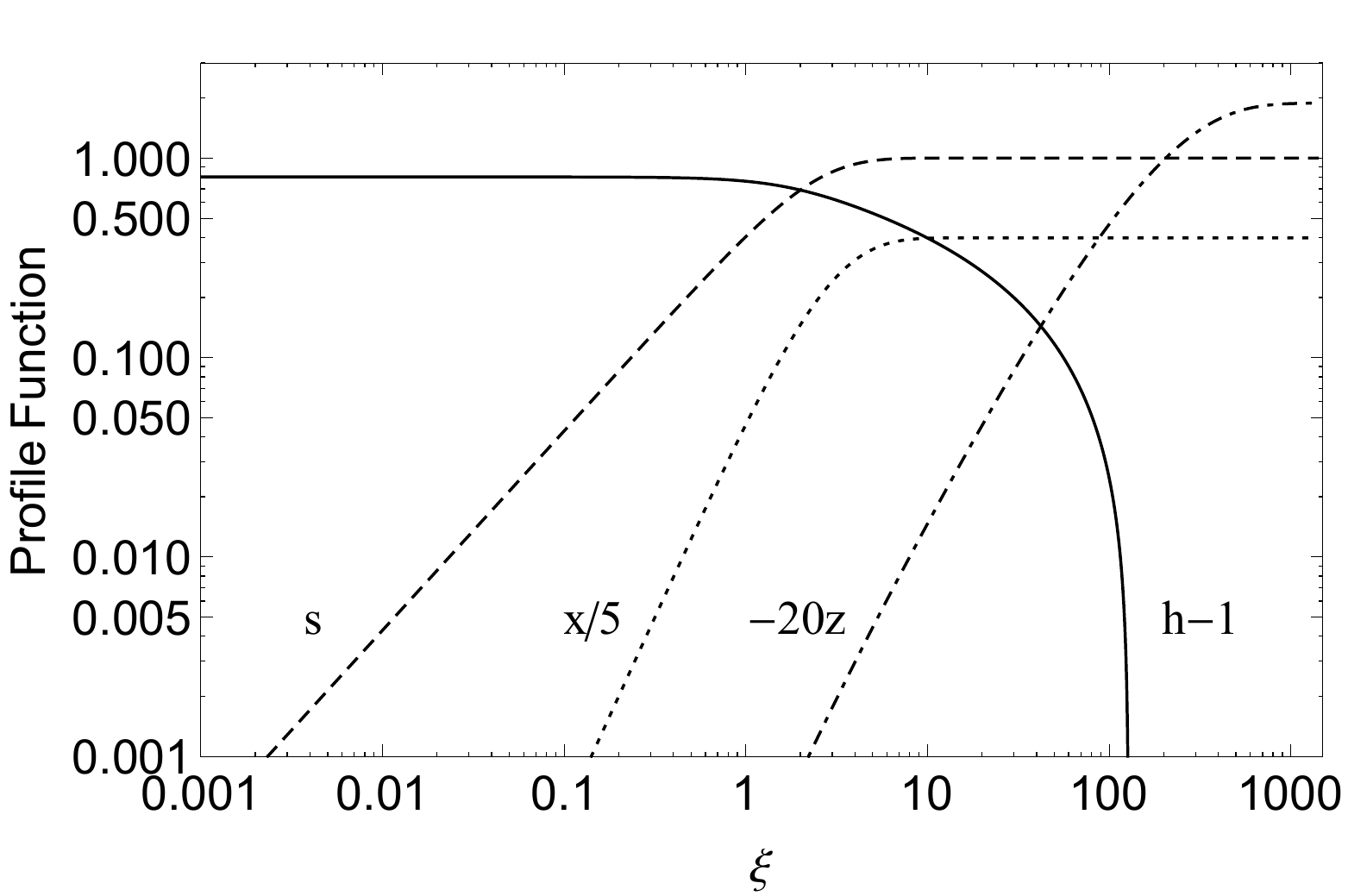}
		\caption{\label{fig:pr01_10tev_log}Profile functions for (0,1) string.}
	\end{subfigure}%
	\begin{subfigure}[b]{0.5\textwidth}
		\includegraphics[width=\textwidth]{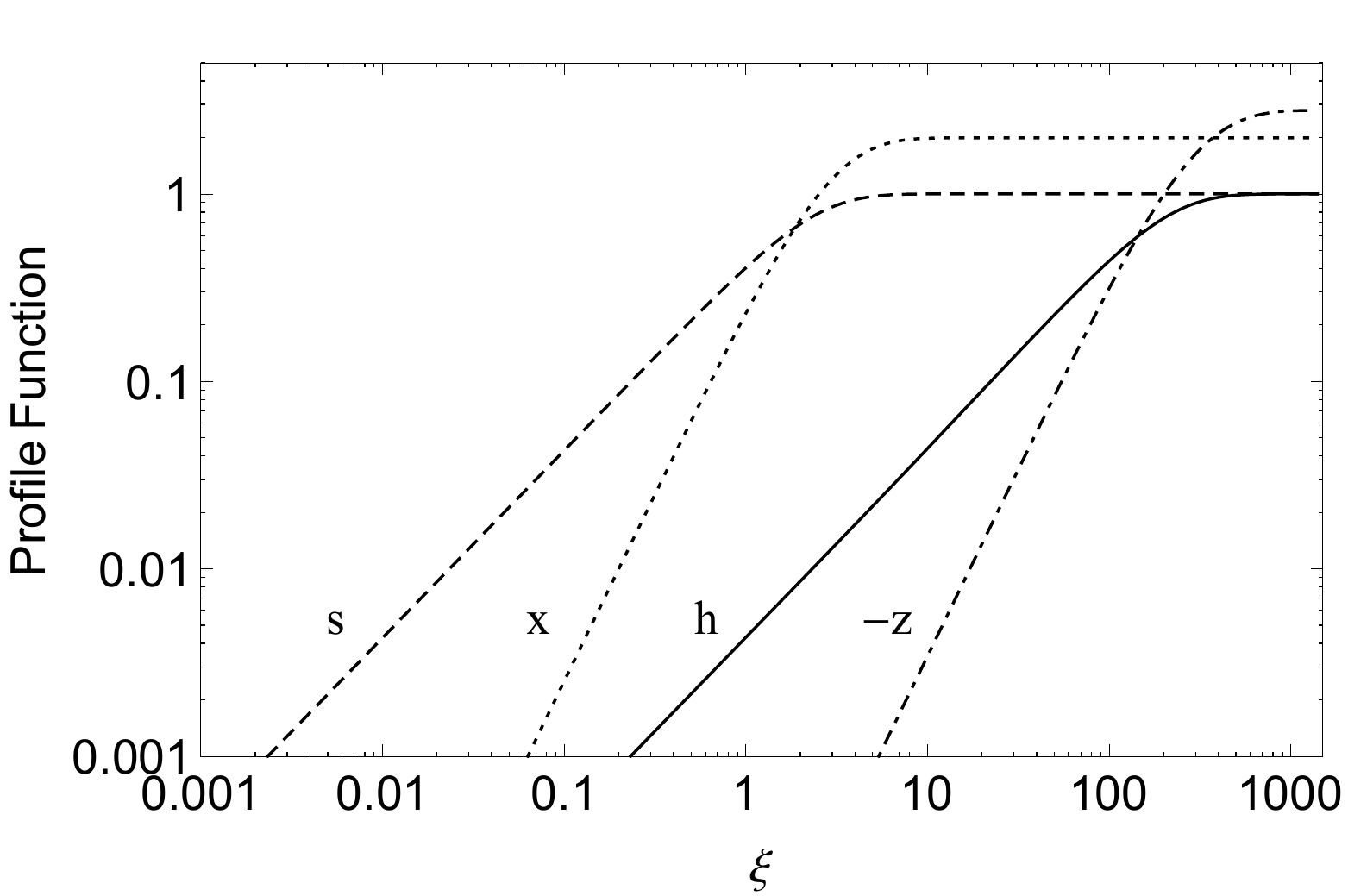}
		\caption{\label{fig:pr11_10tev_log}Profile functions for (1,1) string.}
	\end{subfigure}
	\begin{subfigure}[b]{0.5\textwidth}
		\includegraphics[width=\textwidth]{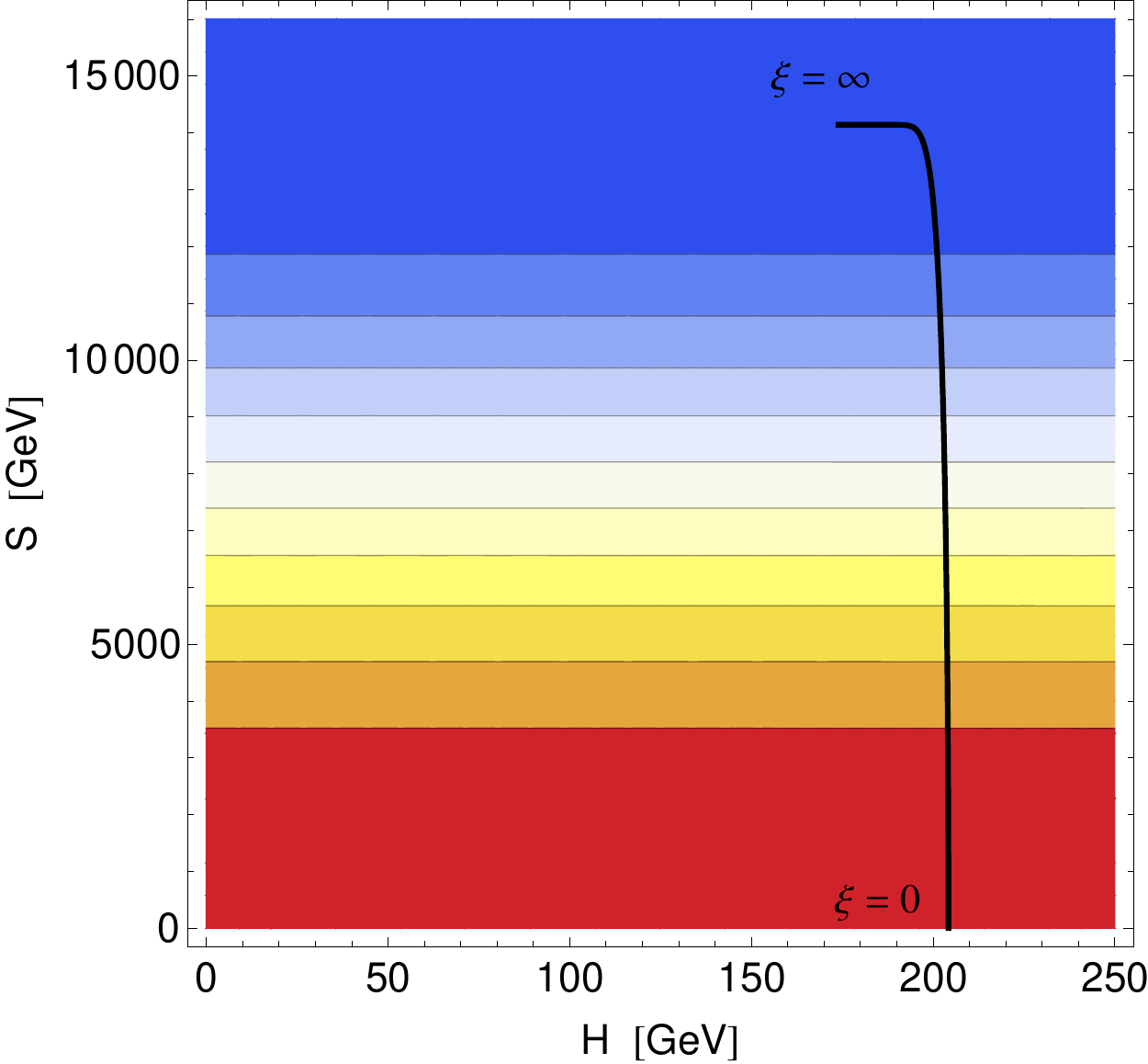}
		\caption{\label{fig:eg01_10tev} (0,1) string solution in (H,S) plane.}
	\end{subfigure}%
	\begin{subfigure}[b]{0.5\textwidth}
		\includegraphics[width=\textwidth]{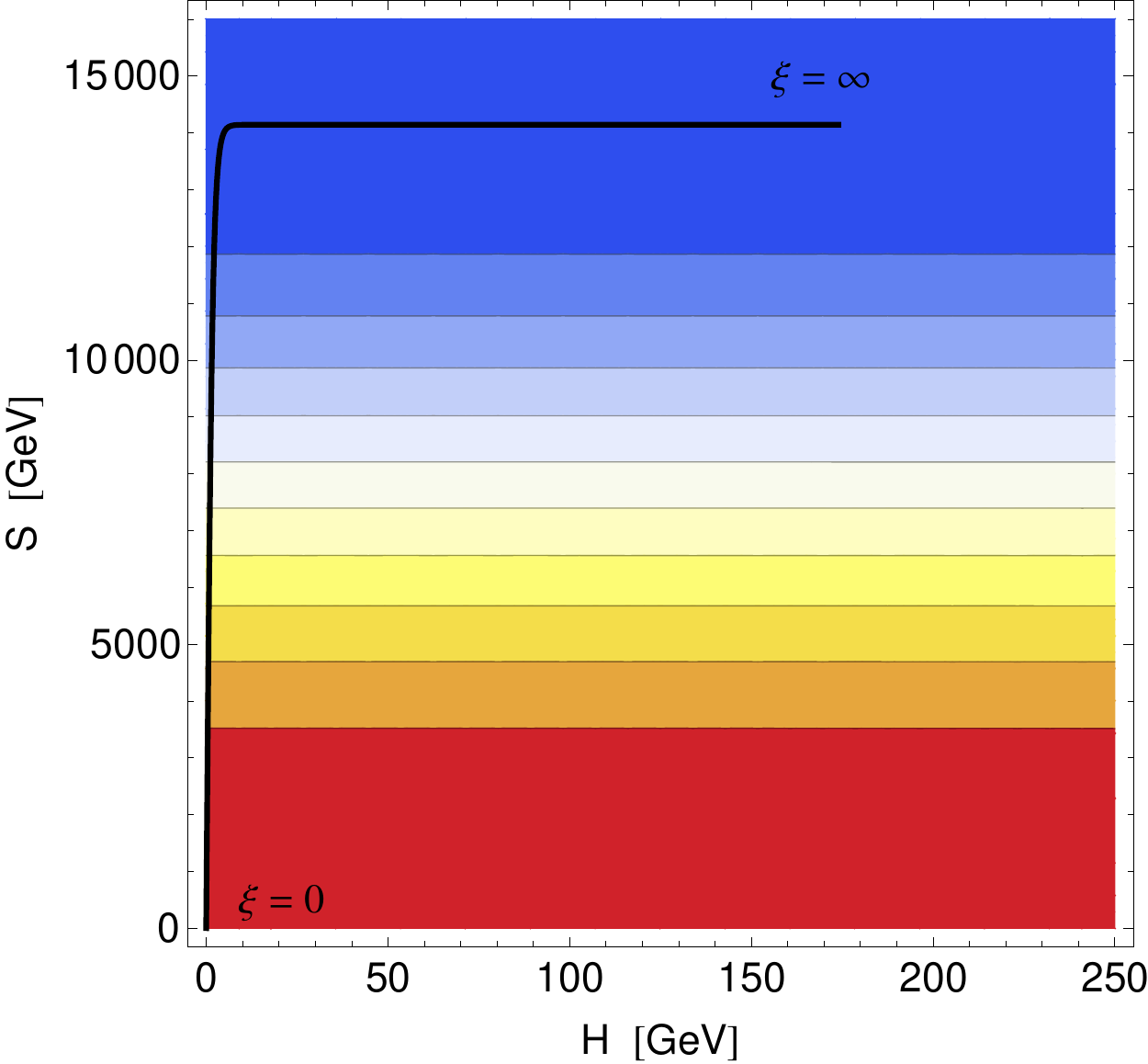}
		\caption{\label{fig:eg11_10tev} (1,1) string solution in (H,S) plane.}
	\end{subfigure}
	\caption{\label{fig:profiles_10tev}
	Same as \fref{fig:profiles} but for $m_X =  M_S = \sigma / \sqrt{2} = 10 \TeV$ and $\alpha = 0.01$.
	}
\end{figure}

In Figures \ref{fig:profiles}, \ref{fig:profiles_1tev}, and \ref{fig:profiles_10tev} we show the profile functions of the $(n,m)=(0,1)$ and $(1,1)$ strings for $M_S = M_X = \sigma / \sqrt{2} = 200 \GeV, 1 \TeV$, and $10 \TeV$.  
In the lower panels, we also show contour plots of the scalar potential, \eref{eq:U1}, where we have overlaid the string trajectories $\left\{ H , S \right\} = \left\{ \eta \, \Hpro(\xi) , \sigma \, \Spro(\xi) \right\}$.  
There are a number of qualitative features which can be seen in these figures that we will discuss at length below.  
First, at the core of the $(0,1)$ strings the Higgs condensate deviates from its vacuum value.
Second, the strings have a tight ``core'' where the gradients of the $S$ and $X^{\mu}$ fields are large, and this core extends out to $\xi = O(1)$ or equivalently the physical length $\rho = O(1 / \sigma)$.  
The $H$ and $Z^{\mu}$ profiles are much wider than the string core.  

\begin{figure}[t!]
	\begin{subfigure}[b]{0.5\textwidth}
		\includegraphics[width=\textwidth]{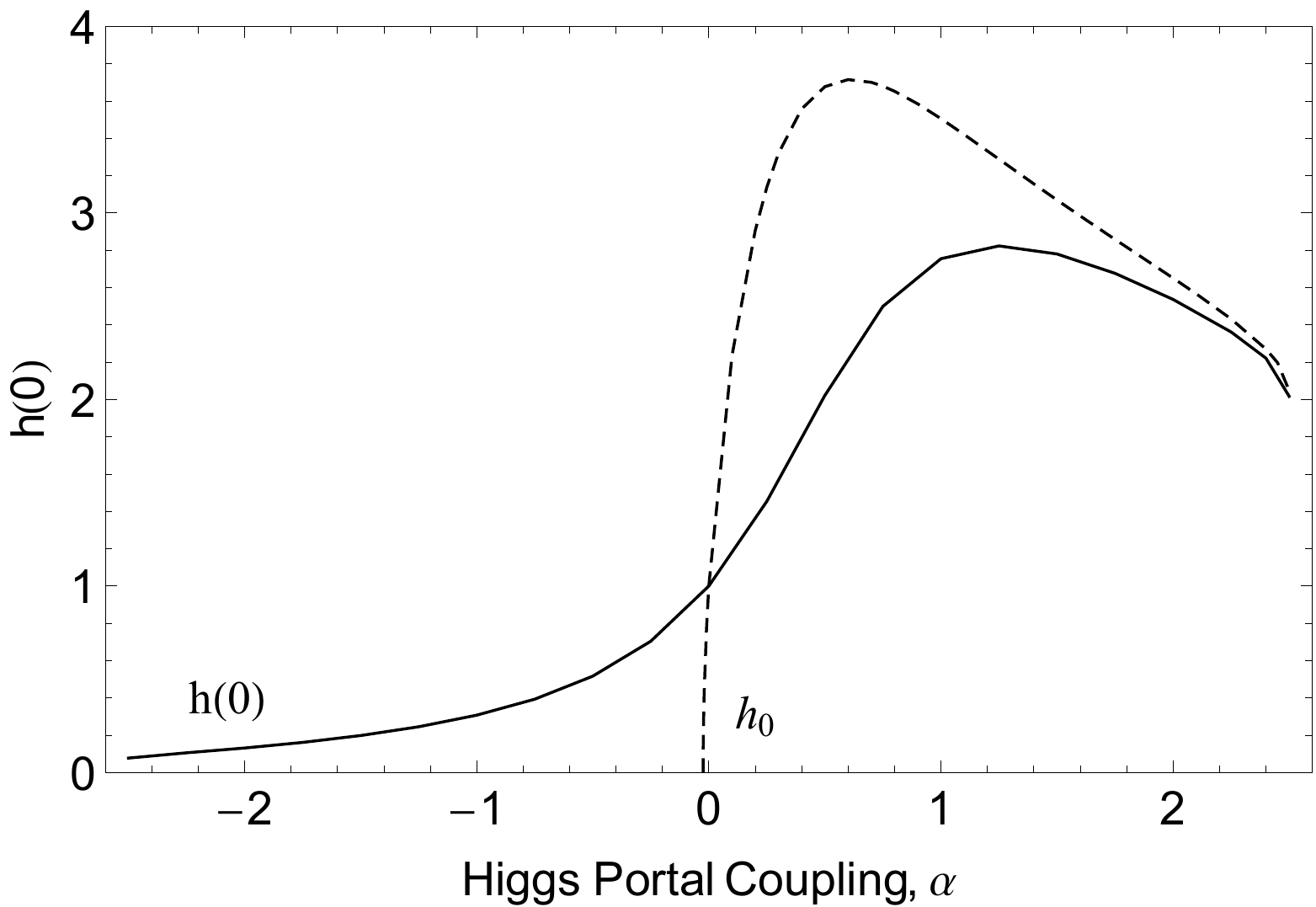}
		\caption{\label{fig:hzero_varyalpha} $M_S = 1 \TeV$, $M_X = 400 \GeV$, $\seps = 0.1$, $\gX=1$.  }
	\end{subfigure}%
	\begin{subfigure}[b]{0.52\textwidth}
		\includegraphics[width=\textwidth]{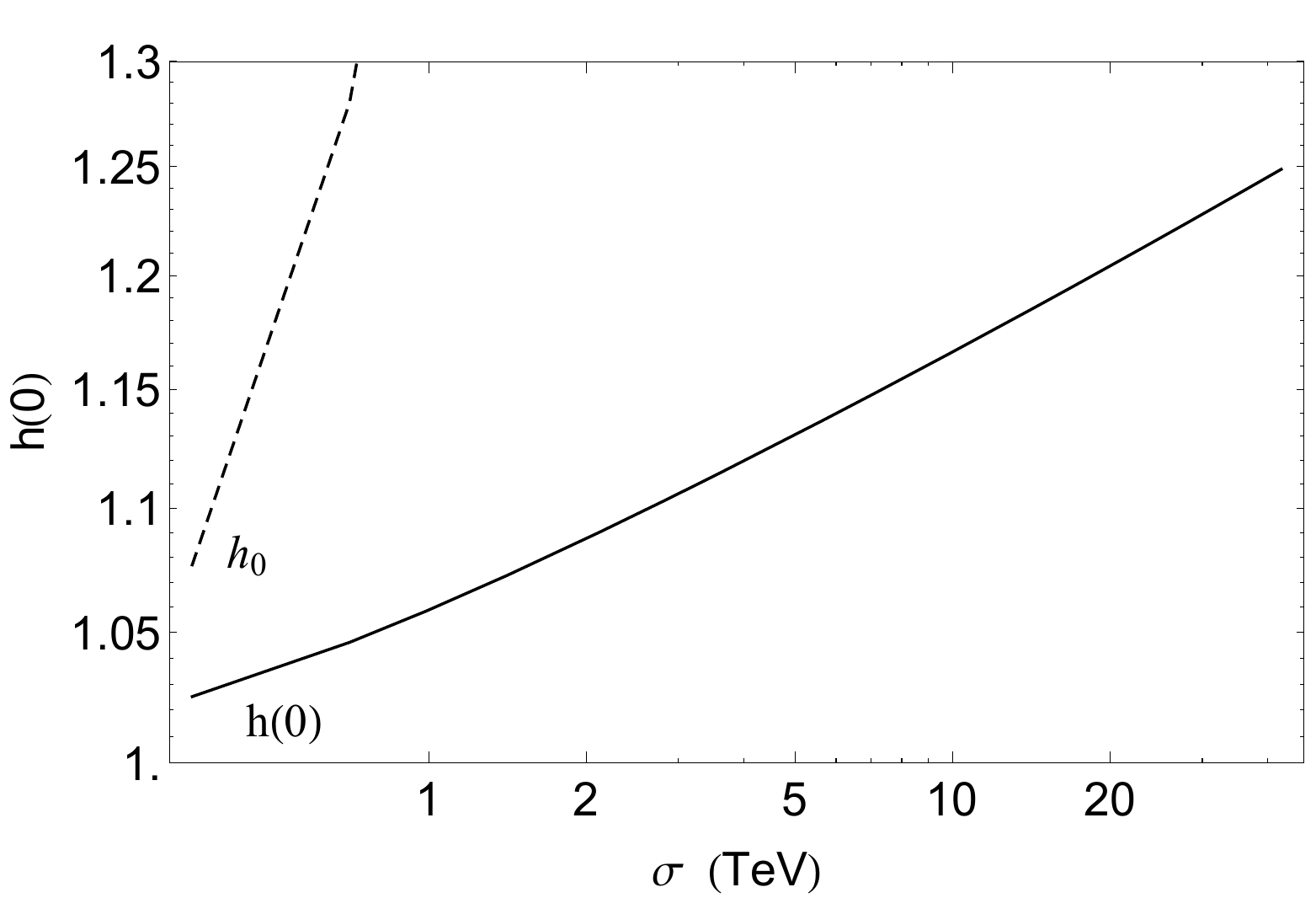}
		\caption{\label{fig:hzero_varysigma} $M_X = M_S = \sigma / \sqrt{2}$, $\alpha = 0.01$, $\seps = 0$, $\gX = 1$}
	\end{subfigure}
		\caption{\label{fig:hzero}
		The Higgs profile at the string core, $\Hpro(\xi = 0)$, for the $(0,1)$ string. For comparison we also show $h_0 = H_0/\eta$ (dashed) where $H_0$ is given by \eref{eq:H0S0_def}. }
\end{figure}

For the $(0,1)$ string, the Higgs field does not wind and satisfies only a Neumann boundary conditions at the origin [see \eref{eq:BC1b}].  
We anticipated in \eref{eq:scalarBC} that the value of the Higgs profile at the core of the $(0,1)$ string should rise or fall toward $\Hpro(0) = \Hpro_0 = \sqrt{1 + \alpha \sigma^2 / 2 \lambda \eta^2}$ depending on the sign of $\alpha$.  
Figures \ref{fig:pr01}, \ref{fig:pr01_1tev}, and \ref{fig:pr01_10tev} reveal that $\Hpro(0) > 1$, indicating that the Higgs condensate is ``attracted'' by the string core in the case $\alpha > 0$.  
Numerically, we find that the magnitude of the deviation is $\abs{\Hpro(0) - 1} \approx O(0.1 - 1)$, depending on the parameter choices.  
In some cases we find $\Hpro(0) \lesssim \Hpro_0$, which confirms the energetic arguments that led to \eref{eq:scalarBC}, whereas in other cases $\Hpro(0) \ll \Hpro_0$ suggesting that the tension is dominated by gradient energy instead of potential energy, and our previous estimate breaks down.  
We compare $\Hpro(0)$ and $\Hpro_0$ in \fref{fig:hzero} where we plot both quantities against $\alpha$ (left panel) and $\sigma$ (right panel).  
For large values of $\alpha$, both $\Hpro(0)$ and $\Hpro_0 = \sqrt{1 + \alpha \sigma^2 / (2 \lambda \eta^2)}$ reach a maximum and turn over.  
To understand this behavior, recall that $\lambda$ is allowed to vary with $\alpha$ according to \eref{eq:lamkap} while $M_H$ and $M_S$ are held fixed, and therefore $\Hpro_0 \sim \sqrt{\alpha / \lambda}$ is not monotonically increasing with $\alpha$.  
For negative values of $\alpha$, \fref{fig:hzero_varyalpha} reveals that $\Hpro(0)$ asymptotes toward zero whereas $\Hpro_0$ vanishes at $\alpha = - 2 \lambda \eta^2 / \sigma^2$.  
In this case, the Higgs condensate is ``repelled'' by the string core.  
We show the behavior of $\Hpro(0)$ and $\Hpro_0$ in the decoupling limit, $\sigma \gg \eta$, in \fref{fig:hzero_varysigma}.  
In this limit, $\Hpro_0 \sim \sigma / \eta$ grows rapidly, but the value of the condensate at the string core, $\Hpro(0)$, rises much more slowly.  

\begin{figure}[t!]
	\begin{subfigure}[b]{0.50\textwidth}
		\includegraphics[width=\textwidth]{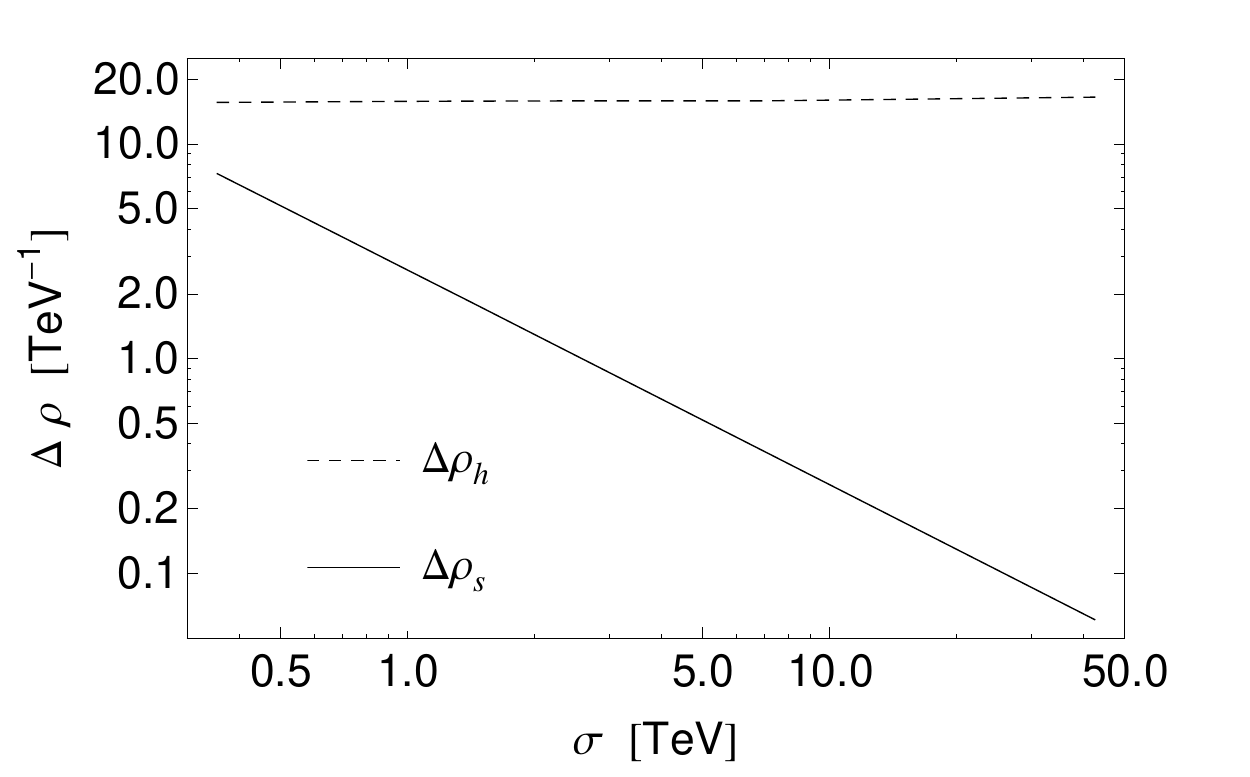}
		\caption{\label{fig:decoupling_11} $(n,m) = (1,1)$  }
	\end{subfigure}%
	\begin{subfigure}[b]{0.50\textwidth}
		\includegraphics[width=\textwidth]{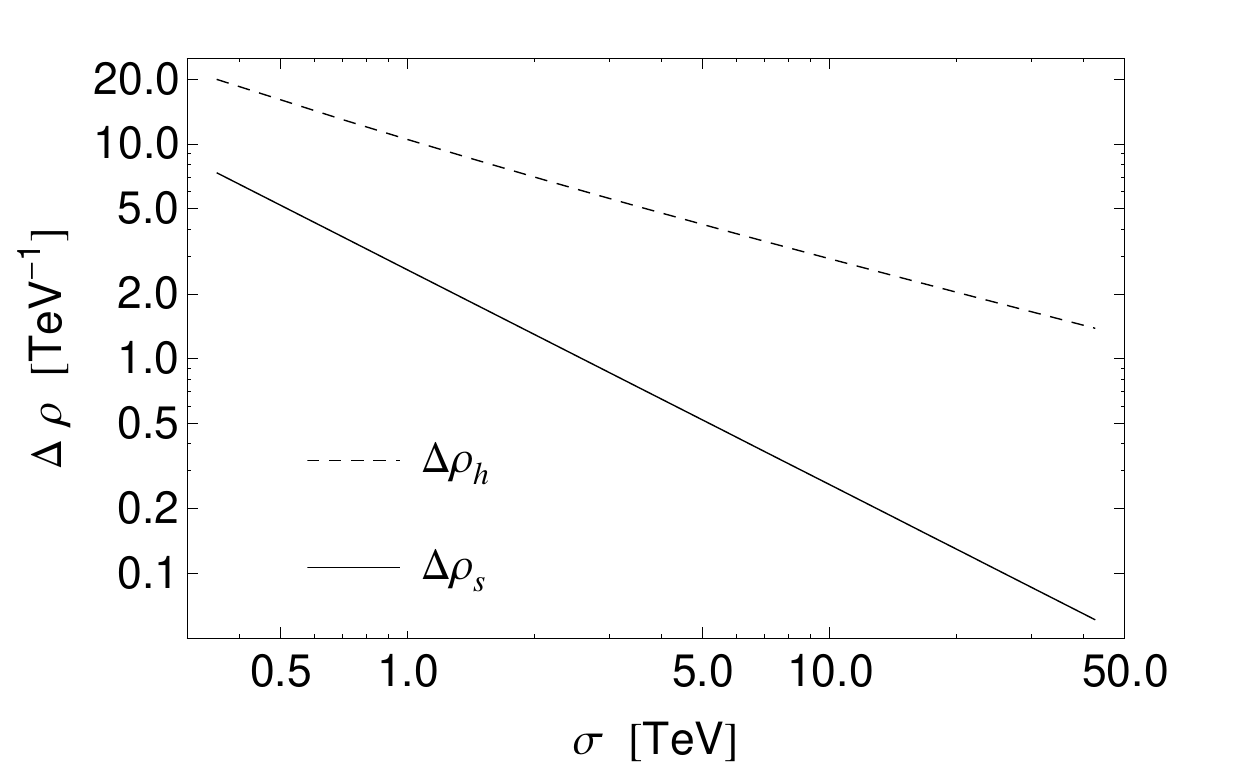}
		\caption{\label{fig:decoupling_01} $(n,m) = (0,1)$  }
	\end{subfigure}%
		\caption{\label{fig:decoupling}  The widths of the scalar field condensates ($H$ dashed; $S$ solid) surrounding the dark string.  See \eref{eq:string_widths}.  The parameters are taken to be $M_X = M_S = \sigma / \sqrt{2}$, $\gX = 1$, $\alpha = 0.01$, and $\seps = 0$.  In both cases, $\Delta \rho_{s} \simeq 2 / M_S$, but $\Delta \rho_{h} \simeq 2 / M_H$ for the $(1,1)$ string, and it decreases gradually for the $(0,1)$ string.  
		}
\end{figure}

In order to better characterize the string solution, we calculate the ``full width at half maximum'' of the scalar profile functions.  
In terms of the dimensionless radial coordinate, these are given by the solutions of $\Hpro(\xi_h / 2) = \Hpro(0) / 2$ and $\Spro(\xi_s / 2) = \Spro(0) / 2$.  
Figure \ref{fig:decoupling} shows the physical widths
\begin{align}\label{eq:string_widths}
	\Delta \rho_h = \rho_0 \xi_h
	\qquad {\rm and} \qquad 
	\Delta \rho_s = \rho_0 \xi_s
\end{align}
for the $(0,1)$ and $(1,1)$ strings as a function of $\sigma$.  
In both cases the width of the $S$ condensate falls off like $\Delta \rho_s \simeq 2 / M_S = 2 \sqrt{2} / \sigma$.  
The Higgs condensate, on the other hand, has a significantly different behavior in the two cases.  
For the $(1,1)$ string the width of the Higgs condensate is insensitive to $\sigma$ and remains approximately equal to $\Delta \rho_h \simeq 2 / M_H \approx 16 \TeV^{-1}$.  
For the $(0,1)$ string the Higgs condensate is narrower, and its width decreases with increasing $\sigma$, but not as fast as $\sigma^{-1}$.  

Let us now take Figures \ref{fig:profiles}--\ref{fig:decoupling} together, and construct a coherent picture of the $(0,1)$ dark string.  
The behavior is similar to what is seen in the familiar case of bosonic superconductivity \cite{Witten:1984eb}.  
When $\seps = \alpha = 0$ the $S$ and $X^{\mu}$ fields form a Nielsen-Olesen string and the Higgs condensate is equal to its vacuum value everywhere.  
Roughy speaking, the Higgs field is unaware of the presence of the string since there is no coupling between them.  
For $\alpha > 0$ ($\alpha < 0$) the Higgs condensate is ``attracted'' (``repelled'') by the string and $\Hpro(0) > 1$ ($\Hpro(0) < 1$).  
In the decoupling limit, $\sigma \sim M_S \gg \eta \sim M_H$, and with $\alpha > 0$, the saddle point moves to $\Hpro_0 \gg 1$, but the tension becomes gradient dominated and $\Hpro(0) \ll \Hpro_0$, contrary to expectations.  
The $S$ and $X$ profiles fall off on a length scale $2 / M_S$, which defines the string core.  
The Higgs condensate, however, forms a wide halo around the core.  
For a $10 \TeV$ scale string, the halo is approximately an order of magnitude wider than the core, but it is still smaller than $2 / M_H$ by another order of magnitude.  

\subsection{Tension}\label{sub:Tension}

The tension of the dark string is defined by $\mu \equiv \int_{0}^{\infty} \rho d \rho \int_{0}^{2\pi} d \varphi \, \tensor{T}{^{0}_{0}}$ where $T^{\mu \nu}$ is the energy-momentum tensor.  
Inserting the dark string ansatz, \eref{eq:Ansatz}, this becomes
\begin{equation}\label{eq:tension}
	\mu = 2 \pi \sigma^2 \int_{0}^{\infty} \mathcal{E} \, \xi \, d\xi
\end{equation}
where 
\begin{equation}\label{eq:radial_energy}
	\mathcal{E} = \mathcal{E}_X + \mathcal{E}_Z + \mathcal{E}_H + \mathcal{E}_S + u 
\end{equation}
is the dimensionless energy density, which consists of contributions from each of the fields:  
\begin{subequations}\label{eq:tension_int}
\begin{align}
\label{eq:tension_x}
	\mathcal{E}_X =
	& \frac{1}{(\sigma\rho_0)^2} \frac{(\Xpro^{\prime})^2}{2\xi^2} \\
\label{eq:tension_z}
	\mathcal{E}_Z =
	& \frac{1}{(\sigma\rho_0)^2} \frac{(\Zpro^{\prime})^2}{2\xi^2} \\
\label{eq:tension_h}
	\mathcal{E}_H =
	& \left(\frac{\eta}{\sigma}\right)^2 \left( (\Hpro^{\prime})^2 + \frac{\Hpro^2}{\xi^2} C_H^2 \right) \\
\label{eq:tension_s}
	\mathcal{E}_S =
	& \left( (\Spro^{\prime})^2 + \frac{\Spro^2}{\xi^2} C_S^2 \right) \\
\label{eq:tension_u}
	u =
	& \lambda (\rho_0\sigma)^2\left(\frac{\eta}{\sigma}\right)^4 \left(\Hpro^2 - 1\right)^2 
	+ \kappa (\rho_0\sigma)^2 \left(\Spro^2 - 1\right)^2
	+ \alpha (\rho_0\eta)^2 \left( \Hpro^2 - 1 \right)\left( \Spro^2 - 1 \right) \per
\end{align}
\end{subequations}
For the special case $\epsilon = \alpha =0$ we have $C_H = n - \gHZ \, z$ and $C_S = m - \gSX \, x$ [see \eref{eq:C_def}].  
Thus, as expected, in the absence of interactions between the SM and dark sector the energy reduces to the sum of energies of two separate Nielsen--Olesen strings. 
In particular, for a $(0,1)$ string with $\epsilon = \alpha = 0$ and $m_S = m_X$, the integral in \eref{eq:tension} numerically evaluates to 1, and we find the tension to be $\mu = 2\pi\sigma^2$. 
From the individual terms in \eref{eq:tension_int}, we can see that with our choice $\rho_0 = \sigma^{-1}$, some terms are independent of $\sigma$ and the rest go as $(\eta/\sigma)^{2}$ or $(\eta / \sigma)^{4}$. 
Thus when $\sigma \sim \eta$, the tension will not follow a simple power law, but when $\sigma \gg \eta$, it will increase as $\sigma^2$. 
The terms that scale as inverse powers of $\sigma$ are more significant for the $(1,1)$ string than for the $(0,1)$ string, so we would expect the $(0,1)$ string tension to essentially scale as $\sigma^2$ even for $\sigma \sim \eta$. 

\begin{figure}[t!]
	\begin{subfigure}[b]{0.45\textwidth}
		\includegraphics[width=\textwidth]{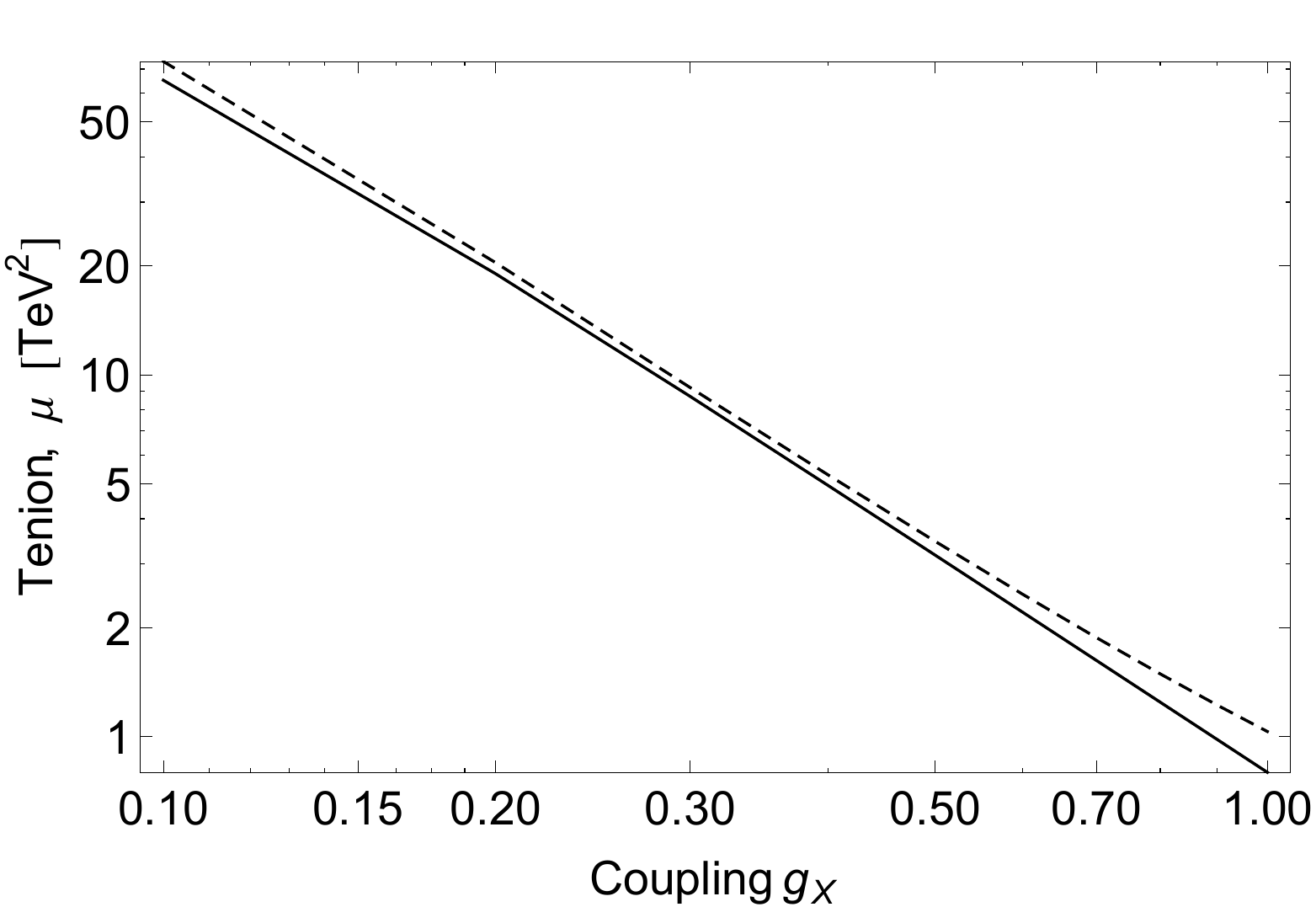}
		\caption{\label{fig:tgx} $3m_X = M_S = 600 \GeV$, $\seps = 0.1$, $\alpha = 0.1$. }
	\end{subfigure}%
	\hfill
	\begin{subfigure}[b]{0.45\textwidth}
		\includegraphics[width=\textwidth]{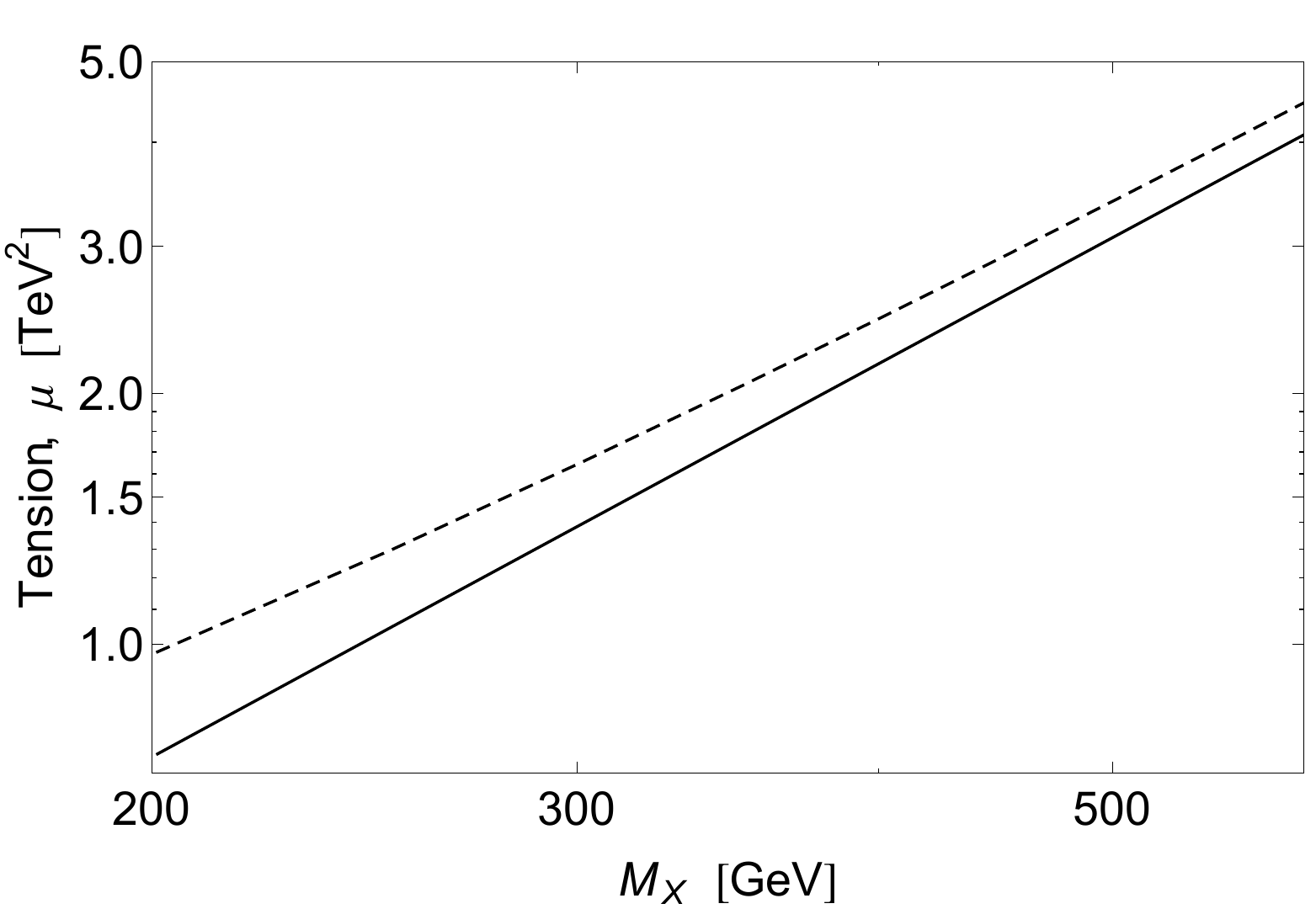}
		\caption{\label{fig:tmx} $M_S = 500 \GeV$, $\seps = 0.1$, $\alpha = 0.1$, $\gX = 1$.}
	\end{subfigure}
	\hfill
	\begin{subfigure}[b]{0.45\textwidth}
		\includegraphics[width=\textwidth]{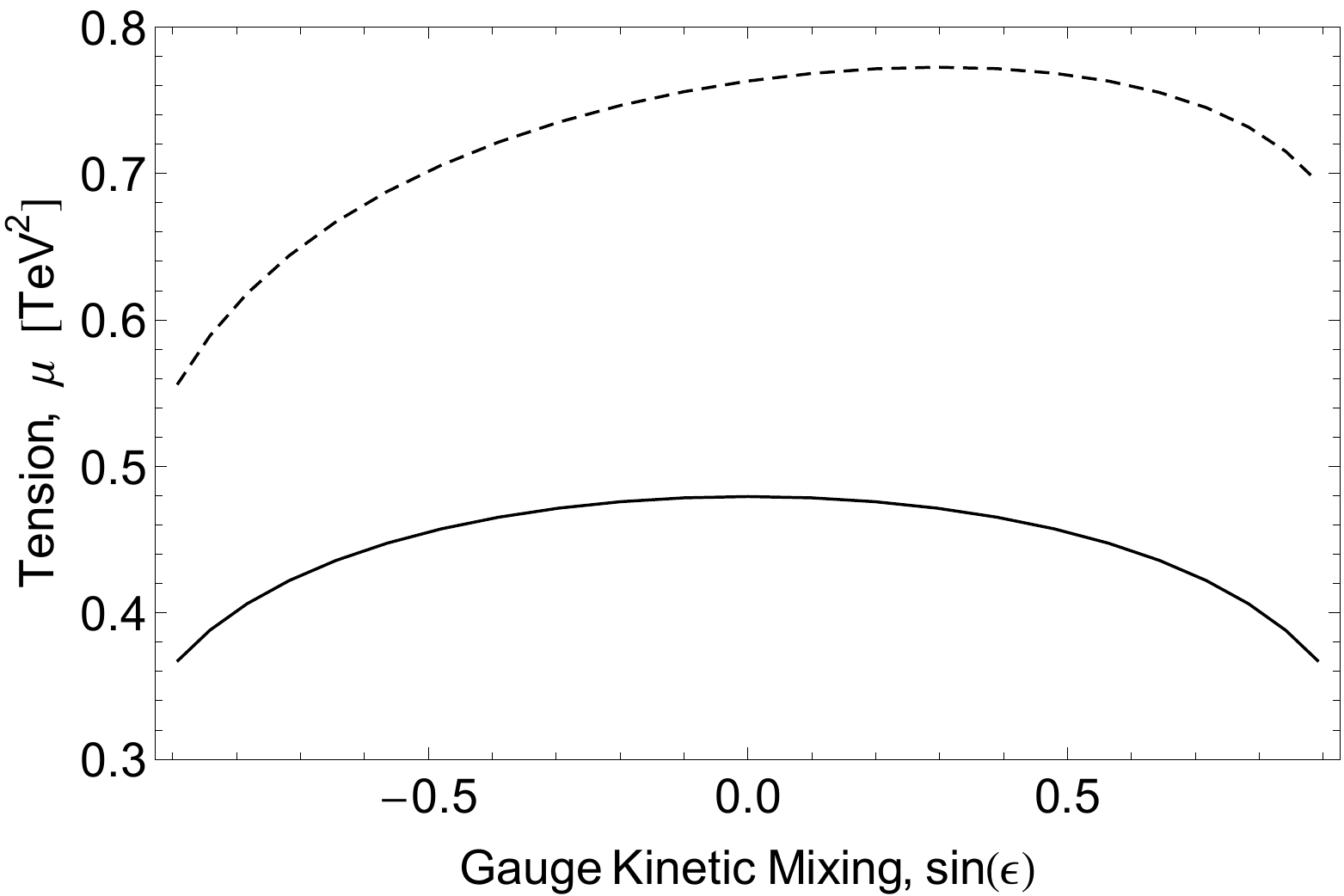}
		\caption{\label{fig:t200200}$m_X = M_S = 200 \GeV$, $\alpha = 0.1$, $\gX = 1$}
	\end{subfigure}%
	\hfill
	\begin{subfigure}[b]{0.45\textwidth}
		\includegraphics[width=\textwidth]{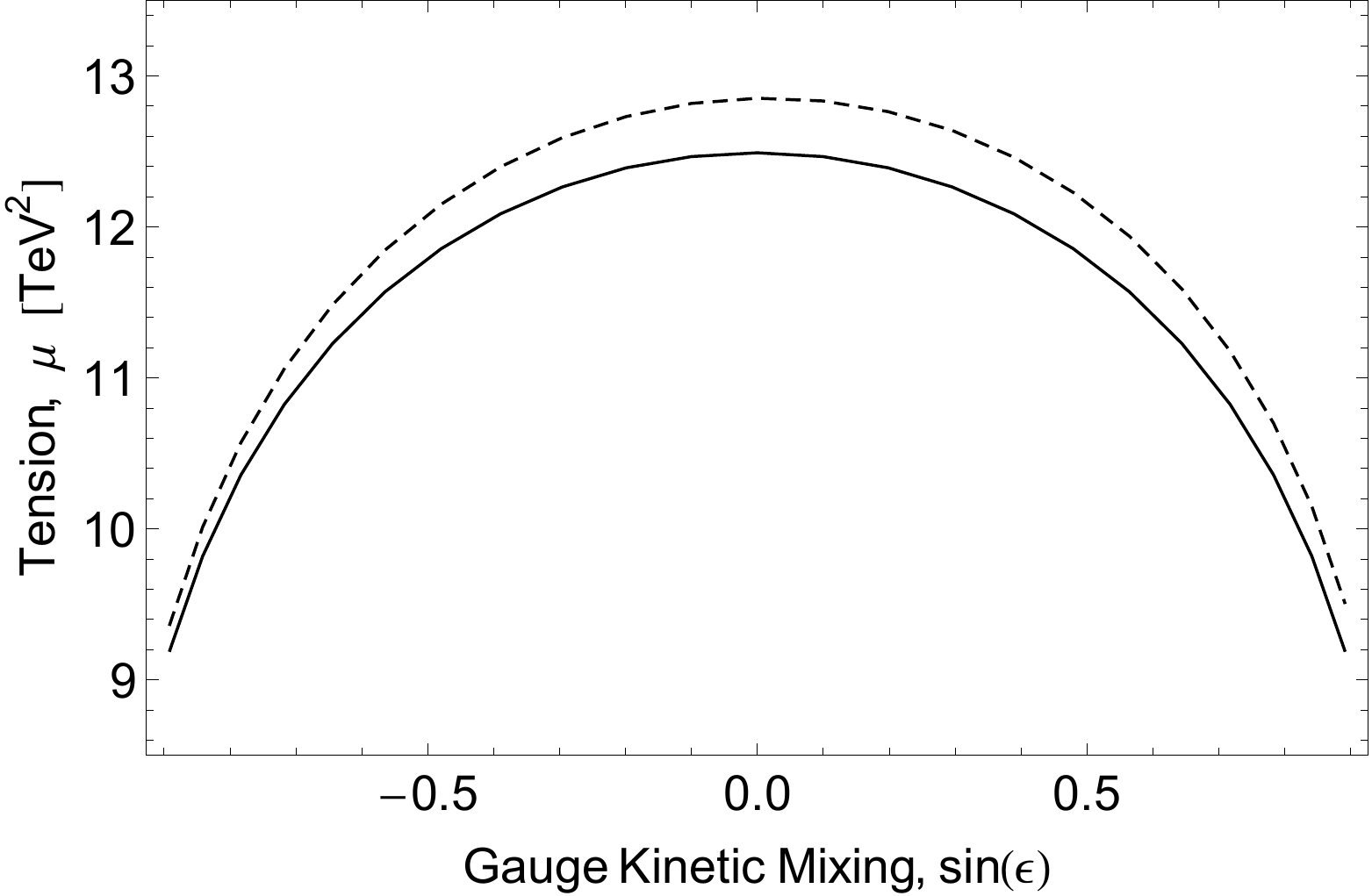}
		\caption{\label{fig:t10001000}$m_X = M_S = 1 \TeV$, $\alpha = 0.1$, $\gX = 1$}
	\end{subfigure}
	\hfill
	\begin{subfigure}[b]{0.45\textwidth}
		\includegraphics[width=\textwidth]{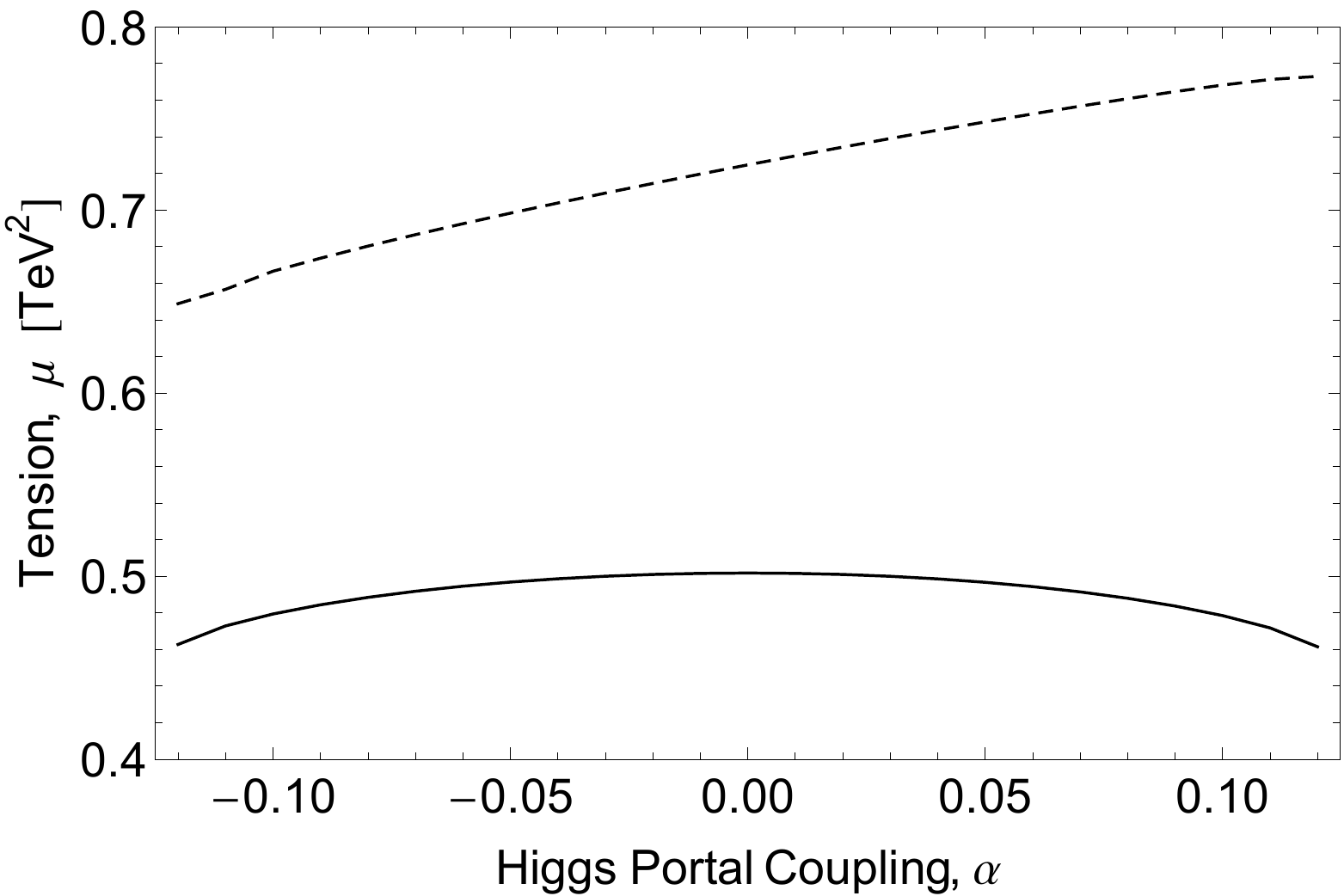}
		\caption{\label{fig:ta} $m_X = M_S = 200 \GeV$, $\seps = 0.1$, $\gX = 1$. }
	\end{subfigure}%
	\hfill
	\begin{subfigure}[b]{0.45\textwidth}
		\includegraphics[width=\textwidth]{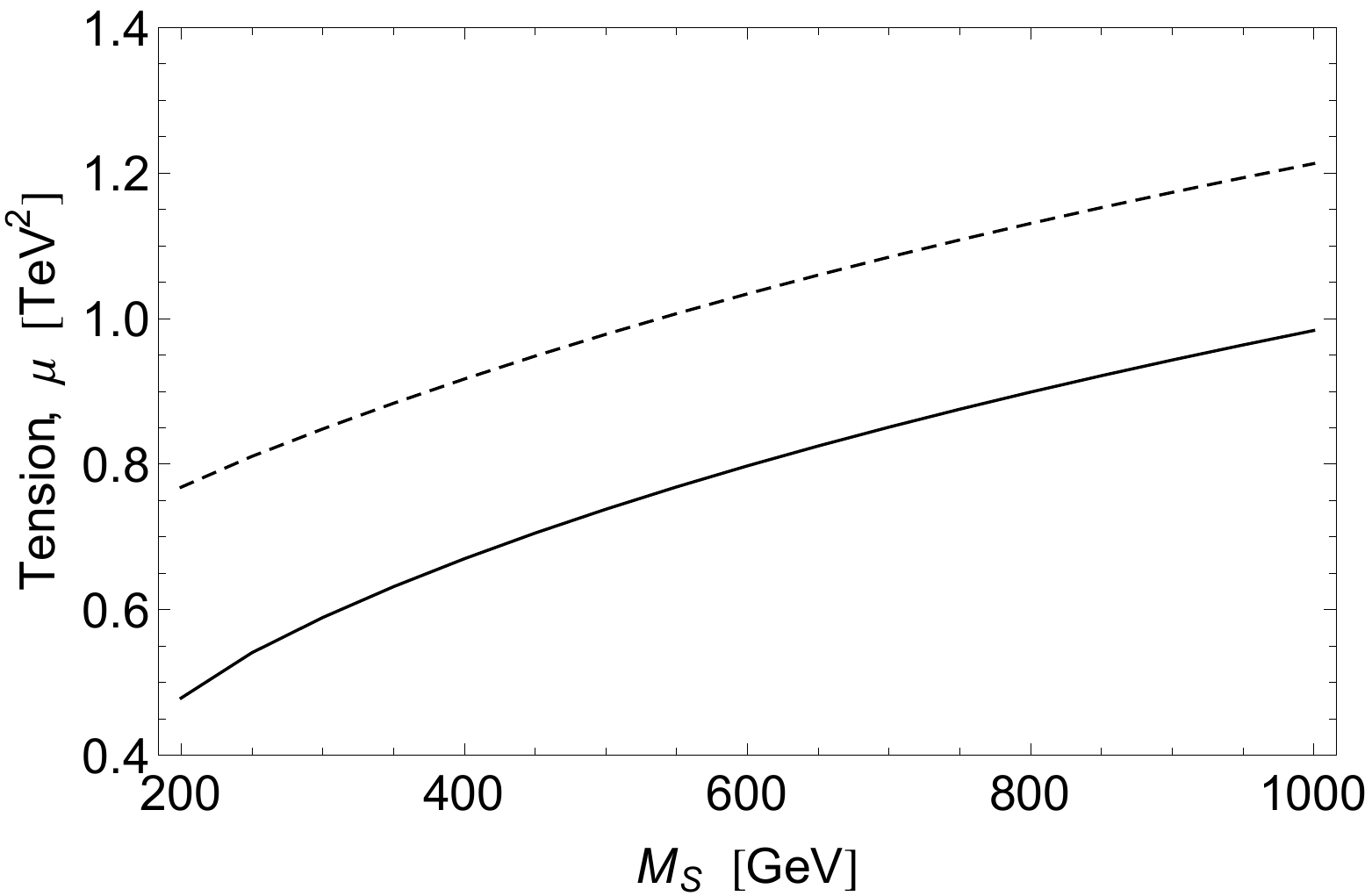}
		\caption{\label{fig:tms} $m_X = 200 \GeV$, $\seps = 0.1$, $\alpha = 0.1$, $\gX = 1$.}
	\end{subfigure}
	\caption{\label{fig:tension2}Tension of $(0,1)$ (solid) and $(1,1)$ (dashed) strings.  }
\end{figure}

Figure \ref{fig:tension2} compares the tension of the $(1,1)$ and $(0,1)$ strings along various slices of parameter space.  
Each subfigure illustrates that the tension of the $(0,1)$ string is always smaller than the tension of the $(1,1)$ string.  
The scaling behavior mentioned above is evident in Figures \ref{fig:tgx} and \ref{fig:tmx}.
Figure \ref{fig:tgx} shows the tension as a function of the $\U{1}_X$ gauge coupling, $\gX$, and it is seen that the tension scales like $\mu \propto \gX^{-2}$.  
This scaling is understood by noting that we hold $m_X = \gX \sigma / \sqrt{2}$ fixed and vary $\sigma \propto \gX^{-1}$.  
Then the figure simply shows that $\mu \propto \sigma^2$.  
Figure \ref{fig:tmx} shows the tension as a function of the mass of the $X$ gauge boson, and since we are now holding $\gX$ fixed and varying $\sigma \propto m_X$, this figure also shows that $\mu \propto \sigma^2$.
In both cases, the $(0,1)$ string tension scales as $\sigma^2$ for all values of $\sigma$, while the $(1,1)$ string tension departs from this behavior at the lower values of $\sigma$.

Figures \ref{fig:t200200} and \ref{fig:t10001000} show how the tension depends on the GKM parameter $\seps$.
From these it can be seen that the tension decreases monotonically with increasing $\abs{\seps}$ for the $(0,1)$ string and almost monotonically for the $(1,1)$ string.  
This behavior can be understood by noting that the gauge kinetic terms of the original Lagrangian, \eref{eq:L1}, can be written as 
\begin{align}
	\Lcal \ni -\frac{1}{4} \left( \frac{1 + \seps}{2} \right) \bigl( Y_{\mu \nu} + \hat{X}_{\mu \nu} \bigr)^2 - \frac{1}{4} \left( \frac{1 - \seps}{2} \right) \bigl( Y_{\mu \nu} - \hat{X}_{\mu \nu} \bigr)^2 \per  
\end{align}
In the limit $\seps \to \pm 1$ it ``costs no energy'' to excite the gauge field $Y_{\mu} \mp \hat{X}_{\mu}$, and the tension of the string is reduced.   
Here it is important to note that we hold fixed the parameter $m_{X} = \gX \sigma / \sqrt{2}$, which differs from the mass eigenvalue $M_X$ for nonzero $\seps$ [see \eref{eq:gauge_evals}].  
In \ref{fig:t200200}, for example, at $\seps = 0$ we have $M_X = 200 \GeV$, while at $\abs{\seps} = 0.9$ it has increased to $M_X = 450 \GeV$.

The dependence of the tension on $\alpha$ is shown in \fref{fig:ta}.  
For the $(1,1)$ string, the tension rises nearly linearly with $\alpha$, whereas for the $(0,1)$ string the tension is symmetric in $\alpha$.  
This parametric behavior is understood by noting that at the core of the $(1,1)$ string the profile functions become $\Spro(0) = \Hpro(0) = 0$, while at the core of the $(0,1)$ string they become $\Spro(0) = 0$ and $\Hpro(0) = 1 + O(\alpha)$ [see \eref{eq:BC1b}].  
The tension depends on $\alpha$ primarily through the potential energy density, $u(h,s)$, given by \eref{eq:tension_u}.  
The parametric behavior of the tension is estimated by $\mu^{(1,1)} \sim u(0,0) = \lambda \eta^4 + \kappa \sigma^4 + \alpha \eta^2 \sigma^2$ for the $(1,1)$ string and by $\mu^{(0,1)} \sim u(1+O(\alpha), 0) = \lambda \eta^4 + \kappa \sigma^4 + O(\alpha^2)$ for the $(0,1)$ string. 
In this way, the dependence on $\alpha$ seen in \fref{fig:ta} is explained.  

Finally, let us remark that our string solutions and tension are consistent with the results available in the literature.  
The authors of \rref{Brihaye:2009fs} considered a model similar to ours, in which they
include a gauge kinetic mixing term but no Higgs portal term. They also take the semilocal
limit $\sw =1$.
Our model reduces to theirs upon setting $\alpha=0$, $\sw = 1$, and $M_H = 125 \GeV$.
For a particular parameter range given in Figure 3 of \rref{Brihaye:2009fs}, we calculate the string tension and find agreement to better than $O(1\%)$.

\subsection{Coupling of the Higgs to the String}\label{sub:Coupling_to_H}

The dark string acts as a source for the scalar fields $H$ and $S$.  
This source causes the fields to locally deviate from their vacuum expectation values and to form a long range ``cloud'' around the string core.  
As discussed in \sref{sub:ScalarSector}, we can parametrize the fields as $H = (\eta + \bar{h} / \sqrt{2}) e^{i a_H}$ and $S = (\sigma + \bar{s} / \sqrt{2}) e^{i a_S}$, and the physical scalars, $\bar{h}$ and $\bar{s}$, mix with one another with a mixing angle $\theta$, given by \eref{eq:mixing}.  
Only the lighter Higgs-like mass eigenstate, $\phi_H = \cos \theta \, \bar{h} - \sin \theta \, \bar{s}$, can be radiated efficiently from the dark string since the $S$-like eigenstate, $\phi_S$, has a mass comparable to the string tension.  
We therefore are only interested in the effective coupling of $\phi_H$ to the dark string.  
 
\begin{figure}[t!]
		\includegraphics[width=0.7\textwidth]{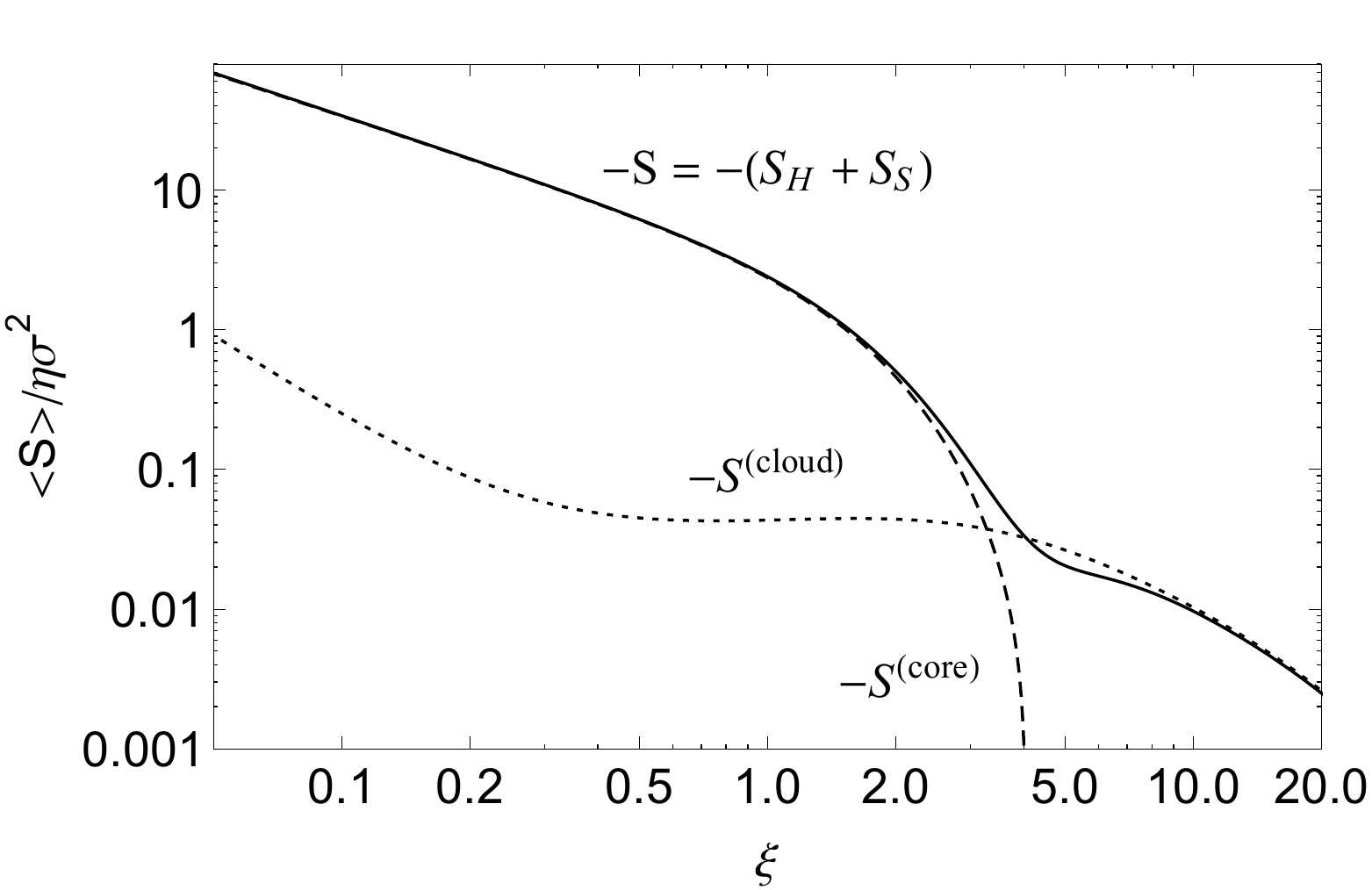}
	\caption{\label{fig:scalar_source} Vacuum expectation value of $\mathcal{S}$ as a function of the scaled radial coordinate $\xi$, where $\mathcal{S}_H$ and $\mathcal{S}_S$ are given by \eref{eq:scalar_source_terms}, and we have defined $\mathcal{S}^{\rm (core)} = \mathcal{S}_{H}^{\rm (core)} + \mathcal{S}_{S}^{\rm (core)}$ and $\mathcal{S}^{\rm (cloud)} = \mathcal{S}_{H}^{\rm (cloud)} + \mathcal{S}_{S}^{\rm (cloud)}$. We have held fixed $M_S = M_X = 1 \TeV$, $\alpha = 0.1$, $\epsilon = 0.1$ and $\gX = 1$.}
\end{figure}

The field equations for $H$ and $S$, given previously by \eref{eq:fieldeqns2}, may be written as follows after expanding out the covariant derivatives:  
\begin{align}
\label{eq:EOM_for_H}
 	\Box H = \ & 
	i \left( \gHZ \partial_{\mu} Z^{\mu} + \gHX \partial_{\mu} X^{\mu} \right) H 
	+ 2 i \left( \gHZ Z^{\mu} + \gHX X^{\mu} \right) \partial_{\mu} H 
	+ \left( \gHZ Z^{\mu} + \gHX X^{\mu} \right)^2 H \nn
	& - 2 \lambda \left( \abs{H}^2 - \eta^2 \right) H 
	- \alpha \left( \abs{S}^2 - \sigma^2 \right) H \\
\label{eq:EOM_for_S}
	\Box S = \ & 
	i \left( \gSZ \partial_{\mu} Z^{\mu} + \gSX \partial_{\mu} X^{\mu} \right) S 
	+ 2 i \left( \gSZ Z^{\mu} + \gSX X^{\mu} \right) \partial_{\mu} S 
	+ \left( \gSZ Z^{\mu} + \gSX X^{\mu} \right)^2 S \nn
	& - 2 \kappa \left( \abs{S}^2 - \sigma^2 \right) S 
	- \alpha \left( \abs{H}^2 - \eta^2 \right) S \per
\end{align}
In the vicinity of the dark string, the fields acquire position-dependent expectation values, and the interactions on the right hand side of these equations become  source terms.  
In order to illustrate the nature of this source, we can evaluate the right hand sides of Eqns.~(\ref{eq:EOM_for_H})~and~(\ref{eq:EOM_for_S}), denoted as $\mathcal{S}_{H}$ and $\mathcal{S}_{S}$ respectively, in the presence of the string background, given by \eref{eq:Ansatz}.  
Doing so we obtain
\begin{align}\label{eq:scalar_source_terms}
	\mathcal{S}_{H} = \mathcal{S}_{H}^{\rm (core)} + \mathcal{S}_{H}^{\rm (cloud)} 
	\qquad {\rm and} \qquad 
	\mathcal{S}_{S} = \mathcal{S}_{S}^{\rm (core)} + \mathcal{S}_{S}^{\rm (cloud)} 
\end{align}
where
\begin{align}
	\mathcal{S}_{H}^{\rm (core)} & \equiv - \frac{\eta}{\rho_0^2} \frac{\Hpro}{\xi^2} \left( \gHZ \Zpro(\infty) + \gHX \Xpro \right)^2 + \alpha \eta \sigma^2 \left( 1 - \Spro^2 \right) \Hpro 
	+ \frac{\eta}{\rho_0^2} \frac{\Hpro}{\xi^2} (\gHZ)^2 \Zpro(\infty)^2 (1 - \Spro^2) \\
	\mathcal{S}_{H}^{\rm (cloud)} & \equiv 
	- 2 \lambda \eta^3 \left( \Hpro^2 - 1 \right) \Hpro 
	- \frac{\eta}{\rho_0^2} \frac{\Hpro}{\xi^2} (\gHZ)^2 \left( \Zpro^2 - \Zpro(\infty)^2 \right) 
	- 2 \frac{\eta}{\rho_0^2} \frac{\Hpro}{\xi^2} (\gHZ \gHX) \Xpro \left( \Zpro - \Zpro(\infty) \right) \nn
	& \qquad - \frac{\eta}{\rho_0^2} \frac{\Hpro}{\xi^2} (\gHZ)^2 \Zpro(\infty)^2 (1 - \Spro^2) \\
	\mathcal{S}_{S}^{\rm (core)} & \equiv - \frac{\sigma}{\rho_0^2} \frac{\Spro}{\xi^2} \left( 1 - \gSZ \Zpro(\infty) - \gSX \Xpro \right)^2 + 2 \kappa \sigma^2 \left( 1 - \Spro^2 \right) \Spro \\
	\mathcal{S}_{S}^{\rm (cloud)} & \equiv - \alpha \sigma \eta^2 \left( \Hpro^2 - 1 \right) \Spro - \frac{\sigma}{\rho_0^2} \frac{\Spro}{\xi^2} \left[ (\gSZ)^2 \left( \Zpro + \Zpro(\infty) \right) - 2\gSZ (1 - \gSX \Xpro ) \right] \left( \Zpro - \Zpro(\infty) \right) \com
\end{align}
and where $\Zpro(\infty)$ is given by \eref{eq:gaugeBC} with $(n,m) = (0,1)$.  
We have added and subtracted the term $- \frac{\eta}{\rho_0^2} \frac{\Hpro}{\xi^2} (\gHZ)^2 \Zpro(\infty)^2 (1 - \Spro^2)$ from $\mathcal{S}_{H}^{\rm (core)}$ and $\mathcal{S}_{H}^{\rm (cloud)}$ in order to keep these functions finite at the origin.  
We show these various contributions to the sources in \fref{fig:scalar_source}.  
The figure confirms that the sources are characterized by a tight core, which drops off on a scale $\xi \gtrsim {\rm few}$ corresponding to $\rho \gtrsim \sigma$, surrounded by a wide tail or cloud, which is smaller in magnitude and drops off more slowly.  
In the decoupling limit, $\sigma \gg \eta$, the cloud can be much wider than the core.  
This motivates our prescription for calculating the effective couplings, which we employ in this section and the following one.  
We will consider fluctuations of the light fields ($\phi_H \approx H$ and $Z^{\mu}$) about their vacuum expectation values in the presence of the background expectation values of the heavy fields ($\phi_S \approx S$ and $X^{\mu}$), which are determined by the long straight string solution.  
Then, we can treat the heavy fields which compose the core as providing the source for the light fields which compose the cloud.  

To implement the above strategy, we will write 
\begin{align}\label{eq:H_expand}
	&S = \left( \sigma \, \Spro(\xi) - \sin \theta \frac{\phi_H}{\sqrt{2}} + \cos \theta \frac{\phi_S}{\sqrt{2}} \right) \, e^{i \varphi}
	\qquad , \qquad
	X^{\mu} = \frac{\Xpro(\xi)}{\rho} V^{\mu} \com \nn
	&H = \eta + \cos \theta \frac{\phi_H}{\sqrt{2}} + \sin \theta \frac{\phi_S}{\sqrt{2}}
	\qquad , \ {\rm and} \qquad
	Z^{\mu} = \frac{\Zpro(\infty)}{\rho} V^{\mu} \per
\end{align}
By taking the appropriate linear combination of Eqns.~(\ref{eq:EOM_for_H})~and~(\ref{eq:EOM_for_S}), we find the field equation for $\phi_H$ to be 
\begin{align}\label{eq:fieldeqn_phiH}
	\bigl( \Box + M_H^2 + \delta M_H^2 \bigr) \phi_H + \delta \mu^2 \phi_S = \mathcal{S} + O(\phi_H^2, \phi_H \phi_S)
\end{align}
where the mass $M_H^2$ was given by \eref{eq:eigenvalues}, the mass shift is defined by 
\begin{align}
	\delta M_H^2(\xi) \equiv \ & 
	\frac{\cos^2 \theta}{\rho_0^2 \xi^2} \left( \gHZ \Zpro(\infty) + \gHX \Xpro \right)^2
	+ \frac{\sin^2 \theta}{\rho_0^2 \xi^2} \bigl( 1 - \gSZ \Zpro(\infty) - \gSX \Xpro \bigr)^2
	- 2 \lambda \eta^2  \frac{\sin^2 2 \theta}{\cos 2 \theta} \nn
	& - 2 \kappa \sigma^2 \left( 1 - 3 \Spro^2 - 2 \sec 2 \theta \right) \sin^2 \theta
	- \alpha \sigma \left( ( 1 - \Spro^2) \sigma \cos \theta + 4 \eta \, \Spro \, \sin \theta \right) \cos \theta \com
\end{align}
the residual mixing is defined by 
\begin{align}
	\delta \mu^2(\xi) \equiv \ & 
	\frac{\sin 2 \theta}{2 \rho_0^2 \xi^2} \left[ \left( \gHZ \Zpro(\infty) + \gHX \Xpro \right)^2 - \bigl( 1 - \gSZ \Zpro(\infty) - \gSX \Xpro \bigr)^2 \right] \nn
	& + 2 \lambda \eta^2 \sin 2 \theta
	+ \kappa \sigma^2 \left( 1 - 3 \Spro^2 \right) \sin 2 \theta
	+ \frac{\alpha}{2} \sigma \left( 4 \eta \Spro \cos 2 \theta - \sigma (1 - \Spro^2) \sin 2 \theta \right)
	\com
\end{align}
and the source term is defined by 
\begin{align}\label{eq:scalar_source_def}
	\mathcal{S}(\xi) \equiv \sqrt{2} \cos \theta \Bigl[ &
	- \frac{\eta}{\rho_0^2} \frac{1}{\xi^2} \left( \gHZ \Zpro(\infty) + \gHX \Xpro \right)^2 
	+ \alpha \eta \sigma^2 \left( 1 - \Spro^2 \right) 
	+ \frac{\eta}{\rho_0^2} \frac{1}{\xi^2} (\gHZ)^2 \Zpro(\infty)^2 (1 - s^2)
	\Bigr] \nn
	- \sqrt{2} \sin \theta \Bigl[ & 
	\frac{\sigma}{\rho_0^2} \Spro^{\prime \prime} - \frac{\sigma}{\rho_0^2} \frac{\Spro}{\xi^2} \bigl( 1 - \gSZ \Zpro(\infty) - \gSX \Xpro \bigr)^2 + 2 \kappa \sigma^3 \left( 1 - \Spro^2 \right) \Spro 
	\Bigr] \per
\end{align}
We have dropped terms in \eref{eq:fieldeqn_phiH} which are higher order in $\phi_H$ and $\phi_S$, because these represent interactions among the quanta of the scalar fields, and they are not relevant for the particle production calculation.  
Near the string core, the spectrum is shifted as compared with far from the string.  
This leads to a residual mixing, $\delta \mu^2(\xi)$, and a position-dependent mass eigenvalue, $M_H^2 + \delta M_H^2(\xi)$.  
Since these shifts vanish rapidly outside of the string, and we are interested in the dynamics
of the long range fields, we can neglect these terms and take $\delta \mu^2 =0 = \delta M_H^2$.  

We would like to reduce the source term, $\mathcal{S}$, down to a single effective coupling parameter $\gHstr$.  
This is accomplished by noting that long wavelength modes of the Higgs field cannot resolve the internal structure of the string, \ie, the core, and for the purposes of studying these modes it is a good approximation to treat the source term as a Dirac delta function: 
\begin{align}\label{eq:S_approx_delta}
	\mathcal{S} \approx \gHstr \, \eta \sigma^2 \, \delta(\sigma x) \delta(\sigma y) \per
\end{align}
The effective, dimensionless coupling constant, $\gHstr \equiv \eta^{-1} \int dx dy~ \mathcal{S}$, is given by 
\begin{align}\label{eq:H_to_string_coupling}
	\gHstr = 2 \pi \sqrt{2} \int_{0}^{\infty} \xi d \xi \, \Biggl( 
	&
	\! \! - \cos \theta \Bigl[ 
	\frac{1}{\xi^2} \left( \gHZ \Zpro(\infty) + \gHX \Xpro \right)^2 
	- \alpha (\rho_0 \sigma)^2 \left( 1 - \Spro^2 \right) 
	- \frac{1}{\xi^2} (\gHZ)^2 \Zpro(\infty)^2 (1 - \Spro^2)
	\Bigr] \nn
	&
	+ \frac{\sigma}{\eta} \sin \theta \Bigl[ 
	- \Spro^{\prime \prime} + \frac{\Spro}{\xi^2} \bigl( 1 - \gSZ \Zpro(\infty) - \gSX \Xpro \bigr)^2 - 2 \kappa (\rho_0 \sigma)^2 \left( 1 - \Spro^2 \right) \Spro 
	\Bigr] 
	\Biggr) \per 
\end{align}
This expression simplifies in the decoupling limit where we can write
\begin{align}\label{eq:SX_cases}
	\Spro(\xi) \approx \begin{cases} \frac{\xi}{\xi_{\rm max}} & \xi \leq \xi_{\rm max} \\ 1 & \xi > \xi_{\rm max} \end{cases}
	\qquad {\rm and} \qquad 
	\Xpro(\xi) \approx \begin{cases} \Xpro(\infty) \left( \frac{\xi}{\xi_{\rm max}} \right)^2 & \xi \leq \xi_{\rm max} \\ \Xpro(\infty) & \xi > \xi_{\rm max} \end{cases}
	\per
\end{align}
Using \eref{eq:string_widths}, the parameter $\xi_{\rm max}$ is related to the profile widths as $\xi_{\rm max} \approx (\Delta \rho_s) \sigma$.  
This can be determined by solving for the full profile functions, but we will take $\xi_{\rm max} = O(1)$ for numerical estimates.  
Then after expanding in the ratio $(\eta^2 / \sigma^2) \ll 1$ we find
\begin{align}\label{eq:gHstr_dec}
	(\gHstr)^{\rm (dec.)} \simeq \left( \frac{e^2 \pi}{\sqrt{2} \, \cw^2 \gX^2} \right) \seps^2 + \frac{\pi}{15 \sqrt{2}} \left( \frac{64}{\kappa} - 17 \xi_{\rm max}^2 \right) \alpha + O \left( \frac{\eta^2}{\sigma^2} \right) \per
\end{align}
Although alternative definitions of the coupling can be proposed, they will differ from
our definition in terms that are suppressed by factors of $O(\eta/\sigma)$ and can be 
ignored in the decoupling limit.

\begin{figure}[t!]
	\begin{subfigure}[b]{0.5\textwidth}
		\includegraphics[width=\textwidth]{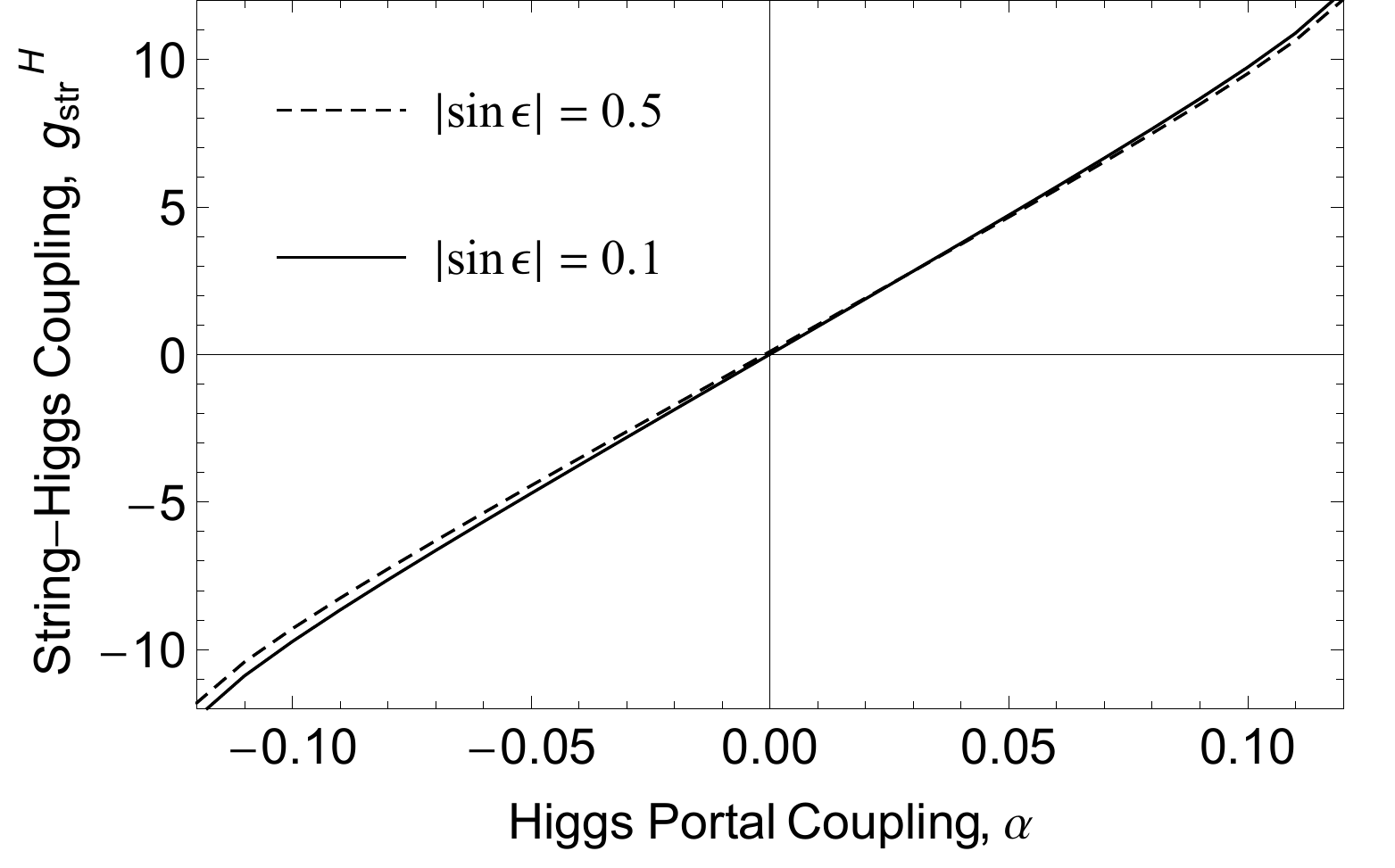}
		\caption{\label{fig:hc200}$m_X =  M_S = \sigma / \sqrt{2} = 200 \GeV$, $\gX = 1$}
	\end{subfigure}%
	\begin{subfigure}[b]{0.5\textwidth}
		\includegraphics[width=\textwidth]{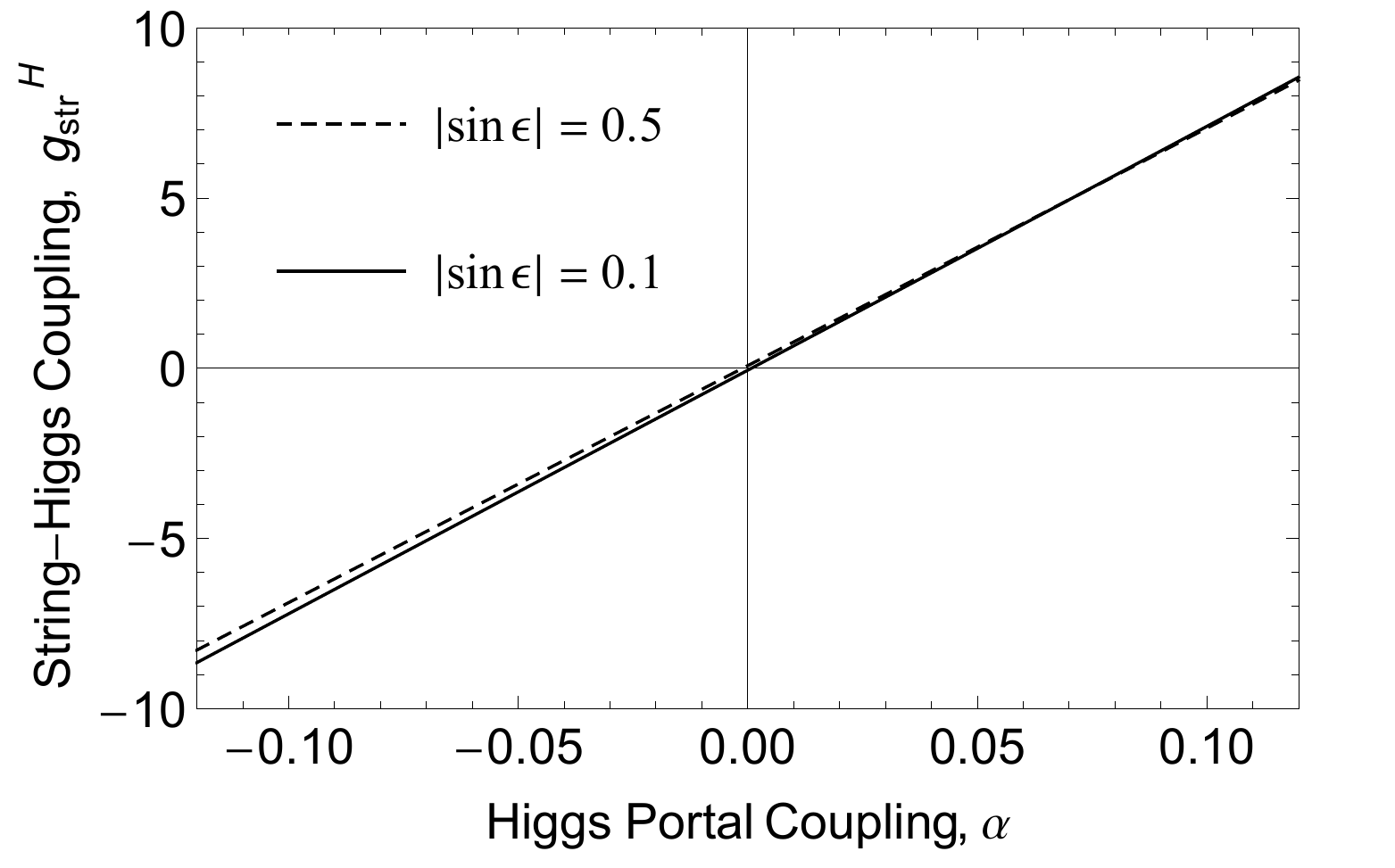}
		\caption{\label{fig:hc1000}$m_X =  M_S = 1 \TeV$, $\gX = 1$.}
	\end{subfigure} \\ 
	\begin{subfigure}[b]{0.5\textwidth}
		\includegraphics[width=\textwidth]{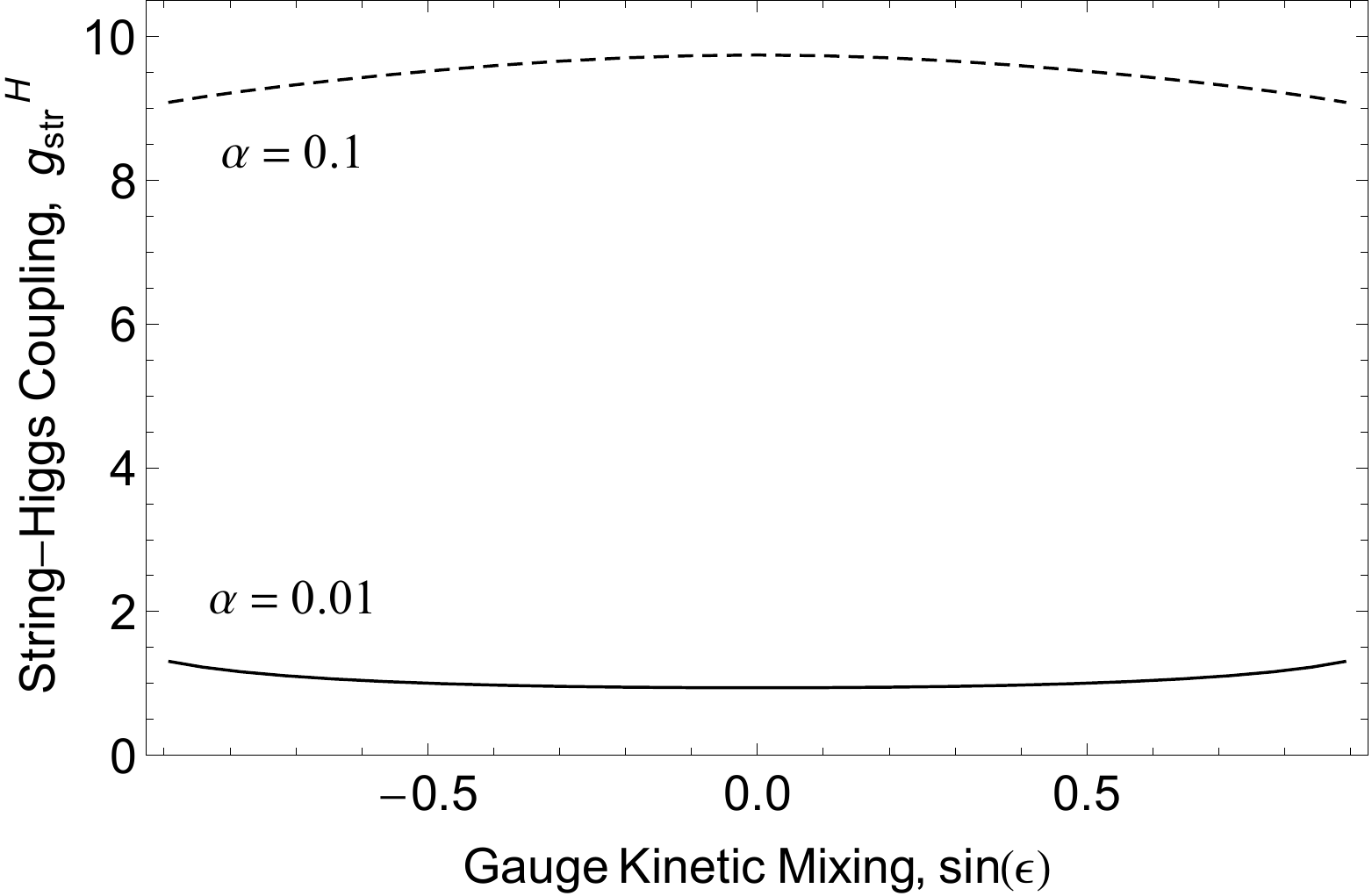}
		\caption{\label{fig:hvseps}$m_X =  M_S = \sigma / \sqrt{2} = 200 \GeV$, $\gX = 1$}
	\end{subfigure}%
	\caption{\label{fig:hc}Effective coupling of the $(0,1)$ string with the Higgs field, given by \eref{eq:H_to_string_coupling}. }
\end{figure}

\begin{figure}[t!]
		\includegraphics[width=0.7\textwidth]{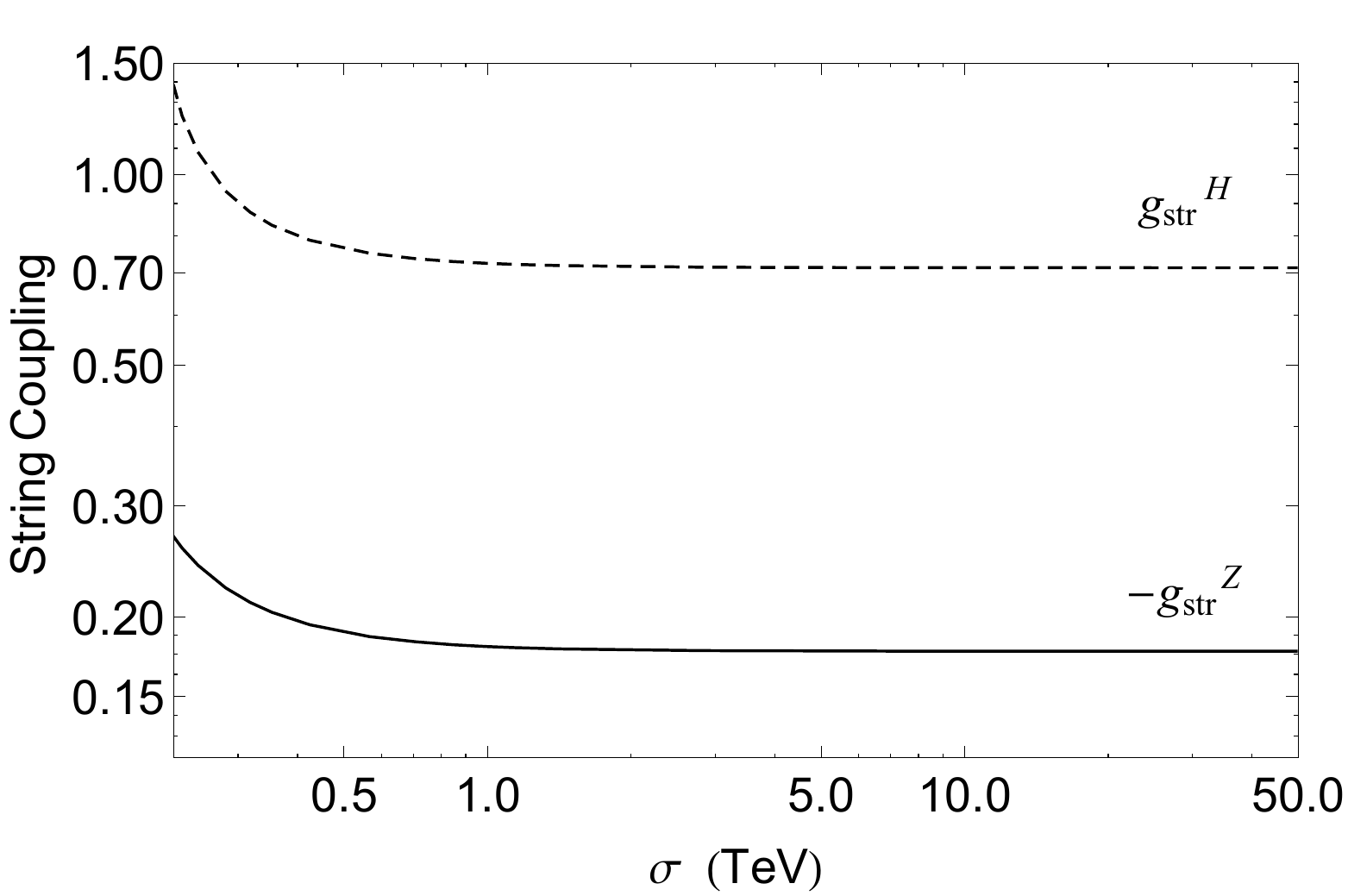}
	\caption{\label{fig:coupling_decoupling} Effective couplings of $(0,1)$ string to $Z$ and $H$ fields as the scale $\sigma$ becomes large. We have held fixed $M_S = M_X = \sigma / \sqrt{2}$, $\alpha = 0.01$, $\seps = 0.1$, and $\gX = 1$.  For comparison, the decoupling limit approximations, given by Eqs.~(\ref{eq:gHstr_dec})~and~(\ref{eq:gZstr_dec}), give $(\gHstr)^{\rm (dec.)} \approx 0.70$ and $(\gZstr)^{\rm (dec.)} \approx -0.17$ for $\xi_{\rm max} = 1.5$ and $2.7$, respectively.  
	}
\end{figure}

Once the string solution is obtained, it is straightforward to perform the integral in \eref{eq:H_to_string_coupling} and evaluate $\gHstr$.  
Figures \ref{fig:hc200} and \ref{fig:hc1000} show the dependence of $\gHstr$ on the Higgs portal coupling, and they suggest the approximate relationship $\gHstr \propto \alpha$.  
This behavior is understood by noting that $\gHstr$ depends explicitly on $\alpha$ through one term in \eref{eq:H_to_string_coupling} and implicitly through the profile functions.  
The explicit dependence dominates at small $\alpha$ and gives $\gHstr \sim \alpha \int \xi d \xi (1-\Spro^2)$, and at larger $\alpha$ the subdominant dependence in $\Spro$ and $\Hpro$ emerges.  
Figure \ref{fig:hvseps} shows that $\gHstr$ has a weak dependence on the gauge kinetic mixing parameter $\gHstr \sim \const - O(\seps^2)$.  
This follows from the relations $\gHX \sim \gSZ \sim \Zpro_{\infty} \sim O(\seps)$ and $\gHZ \sim \gSX \sim \Xpro_{\infty} \sim O(1)$ and $\sin \theta \sim O(\alpha)$ [see \aref{app:Smallepsilon} and \eref{eq:mixing}].  
Finally, \fref{fig:coupling_decoupling} 
shows the dependence of $\gHstr$ on the scale $\sigma$.  
In the decoupling limit, $\sigma \gg \eta$, we see that $\gHstr$ becomes asymptotically independent of $\sigma$, which confirms that dimensionally $ \mathcal{S} \sim \eta \sigma^2$, as given by \eref{eq:S_approx_delta}.  
The appearance of the Higgs VEV, $\eta$, is an important result.  
It reflects the fact that the linear coupling of the Higgs to the string only emerges after electroweak symmetry breaking.  
Prior to electroweak symmetry breaking, the coupling of the Higgs bosons to the string is higher order in powers of the Higgs field, \ie, the string can only radiate Higgs/anti-Higgs pairs.  
This result is not totally obvious since it is possible for the string to carry a Higgs condensate, and thereby break the electroweak symmetry locally, even if the Higgs VEV vanishes outside the string, as in the case of bosonic superconductivity \cite{Witten:1984eb}.  

Thus far we have considered the coupling between the Higgs and the straight static string.  
Now we generalize to the case of an arbitrary Nambu-Goto string with spacetime coordinate 
${\mathbb X}^{\mu}(\tau,\zeta)$ where $\tau$ and $\zeta$ are the world coordinates.  
The source term in \eref{eq:fieldeqn_phiH} derives from the Lagrangian 
$\Lcal = \phi_H \mathcal{S}$.  
Upon approximating the source as a delta function, as in \eref{eq:S_approx_delta}, the action becomes 
\begin{align}\label{eq:SstrH}
	S_{\rm str}^{H} 
	& = \int d^4 x \, \phi_H \, \mathcal{S} \nn
	&= \gHstr \, \eta \int d^4 x \ \phi_H(x)  \, \int d\tau d\zeta \sqrt{-\gamma} \ \delta^{(4)}(x - {\mathbb X}^{\mu}(\tau,\zeta)) \nonumber \\
	& = \gHstr \, \eta   \, \int d\tau d\zeta \sqrt{-\gamma} \ \phi_H({\mathbb X}^\mu) 
\end{align}
where the worldsheet metric is defined by $\gamma_{ab} = g_{\mu\nu} \partial_a {\mathbb X}^\mu \partial_b {\mathbb X}^\nu$ ($a,b=0,1$) and $\gamma = \det(\gamma_{ab}) = (1/2) \epsilon^{ac} \epsilon^{bd} \gamma_{ab} \gamma_{cd}$.

\subsection{Coupling of the Z Boson to the String}\label{sub:Coupling_to_Z}

As in the case of the Higgs field, the string provides a source for the Z field.  
Recall that the Z boson field equation, \eref{eq:fieldeqns2}, was given by 
\begin{align}\label{eq:Zfieldeqn_rewrite}
	\partial_{\nu} Z^{\nu \mu} = &\gHZ \left[ 
	i \left( H \partial^{\mu} H^{\ast} - H^{\ast} \partial^{\mu} H \right) - 2 \left( \gHZ Z^{\mu} + \gHX X^{\mu} \right) \abs{H}^2 
	\right] \nn
	&+ \gSZ \left[
	i \left( S \partial^{\mu} S^{\ast} - S^{\ast} \partial^{\mu} S \right) - 2 \left( \gSZ Z^{\mu} + \gSX X^{\mu} \right) \abs{S}^2 
	\right]
\end{align}
where we have explicitly written out the currents using \eref{eq:J2}.  
As we discussed in \sref{sub:Coupling_to_H}, the decoupling approximation, $\sigma \gg \eta$, allows us to replace the heavy fields with the string background and to expand the light fields about their vacuum expectation values: 
\begin{align}\label{eq:Z_expand}
	&S \to \left( \sigma \, \Spro(\xi) + \bar{S} \right) \, e^{i \varphi}
	\quad , \quad
	X^{\mu} \to \frac{\Xpro(\xi)}{\rho} V^{\mu} \com \nn
	&H = \eta + \bar{H}
	\quad , \ {\rm and} \quad
	Z^{\mu} = \frac{\Zpro(\infty)}{\rho} V^{\mu} + \bar{Z}^{\mu} \com
\end{align}
where $\Zpro(\infty)$ is given by \eref{eq:gaugeBC} with $(n,m) = (0,1)$.  
Since we are now interested in radiation of the Z field, and we are not concerned with its coupling to the scalar fields, we can take $\bar{S} = \bar{H} = 0$.  
Inserting \eref{eq:Z_expand} into \eref{eq:Zfieldeqn_rewrite} yields the field equation for the fluctuation $\bar{Z}^{\mu}$, 
\begin{align}\label{eq:Z_field_eqn}
	\partial_{\nu} \bar{Z}^{\nu \mu} + M_Z^2 \bar{Z}^{\mu} + \delta M_Z^2 \bar{Z}^{\mu} = \mathcal{J}^{\mu} 
	\com
\end{align}
where $\bar{Z}^{\mu \nu} \equiv \partial^{\mu} \bar{Z}^{\nu} - \partial^{\nu} \bar{Z}^{\mu}$, the mass $M_Z$ is given by \eref{eq:gauge_evals}, the position-dependent mass shift is defined as 
\begin{align}\label{eq:dMZ2_def}
	\delta M_Z^2(\xi) \equiv - 2 (\gSZ)^2 \sigma^2 \, \bigl( 1 - \Spro^2 \bigr) \com
\end{align}
and the source current is given by  
\begin{align}\label{eq:Jmu_VEV}
	\mathcal{J}^{\mu} & = \frac{\eta^2}{\rho_0} \, j(\xi) V^{\mu}(\varphi)
\end{align}
where
\begin{align}\label{eq:j_def}
	j(\xi) \equiv 
	2 \gHZ \frac{1}{\xi} \Bigl( - \gHZ \Zpro(\infty) - \gHX \Xpro \Bigr)
	+ 2 \gSZ \frac{\sigma^2}{\eta^2} \frac{\Spro^2}{\xi} \Bigl( 1 - \gSZ \Zpro(\infty) - \gSX \Xpro \Bigr)
\end{align}
for the $(0,1)$ string.  
Despite the factor of $(\sigma^2 / \eta^2)$ in the second term above, both terms in $j(\xi)$ scale like $(\sigma / \eta)^0$ because $\gSZ \sim \eta^2 / \sigma^2$ [see \eref{eq:approx_scalar_gauge_couplings}].  

Using the complete set of orthonormal basis vectors
\begin{align}
	T_{\mu} = \partial_{\mu} t 
	\quad , \quad 
	R_{\mu} = \partial_{\mu} \rho
	\quad , \quad 
	V_{\mu} = \rho \, \partial_{\mu} \varphi 
	\quad , \ {\rm and} \quad
	L_{\mu} = \partial_{\mu} z
\end{align}
the current can also be written as 
\begin{align}\label{eq:Jmu_VEV_v2}
	\mathcal{J}^{\mu} = \eta^2 \, \epsilon^{\mu \alpha \beta \gamma} \partial_{\alpha} \Bigl( k(\xi) T_{\beta} L_{\gamma} \Bigr)
\end{align}
where
\begin{align}\label{eq:k_def}
	k(\xi) \equiv \int_{\infty}^{\xi} \, d\xi^{\prime} \, j(\xi^{\prime}) \per
\end{align}
Note that $j(\xi)$ is approximately equal to the right hand side of the string equation, \eref{eq:eom_1}, and if we were to replace $\Hpro \to 1$ and $\Zpro \to \Zpro(\infty)$, then they would be identical.  
As such, $k(\xi)$ is approximately given by 
\begin{align}\label{eq:k_approx}
	k(\xi) & \approx - \frac{1}{(\rho_0 \eta)^2} \left( \frac{\Zpro^{\prime}}{\xi} - \lim_{\xi \to \infty}\frac{\Zpro^{\prime}}{\xi} \right) = - \frac{1}{\eta^2} B_Z(\xi)
\end{align}
where $B_Z(\xi) \equiv \Zpro^{\prime} / (\rho_0^2 \xi)$ is the magnitude of the Z-magnetic field, $(B_Z)^{i} = (-1/2) \epsilon^{ijk}Z_{jk} = \epsilon^{ijk} \partial_{j} A^{k}$. 

The profile functions $\Spro$ and $\Xpro$ both reach their asymptotic values exponentially fast on a scale $\xi = O(1)$ corresponding to $\rho = O(\rho_0 = \sigma^{-1})$.  
In the decoupling limit, $M_Z \ll \sigma$, long wavelength modes of the Z field cannot resolve the string core, and we can use delta function approximations.  
The mass shift, given by \eref{eq:dMZ2_def}, becomes negligible outside of the narrow string core.  
Therefore it is not relevant for the particle radiation calculation, and we will neglect it by taking $\delta M_Z^2 = 0$.  
The profile function $k(\xi)$ can also be approximated as a delta function
\begin{align}
	k(\xi) \approx \gZstr \, \sigma^{-2} \, \delta(x) \delta(y)
\end{align}
where the effective coupling, $\gZstr \equiv 2\pi \int_{0}^{\infty} \xi d\xi ~ k(\xi)$, is given by 
\begin{align}\label{eq:gZstr}
	\gZstr = 2 \pi \int_{0}^{\infty} \xi d \xi \int_{\infty}^{\xi} d \xi^{\prime} \left[ 2 \gHZ \frac{1}{\xi^{\prime}} \Bigl( - \gHZ \Zpro(\infty) - \gHX \Xpro \Bigr)
	+ 2 \gSZ \frac{\sigma^2}{\eta^2} \frac{\Spro^2}{\xi^{\prime}} \Bigl( 1 - \gSZ \Zpro(\infty) - \gSX \Xpro \Bigr) \right]
\end{align}
after inserting \eref{eq:j_def} into \eref{eq:k_def}.  
Note that the approximation, \eref{eq:k_approx}, would give the effective coupling to be $\gZstr \approx \Phi_Z / (\rho_0 \eta)^2$ where $\Phi_Z \equiv \int dx dy \, B_Z$ is the Z-magnetic flux.  
In the decoupling limit [see \eref{eq:SX_cases}] we find
\begin{align}\label{eq:gZstr_dec}
	(\gZstr)^{\rm (dec.)} \simeq - \left( \frac{11e^2 \pi}{36 \cw^2 \sw \gX} \xi_{\rm max}^2 \right) \seps + O\left( \frac{\eta^2}{\sigma^2} \right) 
\end{align}
where $\xi_{\rm max} = O(1)$.  

\begin{figure}[t!]
	\begin{subfigure}[b]{0.5\textwidth}
		\includegraphics[width=\textwidth]{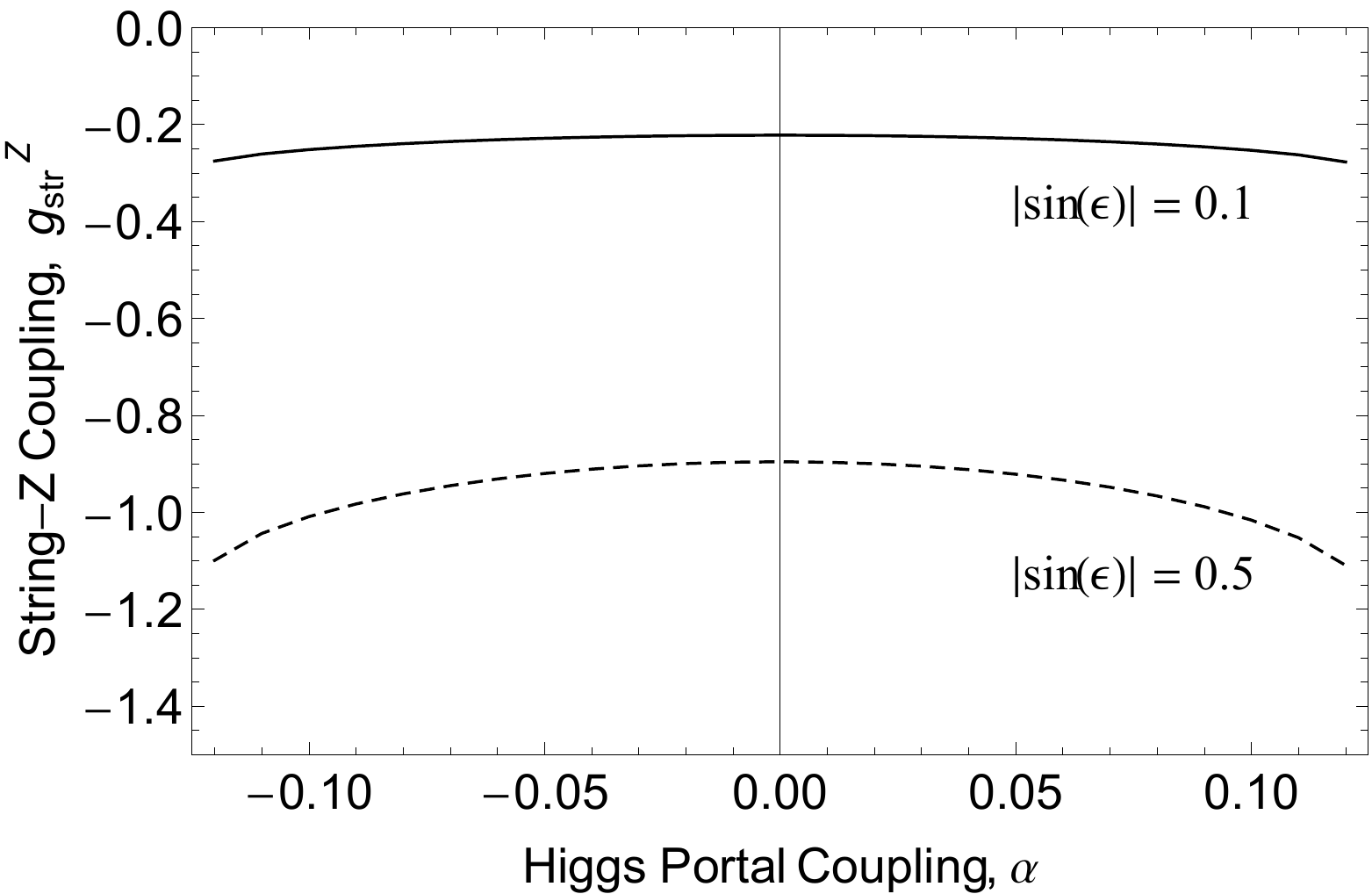}
		\caption{\label{fig:zc200}$m_X =  M_S = \sigma / \sqrt{2} = 200 \GeV$, $\gX = 1$}
	\end{subfigure}%
	\begin{subfigure}[b]{0.5\textwidth}
		\includegraphics[width=\textwidth]{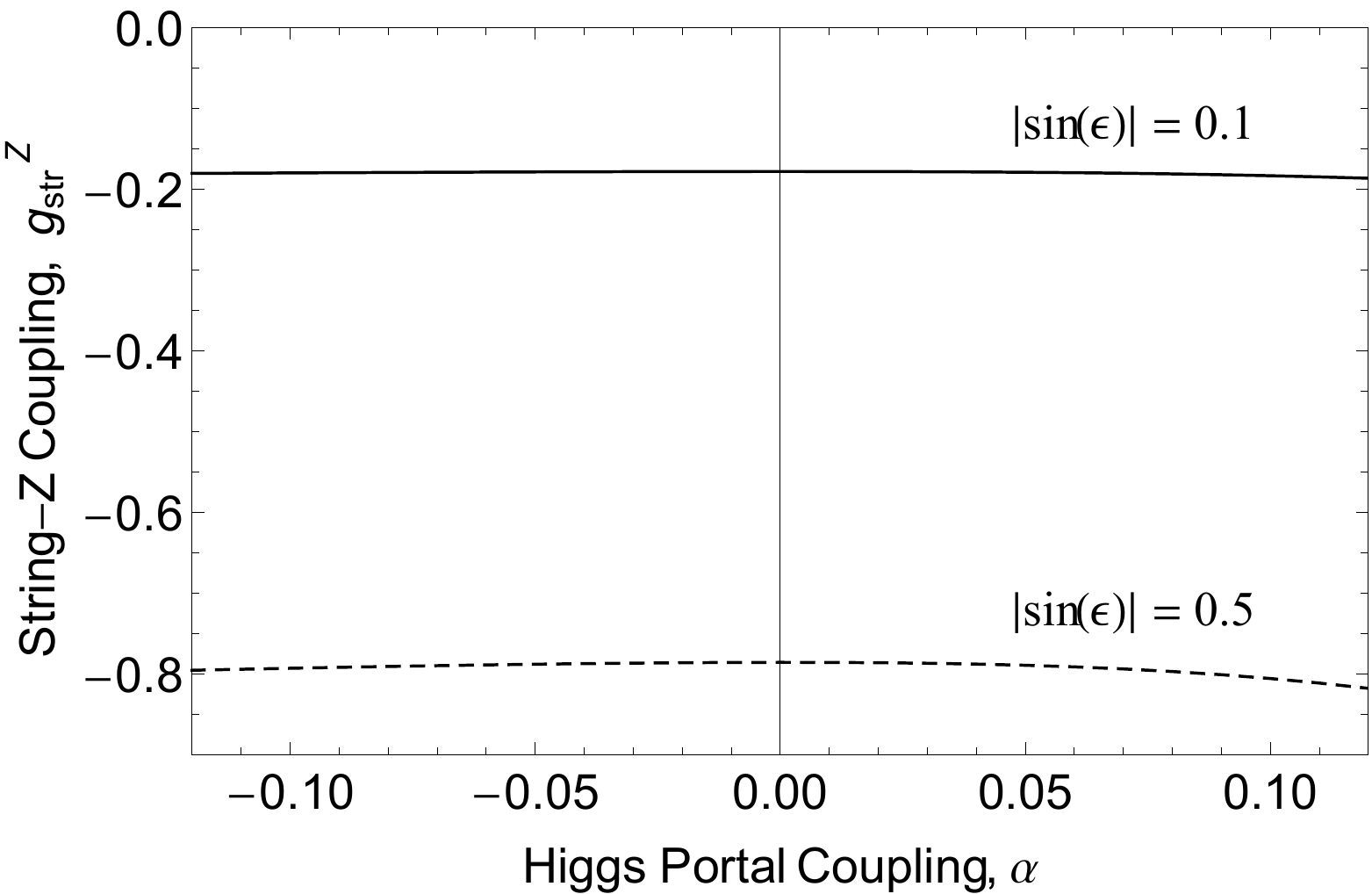}
		\caption{\label{fig:zc1000}$m_X =  M_S = 1 \TeV$, $\gX = 1$.}
	\end{subfigure} \\ 
	\begin{subfigure}[b]{0.5\textwidth}
		\includegraphics[width=\textwidth]{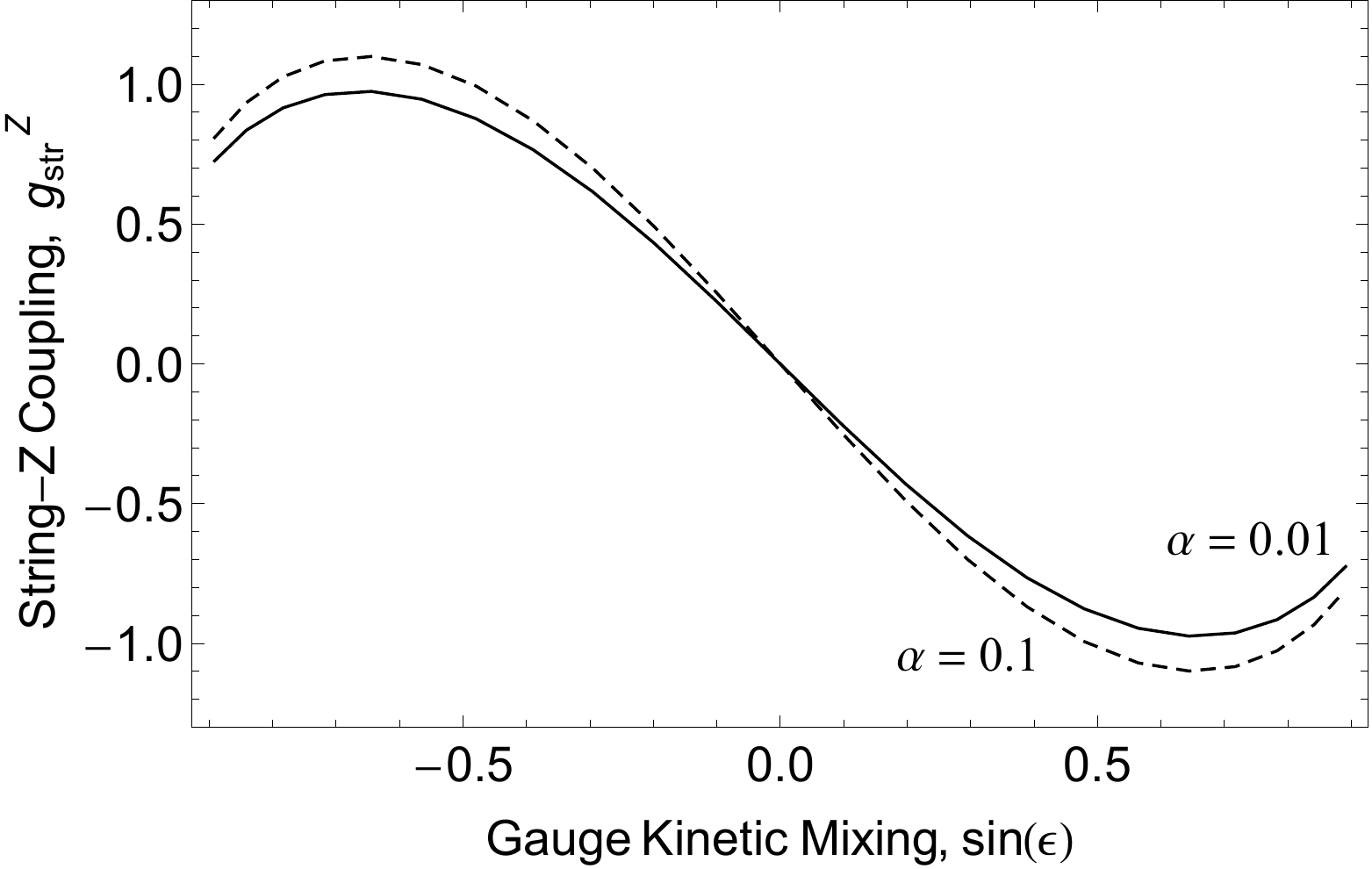}
		\caption{\label{fig:zvseps}$m_X =  M_S = \sigma / \sqrt{2} = 200 \GeV$, $\gX = 1$}
	\end{subfigure}%
	\caption{\label{fig:zc}Effective coupling of the $(0,1)$ string with the Z boson, given by \eref{eq:gZstr}. }
\end{figure}

From \fref{fig:coupling_decoupling} we see that $\gZstr$ asymptotes to a constant in the decoupling limit that is given approximately by \eref{eq:gZstr_dec}.  
Figures \ref{fig:zc200} and \ref{fig:zc1000} show that $\gZstr$ depends weakly on the Higgs portal coupling, with the approximate relationship $\gZstr \propto -\alpha^2$. Since $\alpha$ does not appear explicitly in \eref{eq:gZstr}, the dependence is only through the profile functions.    
Figure \ref{fig:zvseps} shows that $\gZstr$ depends linearly on the gauge kinetic mixing parameter $\gZstr \sim O(\seps)$ for small values of $\seps$, which can be understood from the dependence on $\gSZ$ in \eref{eq:gZstr} and by noting that $\gSZ$ is linear in $\seps$. As $|\seps|$ increases, the terms that are higher order in $\seps$ begin to have an effect.

Thus far we have been assuming that the string is long and straight.  
To generalize to an arbitrary Nambu-Goto string, we can write the source term, \eref{eq:Jmu_VEV_v2}, as 
\begin{align}
	\mathcal{J}^{\mu} = \gZstr (\eta / \sigma)^2 \partial_{\nu} \int d\sigma^{\mu \nu} ~\delta^{(4)}(x - {\mathbb X}(\tau,\zeta) )
\end{align}
where $d\sigma^{\mu\nu} = d \tau d \zeta \epsilon^{\mu\nu\alpha\beta}\epsilon^{ab} \partial_a 
{\mathbb X}_{\alpha} \partial_b {\mathbb X}_{\beta}$ is the areal element of the string worldsheet.  
A source of this form was first given in Ref.~\cite{Alford:1988sj}.
This source can be derived from a term in the effective action
\begin{align}\label{eq:SstrZ}
S_{\rm str}^Z 
	& = \int d^4 x \, Z_{\mu} \mathcal{J}^{\mu} \nn
	&= \frac{\gZstr}{2} \left( \frac{\eta}{\sigma} \right)^2 \int d^4 x \ Z_{\mu \nu} \int d\sigma^{\mu\nu} \delta^{(4)}(x - {\mathbb X}(\tau, \zeta)) \nn
	&= \frac{\gZstr}{2} \left( \frac{\eta}{\sigma} \right)^2 \int d\sigma^{\mu\nu} Z_{\mu\nu}({\mathbb X}^{\mu}) 
\end{align}
where total derivative terms have been dropped.  
We have factored off the $(\eta/\sigma)^2$ scaling such that $\gZstr$ is constant in the limit $\eta \ll \sigma$.

\subsection{Coupling to the Fermions}\label{sub:Coupling_to_Fermions}

Finally, let us turn to the coupling between the dark string and the SM fermions.  
Like the coupling to the bosons, this interaction can give particle radiation from the string \cite{Chu:2010zzb}.
Additionally, as the string passes through the plasma, this interaction induces a drag force that has an important influence on the evolution of the string network as a whole \cite{Martins:1995tg}.  

The interaction that we seek to calculate arises from the kinetic terms for the SM fermions,
\begin{align}
	\Lcal = 
	Q^{\dagger} i \bar{\sigma}^{\mu} D_{\mu} Q + u_R^{\dagger} i \sigma^{\mu} D_{\mu} u_R + d_R^{\dagger} i \sigma^{\mu} D_{\mu} d_R
	+ L^{\dagger} i \bar{\sigma}^{\mu} D_{\mu} L + e_R^{\dagger} i \sigma^{\mu} D_{\mu} e_R 
\end{align}
where we use the two component spinor notation and the doublets are $Q = (u_L \, , \, d_L)$ and $L = (\nu_L \, , \, e_L)$.  
The covariant derivatives are given by 
\begin{align}\label{eq:covarD_fermions}
\begin{array}{l}
	D_{\mu} Q = \left( \partial_{\mu} - i \frac{g}{2} \sigma^a W_{\mu}^{a} - i \frac{g^{\prime}}{2} \yQ Y_{\mu} \right) Q \\
	D_{\mu} u_R = \left( \partial_{\mu} - i \frac{g^{\prime}}{2} \yuR Y_{\mu} \right) u_R \\
	D_{\mu} d_R = \left( \partial_{\mu} - i \frac{g^{\prime}}{2} \ydR Y_{\mu} \right) d_R \\
	D_{\mu} L = \left( \partial_{\mu} - i \frac{g}{2} \sigma^a W_{\mu}^{a} - i \frac{g^{\prime}}{2} \yL Y_{\mu} \right) L \\
	D_{\mu} e_R = \left( \partial_{\mu} - i \frac{g^{\prime}}{2} \yeR Y_{\mu} \right) e_R 
\end{array}
\end{align}
where we have turned off the $\SU{3}$ gauge coupling, since it does not modify the coupling to the dark string, and the hypercharge assignments are 
\begin{align}\label{eq:hypercharges}
	\yQ = \frac{1}{3} \quad , \quad 
	\yuR = \frac{4}{3} \quad , \quad 
	\ydR =- \frac{2}{3} \quad , \quad 
	\yL = -1 \quad , \ \text{and} \quad 
	\yeR = -2 \per
\end{align}
After performing the field redefinition given by \eref{eq:field_trans}, the covariant derivatives become
\begin{align}
\begin{array}{l}
	D_{\mu} Q = \begin{pmatrix}
	D_{\mu} u_L - i \frac{g}{\sqrt{2}} W_{\mu}^{+} d_L \\
	D_{\mu} d_L - i \frac{g}{\sqrt{2}} W_{\mu}^{-} u_L 
	\end{pmatrix} 
	\qquad {\rm with} \qquad 
\begin{array}{l}
	D_{\mu} u_L = \bigl( \partial_{\mu} - i (\guLA A_{\mu} + \guLZ Z_{\mu} + \guLX X_{\mu}) \bigr) u_L \\
	D_{\mu} d_L = \bigl( \partial_{\mu} - i (\gdLA A_{\mu} + \gdLZ Z_{\mu} + \gdLX X_{\mu}) \bigr) d_L 
\end{array} \\
	D_{\mu} u_R = \bigl( \partial_{\mu} - i (\guRA A_{\mu} + \guRZ Z_{\mu} + \guRX X_{\mu}) \bigr) u_L \\
	D_{\mu} d_R = \bigl( \partial_{\mu} - i (\gdRA A_{\mu} + \gdRZ Z_{\mu} + \gdRX X_{\mu}) \bigr) u_L \\
	D_{\mu} L = \begin{pmatrix}
	D_{\mu} \nu_L - i \frac{g}{\sqrt{2}} W_{\mu}^{+} e_L \\
	D_{\mu} e_L - i \frac{g}{\sqrt{2}} W_{\mu}^{-} \nu_L 
	\end{pmatrix} 
	\qquad {\rm with} \qquad 
\begin{array}{l}
	D_{\mu} \nu_L = \bigl( \partial_{\mu} - i (\gnLA A_{\mu} + \gnLZ Z_{\mu} + \gnLX X_{\mu}) \bigr) \nu_L \\
	D_{\mu} e_L = \bigl( \partial_{\mu} - i (\geLA A_{\mu} + \geLZ Z_{\mu} + \geLX X_{\mu}) \bigr) e_L
\end{array} \\
	D_{\mu} e_R = \bigl( \partial_{\mu} - i (\geRA A_{\mu} + \geRZ Z_{\mu} + \geRX X_{\mu}) \bigr) e_R
\end{array}
\end{align}
where
\begin{align}\label{eq:fermion_couplings}
\begin{array}{lclcl}
	\guLA = \frac{2e}{3} & \quad & 
	\guLZ = \cxi \frac{e}{6} \left( \frac{3}{t_w} - t_w \right) - \sxi \frac{e}{6} \frac{t_{\epsilon}}{\cos \theta_w} & \quad & 
	\guLX = - \cxi \frac{e}{6} \frac{t_{\epsilon}}{\cos \theta_w} - \sxi \frac{e}{6} \left( \frac{3}{t_w} - t_w \right) \\
	\guRA = \frac{2e}{3} & \quad & 
	\guRZ = - \cxi \frac{2 e}{3} t_w - \sxi \frac{2 e}{3} \frac{ t_{\epsilon}}{\cos \theta_w}  & \quad & 
	\guRX = - \cxi \frac{2 e}{3} \frac{ t_{\epsilon}}{\cos \theta_w} + \sxi \frac{2 e}{3} t_w \\
	\gdLA = -\frac{e}{3} & \quad & 
	\gdLZ = - \cxi \frac{e}{6} (\frac{3}{t_w} + t_w) - \sxi \frac{e}{6} \frac{t_{\epsilon}}{\cos \theta_w} & \quad &
	\gdLX = - \cxi \frac{e}{6} \frac{t_{\epsilon}}{\cos \theta_w} + \sxi \frac{e}{6} (\frac{3}{t_w} + t_w)\\
	\gdRA = -\frac{e}{3} & \quad & 
	\gdRZ = \cxi \frac{e}{3} t_w + \sxi \frac{e}{3} \frac{t_{\epsilon}}{\cos \theta_w} & \quad & 
	\gdRX = \cxi \frac{e}{3} \frac{t_{\epsilon}}{\cos \theta_w} - \sxi \frac{e}{3} t_w \\
	\geLA = -e & \quad & 
	\geLZ = - \cxi \frac{e}{2} \left( \frac{1}{t_w} - t_w \right) + \sxi \frac{e}{2} \frac{t_{\epsilon}}{\cos \theta_w} & \qquad & 
	\geLX = \cxi \frac{e}{2} \frac{t_{\epsilon}}{\cos \theta_w} + \sxi \frac{e}{2} \left( \frac{1}{t_w} - t_w \right) \\
	\geRA = -e & \quad & 
	\geRZ = \cxi \, e \, t_w + \sxi e \, \frac{t_{\epsilon}}{\cos \theta_w} & \quad & 
	\geRX = \cxi e \frac{t_{\epsilon}}{\cos \theta_w} - \sxi \, e t_w  \\
	\gnLA = 0 & \quad & 
	\gnLZ = \cxi \frac{e}{2} \left( \frac{1}{t_w} + t_w \right)  + \sxi \frac{e}{2} \frac{t_{\epsilon}}{\cos \theta_w} & \quad & 
	\gnLX = \cxi \frac{e}{2} \frac{t_{\epsilon}}{\cos \theta_w} - \sxi \frac{e}{2} \left( \frac{1}{t_w} + t_w \right) 
\end{array} \per
\end{align}
We have included the couplings to the photon field $A_{\mu}$ for completeness, but since the dark string does not contain any electromagnetic flux, these interactions are not relevant for couplings of the string to the SM fermions.  

The dominant interaction between fermions and the dark string is the Aharonov-Bohm (AB) interaction \cite{Aharonov:1959fk, Alford:1988sj}.  
In general when a particle of charge $e$ and momentum ${\bf p}$ (in the rest frame of the string) is incident on a string carrying magnetic flux $\Phi$, it will scatter with a differential cross section per unit length $d \sigma / d \theta$.  
It is useful to define the transport cross section, $\sigma_t \equiv \int_0^{2\pi} d \theta \, (d \sigma / d \theta)(1 - \cos \theta)$, which is given by 
\begin{align}
	\sigma_t = \frac{2}{\abs{\bf p}} \sin^2 \pi \theta
\end{align}
where $\theta \equiv (e / 2 \pi) \Phi$.  
In general these need not be electromagnetic charge and flux, and in fact the dark string carries no electromagnetic flux.  
Instead, the particles scatter off of the Z-flux and X-flux carried by the string.  

The fluxes are defined by 
\begin{align}\label{eq:flux_def}
	\Phi_{Z} \equiv \int {\bf B}_Z \cdot d {\bf A}
	\qquad {\rm and } \qquad
	\Phi_{X} \equiv \int {\bf B}_X \cdot d {\bf A} 
\end{align}
where the integral extends over the plane normal to the string and the magnetic fields are given by ${\bf B}_Z = {\bm \nabla} \times {\bf Z}$ where ${\bf Z}_i = Z^{i}$ and similarly for $X_{\mu}$.  
Using Stokes theorem along with the boundary conditions \eref{eq:gaugeBC}, the fluxes are easily found to be 
\begin{align}
	\Phi_Z &= 2 \pi \frac{\gSX n - \gHX m}{\gSX \gHZ - \gHX \gSZ} \label{eq:PhiZ} \\
	\Phi_X &= 2 \pi \frac{\gHZ m - \gSZ n}{\gSX \gHZ - \gHX \gSZ} \label{eq:PhiX}
\end{align}
Note that $\Phi_Z$ is nonzero even for the $(0,1)$ string for which the Higgs field does not wind, but instead $\Phi_Z \propto s_{\epsilon}$ due to the gauge kinetic mixing.  

As a particle moves around the string, its phase changes due to both fluxes.  
Therefore to calculate the transport cross section for a particle of species $i$ we sum the phases:  
\begin{align}\label{eq:sigmat_def}
	\sigma_t \Bigr|_{i}= \frac{2}{\abs{\bf p}} \sin^2 \pi \theta_i
\end{align}
where
\begin{align}\label{eq:phase_def}
	\theta_i \equiv  \frac{\gZ^i \Phi_Z}{2\pi} + \frac{\gX^i \Phi_X}{2\pi} 
\end{align}
and the $\gZ^i$ and $\gX^i$ are given by \eref{eq:fermion_couplings}.  
Upon performing the sum in \eref{eq:phase_def} a remarkable simplification occurs, and we are left with 
\begin{align}\label{eq:ABphases}
	\theta_i = (\yi - 2 \cw^2 \giA) n + \left( - 2 \frac{\cw e \seps}{\gX} \giA \right) m
\end{align}
where $\yi$ and $\giA$ are the hypercharge and electromagnetic charges of species $i$ given by Eqns.~(\ref{eq:hypercharges})~and~(\ref{eq:fermion_couplings}).  
Specifically, for the case $(n,m) = (0,1)$ we find
\begin{align}\label{eq:ABphases}
	\theta_i = q_i \Theta
	\qquad {\rm with} \qquad 
	\Theta \equiv - 2 \frac{\cw \seps}{\gX}  
\end{align}
and $q_i = e \, \giA$ is the electromagnetic charge.  
Note that we have {\it not expanded} in $\seps \ll 1$; these expressions are exact.  
It is remarkable that the phases $\theta_i$ are independent of the ratio of mass scales $R = m_X / m_Z$, even though $\gZ^i \Phi_Z$ and $\gX^i \Phi_X$ separately depend upon $R$.  
This has the important and interesting implication that the scattering of particles from the string is unchanged in the decoupling limit $R \gg 1$.  

As an example, let us consider the scattering of a few elementary particles from the $(0,1)$ dark string.  
Upon setting $n=0$ and $m=1$ in \eref{eq:ABphases} we see that the left- and right-chiral components have identical AB phases, \eg, $\thetauL = \thetauR \equiv \theta_u$.  
We calculate the transport cross section for the electron, proton, neutron, hydrogen atom, and neutrino as
\begin{align}
\label{eq:sigmat_e}
	\sigma_t \Bigr|_{e} 
	=& \frac{2}{\abs{\bf p}} \sin^2 \pi \theta_{e} 
	\approx \frac{1}{\abs{\bf p}} \frac{8 \pi^2 \cw^2 e^2}{\gX^2} \seps^2 + O(\seps^4)  \\
\label{eq:sigmat_p}
	\sigma_t \Bigr|_{p} 
	=& \frac{2}{\abs{\bf p}} \sin^2 \pi (2\theta_{u} + \theta_d)
	\approx \frac{1}{\abs{\bf p}} \frac{8 \pi^2 \cw^2 e^2}{\gX^2} \seps^2 + O(\seps^4)  \\
\label{eq:sigmat_n}
	\sigma_t \Bigr|_{n} 
	=& \frac{2}{\abs{\bf p}} \sin^2 \pi (\theta_{u} + 2 \theta_{d})
	= 0 \\
\label{eq:sigmat_H}
	\sigma_t \Bigr|_{H} 
	=& \frac{2}{\abs{\bf p}} \sin^2 \pi (2\theta_{u} + \theta_{d} + \theta_{e})
	= 0 \\
\label{eq:sigmat_nu}
	\sigma_t \Bigr|_{\nu} 
	=& \frac{2}{\abs{\bf p}} \sin^2 \pi \thetanL
	= 0 \com
\end{align}
respectively.  
In the second equalities of Eqns.~(\ref{eq:sigmat_e})~and~(\ref{eq:sigmat_p}) we have expanded for $
\seps \ll 1$.  
In performing this expansion, both terms in \eref{eq:phase_def} are of the same order because $\Phi_Z \sim \geLX \sim \geRX = O(\seps^1)$ and $\Phi_X \sim \geLZ \sim \geRZ = O(\seps^0)$.  
After recombination, when the SM particle content of the universe consists mainly of neutral hydrogen and neutrinos, the AB interactions vanish.  
Then, scattering arises from the typically subdominant hard-core interaction between the fermions and the Higgs and Z boson condensates on the string.  
If additionally $\alpha \to 0$, then even this interaction vanishes and the string does not feel the SM fermions at all.  

If the original model had contained fermion fields charged under the $\U{1}_X$, for example a dark matter candidate, then the interactions of these particles with the string would not vanish even as $\seps, \alpha \to 0$.  
For example, let $\Psi$ be a Dirac spinor field with gauge interactions specified by the covariant derivative 
\begin{align}
	D_{\mu} \Psi 
	= \bigl( \partial_{\mu} - i \gX \frac{\qX}{2} \hat{X}_{\mu} \bigr) \Psi
	= \bigl( \partial_{\mu} - i (\gdmA A_{\mu} + \gdmZ Z_{\mu} + \gdmX X_{\mu} ) \bigr) \Psi
\end{align}
where
\begin{align}\label{eq:DM_couplings}
	\gdmA = 0
	\qquad , \qquad
	\gdmZ = \frac{\gX \qX}{2 \ceps} \sxi
	\qquad , \quad {\rm and} \qquad
	\gdmX = \frac{\gX \qX}{2 \ceps} \cxi \per 
\end{align}
Its AB phase is simply $\theta_{\Psi} = m \qX$ and the AB interaction is found to be 
\begin{align}
	\sigma_t \Bigr|_{\Psi} 
	= \frac{2}{\abs{\bf p}} \sin^2 \pi \qX \per
\end{align}
If $\qX$ is an integer, then the transport cross-section vanishes and there is no AB interaction between the dark string and $\Psi$.  

\section{Conclusion}\label{sec:Conclusion}
\numberwithin{equation}{section}


We have studied the properties and couplings of the dark string including, for the first time, 
the full electroweak gauge sector, the gauge kinetic mixing, and Higgs portal interaction.  

The dark string solution field profiles are discussed in Sec.~\ref{sub:StringSoln}. 
The ansatz of the dark string can include a non-topological winding of the electroweak Higgs,
labeled by an integer $n$,
in addition to the topological winding of the new scalar field, $S$, given by an integer
$m$. We have evaluated $(n,m)=(0,1), (1,1)$ classes of solutions. Since the (0,1) string
is lighter, and there is no topology protecting the (1,1) solution, we expect that the (1,1)
solution will be unstable to decay into the (0,1) solution. Hence, we mainly focus on the
(0,1) string which we have also referred to as the ``dark string''.

In Sec.~\ref{sub:Tension} we have evaluated the tension of the dark string and the results 
can be summarized in the formula
\begin{equation}
	\mu \approx  2\pi \frac{\kappa^{1/4}}{\gX^{1/2}} \sigma^2 \left[1 +  \frac{\eta^2}{\sigma^2} O ( \alpha^2 , \seps^2 ) \right]
\end{equation}
where the approximate dependencies are derived from the plots in 
Fig.~\ref{fig:tension2} for small values of the hidden sector scalar self-coupling, $\kappa$,
the gauge kinetic mixing parameter, $\seps$, the Higgs portal coupling $\alpha$, and the dark gauge coupling, $g_X$.  
In the decoupling limit when the electroweak VEV is much less than the hidden sector VEV, $\eta \ll \sigma$, the expression reduces to that of a Nielsen-Olesen string.

\begin{figure}
	\includegraphics[width=0.2\textwidth]{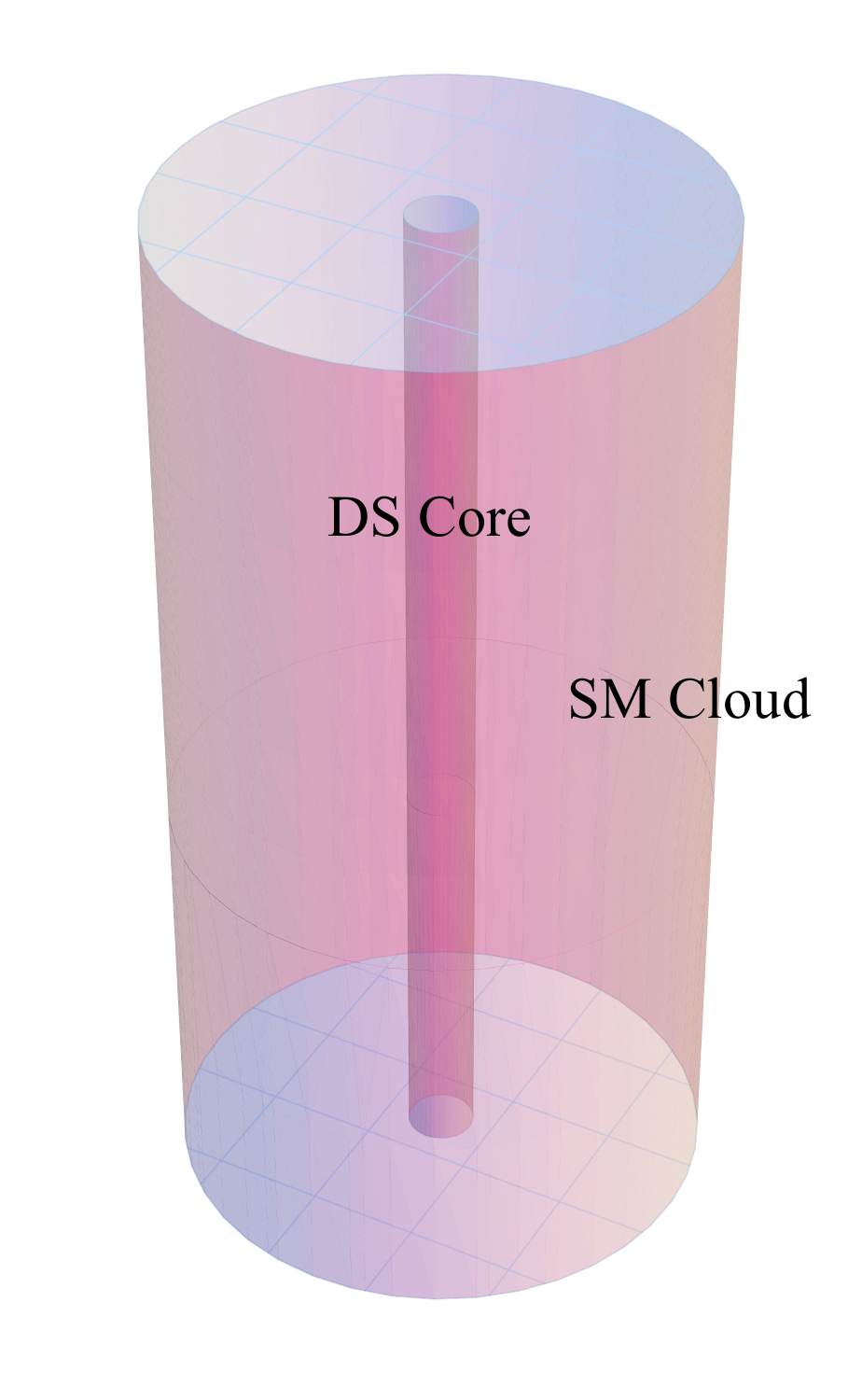}
	\caption{\label{fig:structure} Structure of the dark string.  }
\end{figure}

A novel feature of the dark string is that it also carries a condensate of the
electroweak Higgs and $Z$ fields. The structure of the string is a core of size
$\sim M_X^{-1}$ that contains flux of the dark gauge field $X$ and in which the new scalar
$S$ departs from its VEV. This is just as in the case of the Nielsen-Olesen string.
Around the Nielsen-Olesen core we also have a ``cloud'' or ``dressing'' of Higgs and $Z$ 
fields that extend out to a radius $\sim M_H^{-1}$ as illustrated in Fig.~\ref{fig:structure}.

The presence of the electroweak cloud can be of phenomenological importance because
it connects a topological defect in the dark sector to the matter content of the visible
sector. In particular, an oscillating loop of dark string may be expected to copiously 
radiate Higgs bosons \cite{Vachaspati:2009kq} and $Z$ gauge bosons (similar to 
Goldstone boson radiation discussed in \cite{Vilenkin:1986ku}).  
With these effects in mind, we have proposed effective interactions of the dark string with the Higgs excitations, $\phi_H$, and $Z$ bosons that take the form
\begin{equation}
S_{\rm int} = \gHstr \, \eta   \, \int d^2\sigma \sqrt{-\gamma} \ \phi_H({\mathbb X}^\mu) 
                    + \frac{\gZstr}{2} \left( \frac{\eta}{\sigma} \right)^2 \int d\sigma^{\mu\nu} Z_{\mu\nu}({\mathbb X}^{\mu})
\end{equation}
given by Eqs.~(\ref{eq:SstrH})~and~(\ref{eq:SstrZ}).  
The first term carries a factor of $\eta$ because the emission of a single Higgs boson can only occur after electroweak symmetry breaking.  
The factor of $(\eta / \sigma)^2$ in the second term reflects the suppressed interaction of the Z boson with the hidden sector fields in the decoupling limit where the gauge sector mixing is small.  
The coupling constants in these interactions are shown in Figs.~\ref{fig:hc}, \ref{fig:coupling_decoupling}, and \ref{fig:zc}.  
In the decoupling limit they can be approximated as in Eqs.~(\ref{eq:gHstr_dec})~and~(\ref{eq:gZstr_dec}) by 
\begin{align}
	(\gHstr)^{\rm (dec.)} & \simeq \left( \frac{e^2 \pi}{\sqrt{2} \, \cw^2 \gX^2} \right) \seps^2 + \frac{\pi}{15 \sqrt{2}} \left( \frac{64}{\kappa} - 17 \xi_{\rm max}^2 \right) \alpha \nn
	(\gZstr)^{\rm (dec.)} & \simeq - \left( \frac{11e^2 \pi}{36 \cw^2 \sw \gX} \xi_{\rm max}^2 \right) \seps \per
\end{align}
up to terms of order $\eta^2 / \sigma^2$.  The parameter $\xi_{\rm max} = O(1)$ is the rescaled width of the profile functions.  

The gauge kinetic mixing term in the model also leads to an Aharonov-Bohm interaction
between fermions and the dark string \cite{Vachaspati:2009kq}. These interactions
are important since, in a cosmological setting, the strings are surrounded by a plasma
of fermions that can scatter and affect the evolution of the string network. In addition,
the Aharonov-Bohm interaction will allow for dark string loops to radiate standard model
fermions \cite{Chu:2010zzb}. We give the Aharonov-Bohm phases
for the fermions in Eq.~(\ref{eq:ABphases}), where we should set $n=0$ for the
(0,1) string. The result is simply that the Aharonov-Bohm phase of a fermion with
electric charge $q$ is
\begin{equation}
\theta_q = - \frac{2 c_w s_\epsilon}{g_X} q .
\end{equation}
Following Ref.~\cite{Alford:1988sj}, 
we have also calculated the transport cross sections for fermions scattering
off dark strings in Sec.~\ref{sub:Coupling_to_Fermions}.

Having mapped out the properties of the dark string, we plan to explore their 
cosmological consequences and phenomenological connections in future work.

\acknowledgments
We are very grateful to Eray Sabancilar for many helpful discussions.  
This work was supported by the Department of Energy at ASU.  

\begin{appendix}
\numberwithin{equation}{section}

\section{Limit of Small Gauge Kinetic Mixing}\label{app:Smallepsilon}

In this appendix we consider the limit that the GKM coupling is small, $s_{\epsilon} \ll 1$, for various quantities in the text:  
The mixing angle [\eref{eq:def_tan2xi}]
\begin{align}
	\tan 2 \zeta & \approx -\frac{2 \sw}{R^2 -1} s_{\epsilon} + \frac{\sw}{R^2-1} \left( 1 + \frac{2(1+\sw^2)}{R^2-1} \right) s_{\eps}^3 + O(s_{\epsilon}^5) \\
	\txi & \approx - \frac{\sw}{R^2-1} \seps + \frac{\sw}{2(R^2-1)} \left( 1 + \frac{2(1+\sw^2)}{R^2-1} + \frac{2 \sw^2}{(R^2-1)^2} \right) \seps^3 + O(\seps^5) \\
	\sxi & \approx - \frac{\sw}{R^2-1} \seps + \frac{\sw}{2(R^2-1)} \left( 1 + \frac{2(1+\sw^2)}{R^2-1} + \frac{3 \sw^2}{(R^2-1)^2} \right) \seps^3 + O(\seps^5) \\
	\cxi & \approx 1 - \frac{\sw^2}{2(R^2-1)^2} \seps^2 + \frac{\sw^2}{2(R^2-1)^2} \left(1 + \frac{2(1+\sw^2)}{R^2-1} + \frac{11 \sw^2}{4(R^2-1)^2} \right) \seps^4 + O(\seps^6)  \com
\end{align}
the gauge boson mixing matrix [\eref{eq:Mdef}] 
\begin{align}
	{\bf M} \approx \begin{pmatrix} \cw & -\sw - \frac{\sw \left( 1 + \cw^2 - 2 R^2 \right)}{2(R^2-1)^2} \seps^2 & - \frac{R^2 - \cw^2}{R^2 - 1} \seps \\ 
	\sw & \cw - \frac{\sw^2 \cw}{2(R^2-1)^2} \seps^2 & \frac{\sw \cw}{R^2-1} \seps \\ 
	0 & - \frac{\sw}{R^2-1} \seps & 1 + \frac{\cw^2 - 2 R^2 + R^4}{2(R^2-1)^2} \seps^2 \end{pmatrix} + O(\seps^3) \com 
\end{align}
the gauge couplings of the scalars [\eref{eq:scalar_gauge_couplings}]
\begin{align}\label{eq:approx_scalar_gauge_couplings}
\begin{array}{lclcl}
	\gPhipA = e
	& \qquad & 
	\gPhipZ \approx \frac{e}{2} \left( \frac{1}{t_w} + t_w \right) + \frac{e}{2} \frac{t_w}{R^2-1} \left( 1 + \frac{c_w^2 - s_w^2}{2(R^2-1)} \right) \seps^2
	& \qquad & 
	\gPhipX \approx -\frac{e}{2} \frac{1}{\cw} \left( 1 - \frac{\cw^2-\sw^2}{R^2-1} \right) \seps \\
	\gHA =0
	& \qquad & 
	\gHZ \approx -\frac{e}{2} \frac{1}{\cw \sw} + \frac{e}{2} \frac{\tw}{R^2-1} \left( 1 + \frac{1}{2(R^2-1)}  \right) \seps^2
	& \qquad & 
	\gHX \approx - \frac{e}{2} \frac{1}{\cw} \left( 1 + \frac{1}{R^2-1} \right) \seps \\
	\gSA = 0
	& \qquad & 
	\gSZ \approx -\frac{\gX}{2} \frac{\sw}{R^2-1} \seps 
	& \qquad & 
	\gSX \approx \frac{\gX}{2} + \frac{\gX}{4} \left( 1 - \frac{\sw^2}{(R^2-1)^2} \right) \seps^2 
\end{array} 
\end{align}
up to order $O(\seps^3)$ corrections, the gauge boson mass eigenvalues [\eref{eq:gauge_evals}]
\begin{align}
\begin{array}{l}
	M_{Z}^2 \approx m_{Z}^2 \left( 1 - \frac{\sw^2}{R^2-1} \seps^2 + O(\seps^4) \right) \\
	M_{X}^2 \approx m_{X}^2 \left( 1 + \frac{R^2 - \cw^2}{R^2-1} \seps^2 + O(\seps^4) \right) 
\end{array} \com
\end{align}
the string profile boundary conditions [\eref{eq:gaugeBC}],
\begin{align}
\begin{array}{l}
	\Zpro(\infty) \approx - \frac{\sqrt{2} \eta}{m_Z} n - \frac{\sqrt{2} \sw R^2 \sigma}{m_X(R^2-1)} m \, \seps + \frac{\sw^2 \eta}{\sqrt{2} m_Z (R^2-1)^2} n \, \seps^2 + O(\seps^3) \\
	\Xpro(\infty) \approx \frac{\sqrt{2} \sigma}{m_X} m - \frac{\sqrt{2} \sw \eta}{m_Z(R^2-1)} n \, \seps + \frac{(-\cw^2 + 2 \cw^2 R^2 - R^4)\sigma}{\sqrt{2}m_X(R^2-1)^2} m \, \seps^2 + O(\seps^3) \com
\end{array}
\end{align}
the gauge couplings of the fermions [\eref{eq:fermion_couplings}]
\begin{align}
\hspace{-1cm}
\begin{array}{lclcl}
	\guLA = \frac{2e}{3} \com 
	& \hspace{0.02cm} & 
	\guLZ \approx \frac{e}{6} \left( \frac{3}{t_w} - t_w \right) + \frac{e}{6} \frac{s_w}{R^2-1} \left[ \frac{1}{c_w} - \frac{s_w}{2(R^2-1)} \left( \frac{3}{t_w} - t_w \right) \right] \seps^2 \com
	& \hspace{0.02cm} & 
	\guLX \approx -\frac{e}{6} \left[ \frac{1}{c_w} - \frac{s_w}{R^2-1} \left( \frac{3}{t_w} - t_w \right) \right] \seps \com \\
	\guRA = \frac{2e}{3} \com
	& \hspace{0.02cm} & 
	\guRZ \approx - \frac{2e}{3} t_w + \frac{2e}{3} \frac{t_w}{R^2-1} \left[ 1 + \frac{\sw^2}{2(R^2-1)} \right] \seps^2 \com
	& \hspace{0.02cm} & 
	\guRX \approx -\frac{2e}{3} \left( \frac{1}{\cw} + \frac{\sw^2}{R^2-1} \right) \seps \com \\
	\gdLA = -\frac{e}{3} \com
	& \hspace{0.02cm} & 
	\gdLZ \approx -\frac{e}{6} \left( \frac{3}{t_w} + t_w \right) + \frac{e}{6} \frac{\sw}{R^2-1} \left[ \frac{1}{c_w} + \frac{\sw}{2(R^2-1)} \left( \frac{3}{\tw} + \tw \right) \right] \seps^2 \com
	& \hspace{0.02cm} &
	\gdLX \approx - \frac{e}{6} \left[ \frac{1}{c_w} + \frac{s_w}{R^2-1} \left( \frac{3}{t_w} + t_w \right) \right] \seps \com \\
	\gdRA = -\frac{e}{3} \com
	& \hspace{0.02cm} & 
	\gdRZ \approx \frac{e}{3} t_w - \frac{e}{3} \frac{\tw}{R^2-1} \left[ 1 + \frac{\sw^2}{2(R^2-1)} \right] \seps^2 \com
	& \hspace{0.02cm} & 
	\gdRX \approx \frac{e}{3} \left( \frac{1}{c_w} + \frac{\sw^2}{R^2-1} \right) \seps \com \\
	\geLA = -e \com
	& \hspace{0.02cm} & 
	\geLZ \approx -\frac{e}{2} \left( \frac{1}{t_w} - t_w \right) - \frac{e}{2} \frac{\sw}{R^2-1} \left[ \frac{1}{\cw} - \frac{\sw}{2(R^2-1)} \left( \frac{1}{\tw} - \tw  \right)  \right] \seps^2 \com
	& \hspace{0.02cm} & 
	\geLX \approx \frac{e}{2} \left[ \frac{1}{c_w} - \frac{\sw}{R^2-1} \left( \frac{1}{t_w} - t_w \right) \right] \seps \com \\
	\geRA = -e \com
	& \hspace{0.02cm} & 
	\geRZ \approx e t_w - e \frac{\tw}{R^2-1} \left[ 1 + \frac{\sw^2}{2(R^2-1)} \right] \seps^2 \com
	& \hspace{0.02cm} & 
	\geRX \approx e \left( \frac{1}{c_w} + \frac{\sw^2}{R^2-1} \right) \seps \com \\
	\gnLA = 0 \com
	& \hspace{0.02cm} & 
	\gnLZ \approx \frac{e}{2} \left( \frac{1}{t_w} + t_w \right) - \frac{e}{2} \frac{\sw}{R^2-1} \left[ \frac{1}{\cw} + \frac{\sw}{2(R^2-1)} \left( \frac{1}{\tw} + \tw \right)  \right] \seps^2 \com
	& \hspace{0.02cm} & 
	\gnLX \approx \frac{e}{2} \left[ \frac{1}{c_w} + \frac{s_w}{R^2-1} \left( \frac{1}{t_w} + t_w \right) \right] \seps 
\end{array}
\end{align}
up to order $O(\seps^3)$ corrections, and the magnetic flux [Eqs.~(\ref{eq:PhiZ}) and (\ref{eq:PhiX})]
\begin{align}
	\Phi_Z &\approx - 2 \pi \frac{\sqrt{2} \eta}{m_Z} n - 2 \pi \frac{\sqrt{2} \sw R^2 \sigma}{m_X(R^2-1)} m \, \seps + O(\seps^2) \\
	\Phi_X &\approx 2 \pi \frac{\sqrt{2} \sigma}{m_X} m - 2 \pi \frac{\sqrt{2} \sw \eta}{m_Z(R^2-1)} n \, \seps + O(\seps^2) \per
\end{align}

\section{Numerical Solution of Field Equations}\label{app:numerical_method}

The dark string is the solution of the system of equations given by \ref{eq:darkstring_eqns} along with the boundary conditions in Eqs.~(\ref{eq:BC1b}) and (\ref{eq:gaugeBC}).  
We solve these equations numerically using the Fortran solver {\it Colnew}, which implements collocation to solve boundary value problems (BVPs) involving systems of ordinary differential equations (ODEs) \cite{Bader:1987}.  
In order to obtain convergence, nonlinear BVPs frequently require a very good initial guess as input to an iterative method of solution, and this is the case with our problem.\\

We obtain this using the method of continuation \cite{Ascher:1987}. In the absence of the HP and GKM operators, the dark and standard model sectors decouple. In the $(1,1)$ case this reduces to two independent Nielsen--Olesen strings, and in the $(0,1)$ case this reduces to a Nielsen--Olesen string along with a vacuum solution. In either case, their solution is straightforward. We then use continuation, which relies on the following observation: given two sets of model parameters whose values are very close, we expect the corresponding solutions of \eref{eq:darkstring_eqns} to be nearly identical. Thus, we begin with the solution to the decoupled problem and then solve the system of equations with the HP or GKM small but nonzero. This is the beginning of a series of problems, each using the previous solution as {\it Colnew's} initial guess and returning a solution for incrementally larger HP and GKM. The final step in this procedure solves \eref{eq:darkstring_eqns} for the desired choice of parameters.\\

We impose the $\xi=\infty$ boundary conditions at some $\xi_{\infty}$ and solve numerically on $[0,\xi_{\infty}]$. When $\sigma$ and $\eta$ are comparable, $\xi_{\infty}$ of 200 to 400 is typically sufficient to ensure that the profiles and relevant integrals (tension and couplings) are insensitive to the value of $\xi_{\infty}$. In the $(1,1)$ case and for $\sigma \gg \eta$ we begin with two Nielsen--Olesen strings varying on significantly different scales (as in \fref{fig:profiles_10tev}, for example), and we use $\xi_{\infty}$ of order 1000.
\end{appendix}


\end{document}